\documentclass[aps,prd,11pt,a4paper,nofootinbib,oneside,superscriptaddress]{revtex4-1}
\pdfoutput = 1

\usepackage{amsmath,amssymb,amsfonts,color}
\usepackage{tensor,slashed,paralist,cases,mathrsfs}
\usepackage{float,cancel,xcolor}
\usepackage{graphicx}
\usepackage{dcolumn}
\usepackage{bm}

\usepackage[colorlinks=true,urlcolor=blue,linkcolor=blue,citecolor=blue,linktocpage=true]{hyperref}

\newcommand{\non}{\nonumber}
\newcommand{\beq}{\begin{eqnarray}}
\newcommand{\eeq}{\end{eqnarray}}
\newcommand{\bpmatrix}{\begin{pmatrix}}
\newcommand{\epmatrix}{\end{pmatrix}}
\newcommand{\ba}{\begin{array}}
\newcommand{\ea}{\end{array}}

\renewcommand{\eqref}[1]{Eq.~(\ref{#1})}

\newcommand{\bc}{\begin{center}}
\newcommand{\ec}{\end{center}}

\newcommand*{\defeq}{\stackrel{\small{\mathsf{def}}}{=}}

\begin{document}

\vspace*{1.5em}

\title{Thermodynamic Evolution of Secluded Vector Dark Matter: Conventional WIMPs and Nonconventional WIMPs}

\author{Kwei-Chou Yang}
\email{kcyang@cycu.edu.tw}

\affiliation{Department of Physics and Center for High Energy Physics, Chung Yuan Christian University, Taoyuan 320, Taiwan}


\begin{abstract}
The secluded dark matter resides within a hidden sector and self-annihilates into lighter mediators which subsequently decay to the Standard Model (SM) particles. Depending on the coupling strength of the mediator to the SM, the hidden sector can be kinetically decoupled from the SM bath when the temperature drops below the mediator's mass, and the dark matter annihilation cross section at freeze-out is thus possible to be boosted above the conventional value of weak interacting massive particles.
We present a comprehensive study on thermodynamic evolution of the hidden sector from the first principle, using the simplest secluded vector dark matter model. 
Motivated by the observation of Galactic center gamma-ray excess, we take two mass sets $\sim{\cal O}(80\, \text{GeV})$ for the dark matter and mediator  as examples to illustrate the thermodynamics.
The coupled Boltzmann moment equations for number densities and temperature evolutions of the hidden sector are numerically solved. The formalism  can be easily extended to a general secluded dark matter model. 
We show that a long-lived mediator can result in a boosted dark matter annihilation cross section to account for the relic abundance.
We further show the parameter space which provides a good fit to the Galactic center excess data and is compatible with the current bounds and LUX-ZEPLIN  projected sensitivity.   We find that the future observations of dwarf spheroidal galaxies offer promising reach to probe the most relic allowed parameter space relevant to the boosted dark matter annihilation cross section.

\end{abstract}
\maketitle
\newpage

\section{Introduction}

Motivated by particle physics, the theoretical studies and experimental searches have for many decades focused on the popular class of the dark matter (DM) candidates, called the weakly interacting massive particles (WIMPs). In the WIMP scenario, when the dark matter becomes nonrelativistic, its comoving number density is exponentially depleted through Boltzmann suppression and keeps the thermal equilibrium with the bath until freeze-out. 
The resulting DM with the weak scale interaction and mass can provide the correct relic abundance today.   

Many  DM experiments are thus motivated by the WIMP scenario.
Nevertheless, no conclusive observations have been made by the direct detection searches, Large Hadron Collider (LHC),  and other collider experiments. 
Several groups have reported the GeV gamma-ray excess around the Galactic center (GC)  \cite{Goodenough:2009gk, Hooper:2010mq, Hooper:2011ti, Abazajian:2012pn, Gordon:2013vta, Huang:2013pda, Daylan:2014rsa, Calore:2014xka, Calore:2014nla,Karwin:2016tsw,TheFermi-LAT:2017vmf}, for which, however,  the allowed WIMP dark matter models have been also severely constrained by the current null results of the direct detection \cite{Aprile:2017iyp,Cui:2017nnn,Akerib:2016vxi,Akerib:2018lyp} and collider experiments. In light of these measurements, an interesting paradigm that goes beyond the ``conventional" WIMP scenario and becomes more and more popular is known as ``secluded (WIMP) dark matter".  In this paradigm, the dark matter candidate may reside within one of the hidden sectors and  communicates with the visible sector through a lighter metastable mediator, which weakly couples the standard model (SM) to the WIMP.
As such, the DM signals, suppressed at the direct detection and colliders, could be observable in indirect measurements \cite{Pospelov:2007mp,Ko:2014gha,Berlin:2014pya,Escudero:2017yia,Ko:2014loa,Abdullah:2014lla,Martin:2014sxa,Kim:2016csm,Yang:2017zor,Profumo:2017obk}.

The mechanism for the secluded WIMP dark matter was discussed by Pospelov, Ritz, and Voloshin \cite{Pospelov:2007mp}. 
In this mechanism, the WIMP can still be a thermal relic, and the dominant DM annihilation channel is into a pair of unstable mediators which ultimately decay into SM particles. 
Basically, for this model, as long as the mediator decays before the beginning of the big bang nucleosynthesis (BBN), the effective number of neutrino species and abundance of helium and deuterium will not be modified, as compared with the standard BBN, so that the result can be easily compatible with the current Planck measurement  \cite{Ade:2015xua}.

As for building secluded DM models, many people restricted their works to the parameter space relevant to the WIMP scenario where the hidden sector is in chemical and thermal equilibrium with the bath prior to freeze-out ~\cite{Ko:2014gha,Berlin:2014pya,Escudero:2017yia,Ko:2014loa,Abdullah:2014lla,Martin:2014sxa,Kim:2016csm,Yang:2017zor}. 
However, for the case that the dark sector has kinetically decoupled from the bath,  due to its weak couplings to the SM particles, before it becomes nonrelativistic, if the secluded DM annihilates into nearly degenerate mediators which later decay out-of-equilibrium with the bath, the DM density will be exponentially depleted through the decay process of the mediator, instead of following Boltzmann suppression \cite{Dror:2016rxc}.  Moreover,  during the period of time in which the dark sector is out of thermal equilibrium with the bath, if the $3\to 2$ number changing interactions are allowed and efficiently active, the hidden sector can first  undergo an epoch called ``cannibalism". See the related discussions in Refs. \cite{Farina:2016llk,Pappadopulo:2016pkp,Yang:2018fje,Berlin:2016gtr}. Alternatively, $3\to 2$ DM annihilation mechanism is  also relevant to the strongly-interacting massive particles (or called SIMP) \cite{Hochberg:2014dra} and elastically decoupling relic (or called ELDER) \cite{Kuflik:2015isi,Kuflik:2017iqs} scenarios.

In this paper, to have a thorough understanding about the thermodynamics of the secluded  dark matter from the first principle, we will study the simplest secluded vector dark  matter model, taken as an example in which the vector dark matter and the mediator within the hidden sector are in thermal equilibrium with each other before freeze-out, but may be kinetically decoupled from the SM bath at temperature $T\sim m_{X,S}$, depending on the couplings to the SM, where $m_X$ and $m_S$ are the masses of the DM and hidden scalar, respectively.

We separately obtain the evolution equations of number densities and temperatures for the hidden species, by taking suitable moments of the Boltzmann equation.
We will give a detailed result of describing chemical and kinetic decouplings of the hidden sector from the thermal bath. 
We will show that, depending on the coupling strength of the mediator to the SM, the relic annihilation cross section is likely to be boosted above the conventional WIMP value.
The present study can be easily generalized to a generic case.

Using two mass sets: (i) $m_X=80$~GeV, $m_S=0.8 m_X$, and (ii) $m_X=80$~GeV, $m_S =0.99 m_X$, we numerically solve the thermodynamic evolution of the hidden sector, which can be either in thermal equilibrium or out of equilibrium  with the bath before the DM freezes out, and moreover, is secluded from the visible sector with small interaction rates compatible with colliders and direct detection bounds. Use of the present mass sets of the hidden sector is motivated by the observed GC gamma-ray excess which can be accounted for by this model via one-step cascade annihilation \cite{Ko:2014gha,Escudero:2017yia}. More detailed discussions about the GC allowed region, which are constrained by the astrophysical and cosmological measurements as well as the  LUX-ZEPLIN projected sensitivity \cite{Akerib:2018lyp},  will be presented in Sec.~\ref{sec:discussions}.

The rest of this paper is organized as follows. In Sec.~\ref{sec:model}, we start with an introduction of the vector DM model which is UV-complete. In this model, the hidden sector contains an abelian vector dark matter and a complex scalar. The former is a gauge boson associated with a dark (hidden) gauge symmetry $U_X(1)$, while the latter is charged under  $U_X(1)$.
 In Sec.~\ref{sec:dd-colliders},  the model parameters constrained by direct detection and collider experiments will be described first. In Sec.~\ref{sec:thermodynamics}, we present a general description of Boltzmann equation in the framework of an expanding Universe which is homogeneous and isotropic. We further consider the moments of Boltzmann equations that are relevant to the evolutions of the number densities and temperatures for the hidden species. In Sec.~\ref{sec:analysis},  two sets of mass parameters which can account for the GC gamma-ray excess are used in the numerical analyses. The results are  given and discussed.
The parameter space relevant to the GC gamma-ray excess and concerning the current limits and prospects are further discussed in
Sec.~\ref{sec:discussions}.
 In Sec.~\ref{sec:conclusions}, we draw the conclusions. All technical derivations are collected in Appendices.

\section{The Model}\label{sec:model}

The simplest secluded vector dark matter can be made of the abelian gauge bosons, $X_\mu$'s, which get the mass from the vacuum expectation value (VEV) of the hidden complex scalar field $\Phi_S$ due to the spontaneously dark gauge symmetry $U_X(1)$ breaking, where the $Z_2$ symmetry, $X_\mu \to -X_\mu$ and $\Phi_S \to \Phi_S^*$, is imposed to stabilize the dark matter  \cite{Ko:2014gha}.
The relevant kinetic Lagrangian (${\cal L}_{\text{kinetic}}$) and the scalar potential (${\cal L}_{\text{scalar}})$ are given by
\begin{align}
&{\cal L}_\text{kinetic} \supset -\frac{1}{4} X_{\mu\nu} X^{\mu\nu} + (D_\mu\Phi_S)^\dagger (D^\mu \Phi_S) \,, \\
&{\cal L}_\text{scalar} =
- \mu_{H}^2 |\Phi_H|^2 - \mu_{S}^2 |\Phi_S|^2 
 - \frac{\lambda_H}{2} (\Phi_H^\dagger \Phi_H)^2 
 - \frac{\lambda_S}{2} (\Phi_S^\dagger \Phi_S)^2 
 - \lambda_{HS} (\Phi_H^\dagger \Phi_H) (\Phi_S^\dagger \Phi_S)  \;,
\label{eq:lagrangian}
\end{align}
where $\Phi_H= (H^+, H^0)^{\rm T}$ is the SM Higgs doublet, $X_{\mu\nu} =\partial_\mu X_\nu -\partial_\nu X_\mu$, and the covariant derivative is defined as $
D_\mu \Phi_S = (\partial_\mu + i g_{\rm dm} Q_{\Phi_S} X_\mu )\Phi_S$. Here $g_{\rm dm}$ is the  gauge coupling and $Q_{\Phi_S}$ is the $U_{\rm dm}(1)$ charge of $\Phi_S$.  After spontaneous symmetry breaking, the Higgs fields develop  non-zero VEV's,
\begin{equation}
\Phi_H=\frac{1}{\sqrt{2}} (v_H + \phi_h + i \sigma_h), \quad
\Phi_S=\frac{1}{\sqrt{2}} (v_S + \phi_s + i \sigma_s),
\label{eq:vev}
\end{equation}
where the CP-odd states, $\sigma_h$ and $\sigma_s$, respectively becomes the longitudinal components of the $Z$ boson and $X_\mu$; the dark matter thus obtain a mass, $m_X=g_{\rm dm} Q_{\Phi_S} v_S$. In the present paper, we will simply use $Q_{\Phi_S} =1$.

The scalar fields $(\phi_h, \phi_s)$ can be expressed in terms of mass eigenstates of physical Higgses $(h, S)$ as
\begin{align}
\left(\begin{array}{c}
\phi_h \\  \phi_s
\end{array}\right) 
&=
\left(\begin{array}{cc}
\cos\alpha & -\sin\alpha \\
\sin\alpha & \cos\alpha
\end{array}\right)
\left(\begin{array}{c}
h  \\  S
\end{array}\right),  
\end{align}
and the mass squared matrix in the former basis can be parametrized in terms of masses of the latter and the mixing angle $\alpha$,

\begin{equation}
\left(
\begin{array}{cc}
 \lambda_H v_H^2 & \lambda_{HS} v_S v_H \\ 
\lambda_{HS} v_S v_H &  \lambda_S v_S^2
\end{array}
\right) =
\left(
\begin{array}{cc}
 m_h^2 c_\alpha^2  + m_S^2 s_\alpha^2  & (m_h^2 -m_S^2 )s_\alpha c_\alpha \\ 
(m_h^2 -m_S^2 )s_\alpha c_\alpha & m_S^2 c_\alpha^2 + m_h^2 s_\alpha^2
\end{array}
\right) \,.
\label{eq:mass_matrix}
\end{equation}
Here and throughout the paper, we adopt the abbreviations: $s_\alpha \equiv \sin\alpha$ and $c_\alpha \equiv \cos\alpha$.
Using $v_H\simeq 246$~GeV and $m_h=125.18$~GeV \cite{pdg2018}, we will take $m_X , m_S$, $g_{\rm dm}$ and $\alpha$ as the independent parameters in the following analysis.

\begin{figure}[t!]
\begin{center}
\includegraphics[width=0.45\textwidth]{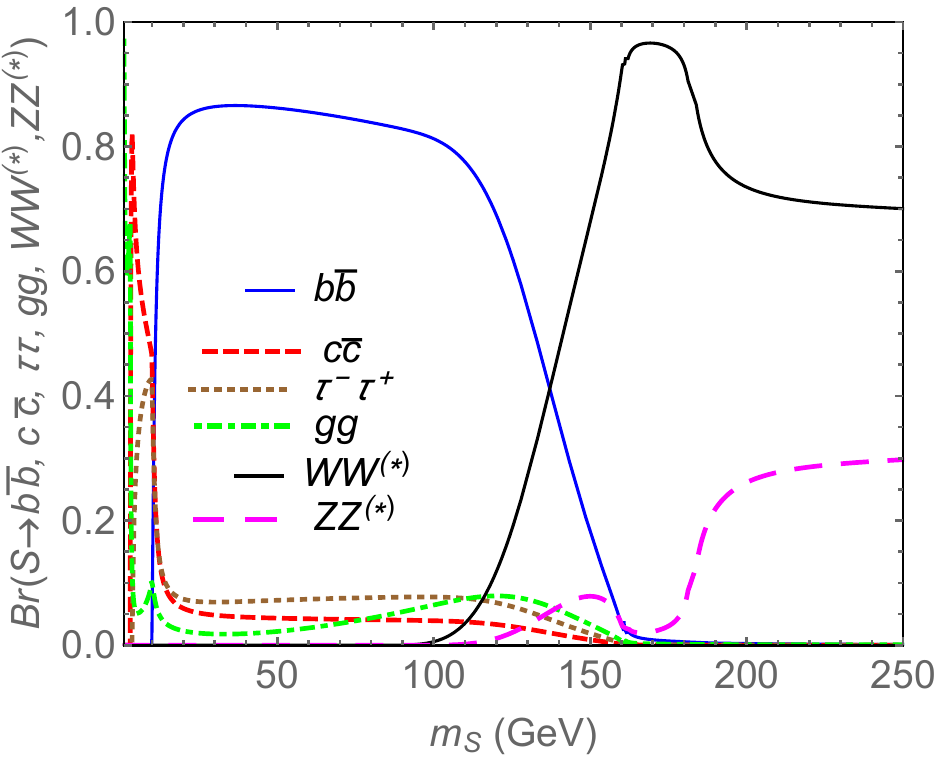}\hskip0.6cm
\includegraphics[width=0.452\textwidth]{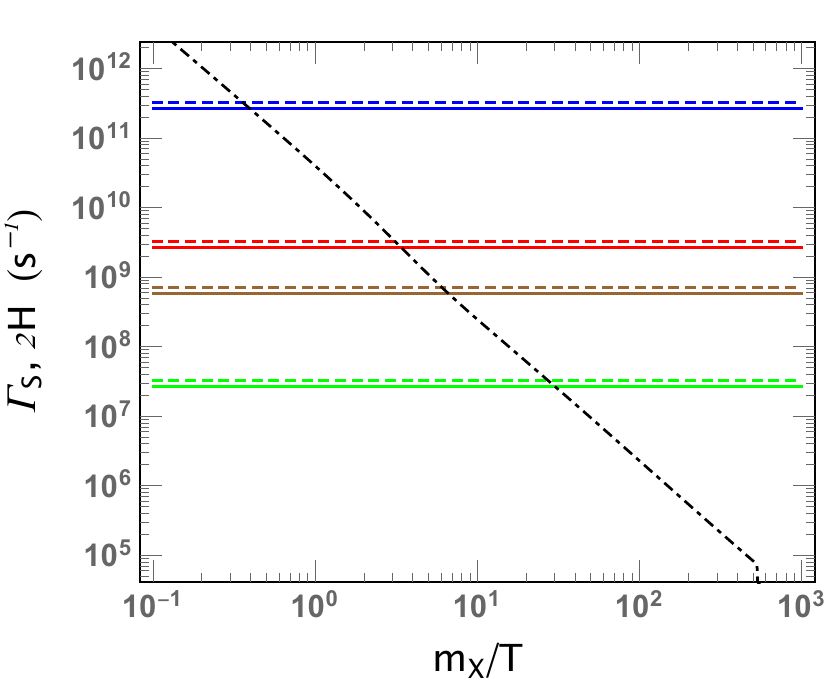}\\

\caption{ Left panel: The main branching ratios of the hidden mediator $S$ with $m_S <250$~GeV.
Right panel: The $S$ decay rate (total width), $\Gamma_S$, and twice of the Hubble rate, $2H$, as functions of $m_X/T$, where $T$ is the temperature of the bath. 
The horizontal lines from up to down with colors blue, red, brown, and green are the $S$ decay rates corresponding to $\alpha = 1\times 10^{-5}, 1\times 10^{-6}, 5\times 10^{-7}$, and $1 \times 10^{-7}$, respectively, while the black dotdashed line stands for the Hubble rate.  In this plot, using $m_X=$80~GeV, the solid and dashed lines correspond to $m_S=0.8\, m_X$ and $0.99\, m_X$, respectively.
}
\label{fig:BrS}
\end{center}
\end{figure}

The branching  ratios of the hidden scalar, $S$, with a mass of $m_S\lesssim 2m_h$, are depicted in the left panel of Fig.~\ref{fig:BrS}, where, in the range giving a good fit to the GC gamma-excess data, the scalar mass satisfies  $m_S \lesssim m_X \lesssim 130$~GeV. The related partial widths of the hidden scalar $S$ are summarized and discussed in Appendix~\ref{app:s-width}, where the results are relevant to the studies of the relic abundance and indirect detection searches.  
In the right panel of  Fig.~\ref{fig:BrS}, we show  several values of the $S$ decay width, $\Gamma_S$, compared with  the evolution of $2H$ (the inverse time interval of  the radiation dominated epoch), where $H$ is the Hubble rate which is given by the Friedmann equation,
\begin{align}
H = \left( \frac{8 \pi G}{3} \rho_t \right)^{1/2} \,,
\end{align}
with the total energy density being
\begin{align}
\rho_t = \frac{\pi^2}{30} \left( g_{\rm eff, SM}(T)\,  T^4 + g_{\rm eff, h}(T_h)\,  T_h^4\right) \,.
\end{align}
Here,  $g_{\rm eff, SM}$ and  $g_{\rm eff, h}$ are the effective relativistic degrees of freedom of the SM and hidden sector at the temperatures $T$ and $T_h$, respectively.
In Fig.~\ref{fig:BrS}, we have simply adopted $T_h= T$.
As will be shown in Eqs.~(\ref{eq:boltz-t}) and (\ref{eq:tde}), and discussed in Sec.~\ref{sec:analysis}(\ref{point:decay}), if the nonrelativistic scalar $S$ is kept in kinetic equilibrium with the thermal bath via its inverse decay $\text{SM}~\text{SM} \to S$, then this kinetic energy injection rate to $S$ will be larger than the Hubble cooling rate, i.e., roughly  $\Gamma_S\gtrsim 2H$.  In the right panel of  Fig.~\ref{fig:BrS}, the $S$ particles with the width corresponding to  $\alpha =1\times 10^{-5}$ or $1 \times 10^{-6}$  can be in thermal equilibrium with the bath
when $m_X/T \gtrsim 0.4$ or 3 (with $m_X$=80 GeV) (see Sec.~\ref{sec:analysis}(\ref{point:decay}), where a more precise estimation is given).
However, for the hidden scalar with a much smaller mixing angle $\alpha =5\times 10^{-7}$ or $1 \times 10^{-7}$, because the ratio of the Hubble cooling rate to heating rate, $\sim 2H n_S(T_S) T_S /  \big( \Gamma_S n_S^{\rm eq}(T) \, T \big)$,  is much larger than 1 due to the fact that $n_S(T_S) T_S  \gg n_S^{\rm eq}(T) T$ for the nonrelativistic $S$  (see Figs.~\ref{fig:relic-not-equal-2} and \ref{fig:relic-equal}), where $n_S(T_S)$ is the number density of $S$ at its temperature $T_S$, and  $n_S^{\rm eq}(T)$ is the equilibrium number density of $S$ at the corresponding bath temperature $T$, the hidden scalar thus starts to undergo out-of-equilibrium decay at the cosmological time $\simeq$ the $S$ lifetime, $(2H)^{-1} \simeq \Gamma_S^{-1}$. As shown in the right panel of  Fig.~\ref{fig:BrS}, the corresponding out-of-equilibrium temperature is about  $m_X/T \sim $ 6 or 30 for $\alpha =5\times 10^{-7}$ or $1 \times 10^{-7}$. The underlying physics and a more precise estimation will be given in  Sec.~\ref{sec:analysis}(\ref{item:decay}).

\section{Direct detection and LHC constraints}\label{sec:dd-colliders}

In this paper, we will use two sets of the masses for the dark matter and mediator: (i) $m_X=80$~GeV, $m_S=0.8m_X=64$~GeV, and (ii) $m_X =80$~GeV, $m_S=0.99 m_X=79.2$~GeV, to study the thermal evolution of the hidden sector.  These two sets can provide a good fit to the GC gamma-ray excess data.  For the first set,  when the hidden sector with a sizable mass gap undergoes the cannibal process, the down-scattering rate,
$XX \to SS$, can be significantly larger than the up-scattering one, $SS \to XX$. For the second set, the hidden sector is nearly degenerate, and can be further constrained by the gamma-line searches at the indirect detection. 
Moreover, because the low-velocity DM annihilation cross section is zero in $m_S \to m_X$ limit, a larger $X$-$S$ coupling is needed to account for the GC data and the DM relic abundance. 

\begin{figure}[t!]
\begin{center}
\includegraphics[width=0.39\textwidth]{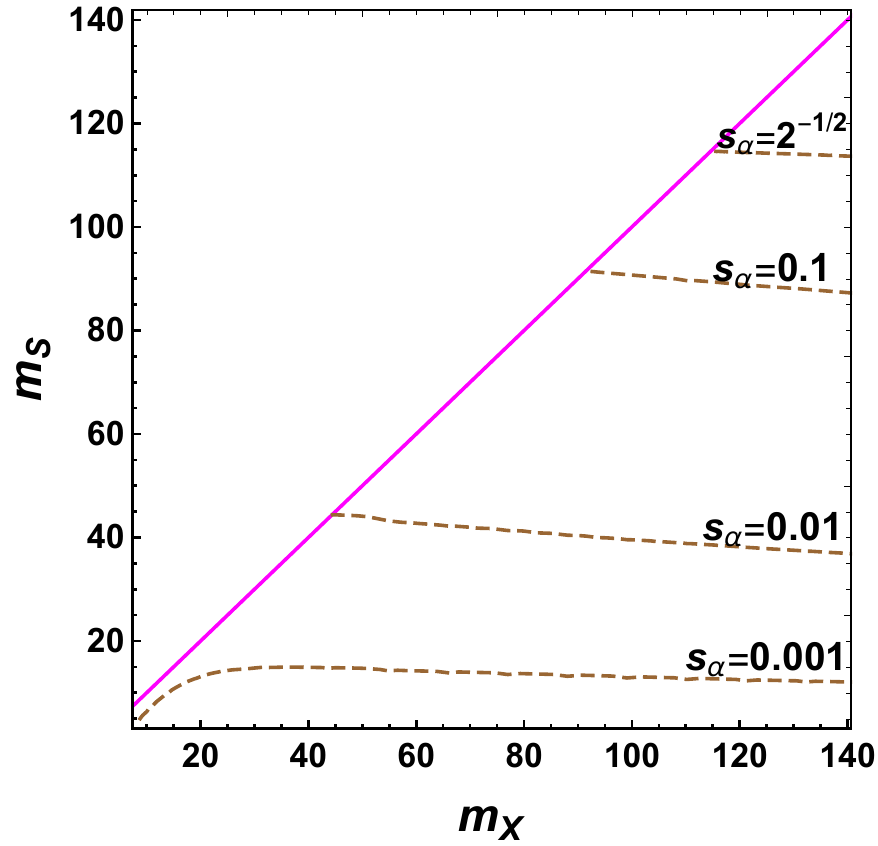}\hskip0.9cm
\includegraphics[width=0.40\textwidth]{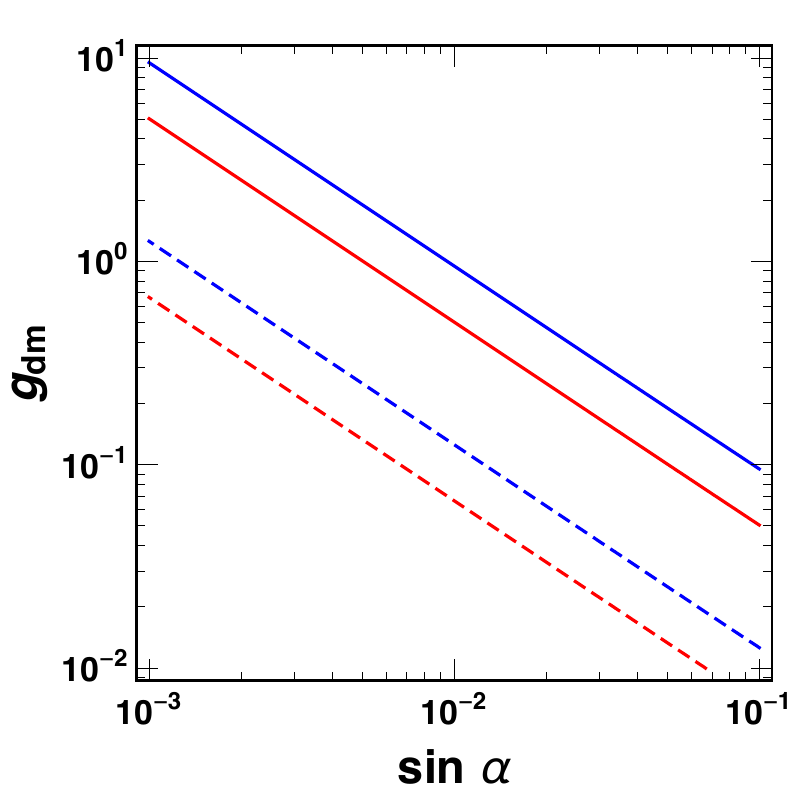}\\
\caption{ Left panel: The direct search limit on the ($m_X,m_S$) parameter space with the requirement $m_X>m_S$ which is the region on the right hand side of the magenta solid line.
For a given value of $s_\alpha (\equiv \sin\alpha)\leq 1/\sqrt{2}$, the region above the corresponding dashed (brown) curve is allowed by XENON1T \cite{Aprile:2017iyp}. Here we use $g_{\rm dm}=0.173$.
Right panel: Upper bounds  of $\sin\alpha$ and $g_{\rm dm}$ from XENON1T and LZ projected sensitivity  \cite{Akerib:2018lyp} denoted by the solid and dashed lines, respectively, where the red line is for $m_X=80$~GeV, $m_S=0.8 m_X$, and the blue for  $m_X=80$~GeV, $m_S=0.99 m_X$. 
}
\label{fig:DDconstraint}
\end{center}
\end{figure}

In the secluded DM model, the direct detection measurements and colliders weakly constrain the parameter region allowed by the GC excess result. The spin-independent cross section for a vector dark matter particle scattering off a single nucleon via a scalar mediator $S$ exchange  is given by
\begin{align}
\sigma_N =  \frac{\mu_{XN}^2 m_N^2 f_N^2 g_{\rm dm}^2}{4\pi} \frac{\sin^2 2\alpha}{v_H^2} \left( \frac{1}{m_S^2} - \frac{1}{m_h^2} \right)^2 \,,
\end{align}
where $\mu_{XN} = m_X m_N/ (m_X + m_N)$ is the reduced mass of the dark matter ($X$) and nucleon ($N$), and $f_N=  \sum_q \langle N | \bar q q |N \rangle m_q/m_N \simeq 0.3$  \cite{Cline:2013gha}.
The parameter space constrained by XENON1T  \cite{Aprile:2017iyp} is shown in Fig.~\ref{fig:DDconstraint}, where in the right panel the bound by the LUX-ZEPLIN (LZ) projected sensitivity \cite{Akerib:2018lyp} is given.  In the left panel of Fig.~\ref{fig:DDconstraint}, the allowed parameter region on the ($m_X, m_S$) plane for a given value of $\alpha$ is above the corresponding dashed line, where we have limited $\alpha \leq \pi/4$ which is suitable for the case with a small $\alpha$. As for $ \pi/4 \leq  \alpha \leq \pi/2$,  the bound is same as that with a mixing angle $=\pi/2 - \alpha$, because $\sin2\alpha =\sin 2(\pi/2 - \alpha)$. 

The paramter constraint from the invisible Higgs decay, which is less than  25\% at the 95\% CL \cite{pdg2018},  is much weaker than that from direct detection. Meanwhile, for the present case,  $h \to S\, S$ is kinematically forbidden.

\section{Thermal evolution of the nonrelativistic hidden particles}\label{sec:thermodynamics}

The evolution of the phase space distribution $f_h$ (with $h\equiv X \text{ or }S$) of the hidden sector particles in the homogeneous isotropic Friedmann-Robertson-Walker Universe is described by  the Boltzmann equation,
\begin{align} \label{eq:boltzmann}
\frac{\partial f_h}{\partial t}  - H p \frac{\partial f_h}{\partial p} =C[ f_h ] \,,
\end{align} 
where $H$ is the Hubble expansion parameter, $p=\sqrt{E_h^2 - m_h^2}$ is the momentum of the hidden particle, and $C[f_h]$ is the collision term. 
During the process of the thermal evolution, the distribution of the hidden sector particles follows Bose-Einstein statistics,
\begin{align}
f_h (E_h, T_h) = \frac{1}{e^{(E_h -\mu_h)/T_h} -1} \,,
\end{align}
with $\mu_h$ the chemical potential of the particle  species $h$.
In the present case, we consider $m_X\gtrsim m_S \sim {\cal O} (10-100~\text{GeV})$ and the thermal evolution that the elastic scattering $X S \leftrightarrow X S$ can keep the $X$ and $S$ particles in thermal  equilibrium ($T_X = T_S$) until kinetic decoupling temperature $T_X^{\rm kd} \equiv T_X(a_X^{\rm kd})$, below that we have $T_X(a)\simeq T_X^{\rm kd} \cdot (a_X^{\rm kd}/a)^2$, where $a$ is the cosmic scale factor and $a_X^{\rm kd}$ is its corresponding value at $T_X^{\rm kd}$.

For a hidden sector particle, $h_1$, the generic form of one of the collision terms described by the interaction  ``$h_1 h_2 \cdots b_1 b_2 \cdots \leftrightarrow h_1' h_2' \cdots b_1' b_2' \cdots $'' can be written as 
\begin{align}
C[f_{h_1}]= 
& \frac{1}{2E_{h_1} g_{h_1} }\int  d\Pi_{h_2} \dots d\Pi_{b_1} d\Pi_{b_2} \dots
   d\Pi_{h_1'} d\Pi_{h_2'} \dots d\Pi_{b_1'} d\Pi_{b_2'} \dots 
\nonumber\\
&
\times (2\pi)^4 \delta^{(4)} (p_{h_1} +p_{h_2} + \cdots +p_{b_1}+p_{b_2} + \cdots 
- p_{h_1'} - p_{h_2'} - \cdots - p_{b_1'} -p_{b_2'} - \cdots )
|M|^2  \nonumber \\
& \times  \frac{\Delta - \Delta'}{S_y\, S_y'}\Big[ f_{h_1'} f_{h_2'} \cdots f_{b_1'} f_{b_2'} \cdots (1 + f_{h_1}) (1+ f_{h_2}) \cdots  (1\pm f_{b_1}) (1\pm f_{b_2}) \nonumber\\
& \  -f_{h_1} f_{h_2} \cdots f_{b_1} f_{b_2} \cdots (1 + f_{h_1'}) (1 + f_{h_2'}) \cdots  (1\pm f_{b_1'}) (1\pm f_{b_2'}) \Big] \,, \label{eq:generic-collision-1}
\end{align}
 where 
 \begin{align}
 d\Pi_i \equiv \frac{d^3p_i}{(2\pi)^3 2E_i} \,,
 \end{align}
 $h_i^{(\prime)}$ is the particle of the hidden sector with temperature $T_X$ for $X$ or $T_S$ for $S$,  $b_i^{(\prime)}$ is the relativistic SM particle with temperature $T$,
 $|M|^2$ invariant under times reversal and reflection is the square of the amplitude summed over the internal degrees of freedom (dof), $g_i$, of all the initial and final particles,
  $S_y$ and $S_y'$ are the symmetric factors in the initial and final states, respectively, $\Delta$ and $\Delta'$ are the numbers of the species which are the same as $h_1$ and participate in the interaction in the initial and final states, respectively; note that if  the particle composition in the initial state is exactly the same as that in the final state 
 ({\it e.g.} elastic scattering $S + \text{SM} \leftrightarrow S + \text{SM}$, and elastic self-scattering $SS \leftrightarrow SS$), the moment result can be non-vanishing (see Eq.~(\ref{eq:collision-moment}) for instance), and an additional factor ``1/2" needs to be added in $C[f_{h_1}]$ to avoid double-counting. Taking $SSS\leftrightarrow XX$ as an example, we have  $S_y\equiv 3!$, and $S_y'\equiv 2!$. Moreover, we have $\Delta \equiv 3, \Delta' \equiv 0$ for considering the Boltzmann equation of the $S$ particles, while $\Delta \equiv 0, \Delta' \equiv 2$ for $X$.
 Here the  $1\pm f_i$ terms  with plus and minus signs encode the influence due to Bose enhancement and Pauli blocking, respectively.

We are interested in reactions dominated by the phase space region where the average number of particles in a single-particle state is much less than 1, i.e.,  $1\pm f_i \simeq 1$, and thus approximate the distributions as
\begin{align}
f_{i} =  e^{-(E_{i}-\mu_i)/T_i} (1\pm f_i) \simeq e^{-(E_{i}-\mu_i)/T_i}  \,.
\end{align}
Basically, this is a good approximation even for high or low temperature.
The collision term can be then given as the following form,
\begin{align}
C[f_{h_1}] 
& = 
 \frac{1}{2E_{h_1} g_{h_1} }\int  d\Pi_{h_2} \dots d\Pi_{b_1} d\Pi_{b_2} \dots
   d\Pi_{h_1'} d\Pi_{h_2'} \dots d\Pi_{b_1'} d\Pi_{b_2'} \dots 
\nonumber\\
\times & (2\pi)^4 \delta^{(4)} (p_{h_1} +p_{h_2} + \cdots +p_{b_1}+p_{b_2} + \cdots 
- p_{h_1'} - p_{h_2'} - \cdots - p_{b_1'} -p_{b_2'} - \cdots )
|M|^2  \nonumber \\
   \times &  \frac{\Delta -  \Delta'}{S_y \, S_y'}\Big[ f_{h_1'} (T_{h_1})  f_{h_2'} (T_{h_2}) \cdots f_{b_1'} (T) f_{b_2'} (T) \cdots 
   -f_{h_1}  (T_{h_1}) f_{h_2} (T_{h_2}) \cdots f_{b_1}(T)  f_{b_2} (T) \cdots  \Big] \,, \label{eq:generic-collision-2}
\end{align}
where $f_i (T_i) \equiv f_i (E_i, T_i)$. It should be noted that the relativistic SM, $X$ and $S$ are defined by the different temperatures, $T, T_X$ and $T_S$, respectively.  In our case, the dark matter and mediator are in thermal equilibrium, i.e. $T_X=T_S$, until their kinetic decoupling; we will further discuss this point in the following sections.

\subsection{The Boltzmann moment equation for the number densities of hidden sector particles}\label{sec:boltz-number-density}

To get the coupled Boltzmann equations for the number densities of $X$ and $S$, we form the moment by multiplying Eq.~(\ref{eq:boltzmann}) with ``1'' and integrating over the momentum space,
\begin{align}
n_X (T_X) =g_X \int \frac{d^3 p_X}{(2\pi)^3} f_X (T_X), \qquad    n_S (T_S) =g_S \int \frac{d^3 p_S}{(2\pi)^3} f_S (T_S) \,.
\end{align}
In our case, the sufficiently large interactions within the hidden sector can maintain thermal equilibrium among the hidden sector particles, i.e.,  $T_X=T_S$,  before the DM freezes out.  The resulting equations of number densities are given by
\begin{align}
\frac{dn_X}{dt} + 3 H n_X=  & -\langle \sigma v \rangle_{XX\to SS} \bigg(n_X^2 -(n_X^{\text{eq}})^2 \frac{n_S^2}{(n_S^{\text{eq}})^2} \bigg) 
    -  \frac{1}{3} \langle \sigma v^2 \rangle_{XXX\to XS} \bigg( n_X^3 -n_X n_S \frac{(n_X^{\text{eq}})^2}{n_S^{\text{eq}}} \bigg)   \nonumber\\
& -  \langle \sigma v^2 \rangle_{XXS\to SS} \bigg( n_X^2 n_S -(n_X^{\text{eq}})^2 \frac{n_S^2}{(n_S^{\text{eq}})} \bigg)
   + \frac{1}{3} \langle \sigma v^2 \rangle_{SSS\to XX} \bigg( n_S^3 -(n_S^{\text{eq}})^3 \frac{n_X^2}{(n_X^{\text{eq}})^2} \bigg) \,, \label{eq:boltz-1} 
\end{align}
\begin{align}
\frac{dn_S}{dt} + 3 H n_S = & -  \Gamma_{S}
        \bigg( \frac{K_1(x_S\cdot m_S/m_X)}{K_2(x_S\cdot m_S/m_X)} n_S -  \frac{K_1(x\cdot m_S/m_X) }{K_2 (x\cdot m_S/m_X)}  n_S^{\text{eq}} (T)  \bigg)  \nonumber\\
  &  -   \bigg( \langle \sigma v \rangle_{SS\to  \sum_{ij} {\rm SM}_i\, {\rm SM}_j} n_S^2
     -   \langle \sigma v \rangle_{SS\to  \sum_{ij}{\rm SM}_i\, {\rm SM}_j} (T) (n_S^{\text{eq}} (T) )^2   \bigg) \nonumber\\
 & + \langle \sigma v \rangle_{XX\to SS} \bigg(n_X^2 -(n_X^{\text{eq}})^2 \frac{n_S^2}{(n_S^{\text{eq}})^2} \bigg)
      + \frac{1}{6} \langle \sigma v^2 \rangle_{XXX\to XS} \bigg( n_X^3 -n_X n_S \frac{(n_X^{\text{eq}})^2}{n_S^{\text{eq}}} \bigg)  \nonumber\\
 & + \frac{1}{2} \langle \sigma v^2 \rangle_{XXS\to SS} \bigg( n_X^2 n_S -(n_X^{\text{eq}})^2 \frac{n_S^2}{(n_S^{\text{eq}})} \bigg)
   - \frac{1}{2} \langle \sigma v^2 \rangle_{XSS\to XS} \bigg( n_X n_S^2 - n_X n_S n_S^{\text{eq}}   \bigg) 
\nonumber\\
 &  - \frac{1}{2} \langle \sigma v^2 \rangle_{SSS\to XX} \bigg( n_S^3 -(n_S^{\text{eq}})^3 \frac{n_X^2}{(n_X^{\text{eq}})^2} \bigg)
  - \frac{1}{6} \langle \sigma v^2 \rangle_{SSS\to SS} \bigg( n_S^3 -n_S^2 n_S^{\text{eq}}   \bigg)  \,, \label{eq:boltz-2}
\end{align}
where $K_i$ is the modified Bessel function of the second kind with $x_S\equiv m_X/T_S$ and $x\equiv m_X/T$, $n_{i}^{\rm eq}$ is the equilibrium number  density with vanishing chemical potential,  $\Gamma_S$ is the total decay width of $S$ into SM final states, and $\langle \sigma v \rangle_i$ and $\langle \sigma v^2 \rangle_i$ are respectively the thermally averaged cross sections for $2\to2$ and $3\to2$ annihilation processes denoted by the subscript ``{\it i}"; the details for these results are given in Appendices \ref{app:annXS} and \ref{app:3-2}. Note that only the terms  involving  $n_S^{\text{eq}}$ and, meanwhile, relevant to $S\to {\rm SM}_i~ {\rm SM}_i$ and $SS\to    {\rm SM}_i~ {\rm SM}_i$ on the right hand side (RHS) of Eq.~(\ref{eq:boltz-2}) are functions of  the bath temperature, ``$T$", which is  explicitly indicated, whereas the remaining ones appearing in Eqs.~(\ref{eq:boltz-1}) and (\ref{eq:boltz-2}) are functions of ``$T_S$" or  ``$T_X$".  Here and in the following analysis, we will use  $T_X=T_S$ due to the fact that the hidden sector particles keep thermal equilibrium before the DM freeze-out.

Because the comoving number density of dark matter is conserved after freeze-out, we  introduce the normalized yields for the hidden sector,
\begin{align}
y_X(x_X; x) &=\sqrt{\frac{\pi}{45 G}} m_X g_*^{1/2}(m_X)   \langle \sigma v \rangle^{(0)}_{XX\to SS}  Y_X (x_X; x) \,, \label{eq:yx} \\
y_S(x_S; x) &=\sqrt{\frac{\pi}{45 G}} m_X g_*^{1/2}(m_X)   \langle \sigma v \rangle^{(0)}_{XX\to SS}  Y_S (x_S; x) \label{eq:ys}\,,
\end{align}
with $x_i \equiv m_X/T_i$ and $x\equiv m_X/T$ being the temperature variables of the hidden sector particles and thermal bath, respectively, the yields $Y_{i} \equiv  n_{i}(T_i)/ s(T)$  being the number density normalized by the total entropy density, and $g_*$  being the effectively total number of relativistic dof; see below for definition. Here $\langle \sigma v\rangle^{(0)}_{XX\to SS}$ is the leading approximation of $\langle \sigma v\rangle_{XX\to SS}$ which is s-wave. In the following discussion, we will simply use $y_{i} (x)\equiv y_{i} (x_i; x)$. We use $x\equiv m_X/T$ as the evolution variable and trade the time derivative in the Boltzmann equations to be
\begin{align}
\frac{d}{dt} =\frac{m_X^2}{x}  \bigg(\frac{8\pi^3 G}{90} \bigg)^{1/2} \frac{h_{\rm eff}}{g_*^{1/2}} \frac{d}{dx} \,,
\end{align}
where the relativistic degrees of freedom, $h_{\rm eff}$ and $g_*^{1/2} \equiv \tilde{h}_{\rm eff}  /g_{\rm eff}^{1/2}$  with 
\begin{align} 
\tilde{h}_{\rm eff}  \equiv h_{\rm eff} [1+(1/3) (d\ln h_{\rm eff} / d\ln T)] \,,
\end{align}  
are defined via
\begin{align}
 s(T)= \frac{2\pi^2}{45} h_{\rm eff}(T) T^3 \,,\qquad
\rho (T)= \frac{\pi^2}{30} g_{\rm eff}(T) T^4 \,.
\end{align}
Thus, we can further rewrite the Boltzmann equations to be
\begin{align}
\frac{d y_X}{dx} =
& - \frac{\delta_{\text{XS}} \delta_{\text{dof}} }{x^2} \bigg( y_X^2 - (y_X^{\text{eq}})^2 \frac{y_S^2}{ (y_S^{\text{eq}})^2} \bigg)  \nonumber\\
     & +  \frac{ \delta_{\text{dof}} }{x^5} \frac{\pi} {\sqrt{90}} \frac{h_{\text{eff}}(T) }{g_*^{1/2}(m_X) } \frac{m_X^2}{M_{\rm pl} } 
              \Bigg[
             - \frac{1}{3} \frac{ \langle \sigma v^2 \rangle_{XXX\to XS}}{  (\langle \sigma v\rangle^{(0)}_{XX\to SS})^2 } 
               \bigg( y_X^3 -  y_X  (y_X^{\text{eq}})^2 \frac{ y_S }{ y_S^{\text{eq}}} \bigg)     
  \nonumber \\
&         + \frac{1}{3} \frac{ \langle \sigma v^2 \rangle_{SSS\to XX}}{  (\langle \sigma v\rangle^{(0)}_{XX\to SS})^2 } 
                 \bigg( y_S^3 - \frac{y_X^2}{(y_X^{\text{eq}})^2} (y_S^{\text{eq}})^3  \bigg) 
                - \frac{ \langle \sigma v^2 \rangle_{XXS\to SS}}{  (\langle \sigma v\rangle^{(0)}_{XX\to SS})^2 } 
                   \bigg( y_X^2 y_S -    \frac{y_S^2}{y_S^{\text{eq}}} (y_X^{\text{eq}})^2  \bigg)  
        \Bigg]
 \,,    \label{eq:boltzmann-1}
 \end{align}
 \begin{align}
\frac{d y_S}{dx} =
 &  - x  \frac{ \sqrt{90}}{\pi} M_{\rm pl} \frac{g_*^{1/2}(T) }{ h_{\text{eff}}(T) } \frac{ \Gamma_{S} }{m_X^2}
           \bigg( \frac{K_1(x_S\cdot m_S/m_X)}{K_2(x_S\cdot m_S/m_X)} y_S
         - \frac{K_1(x\cdot m_S/m_X)}{K_2(x\cdot m_S/m_X)} y_S^{\text{eq}}(x) \bigg)  
   \nonumber\\
 &  -    \frac{\delta_{\text{dof}} }{x^2 \langle \sigma v \rangle^{(0)}_{XX\to SS} } 
           \Big[  \langle \sigma v \rangle_{SS\to \sum_i {\rm SM}_i {\rm SM}_i}  y_S^2 
         - \langle \sigma v \rangle_{SS\to \sum_i {\rm SM}_i {\rm SM}_i} (x) \, ( y_S^{\text{eq}} (x) )^2 \Big]
   \nonumber\\
 &  -    \frac{\delta_{\text{XS}} \delta_{\text{dof}} }{x^2} \bigg[ \frac{(y_X^{\text{eq}})^2}{(y_S^{\text{eq}})^2} y_S^2 -  y_X^2 \bigg]  
         +  \frac{\delta_{\text{dof}} }{x^5}  \frac{\pi} {\sqrt{90}} \frac{h_{\text{eff}}(T) }{g_*^{1/2}(m_X) } \frac{m_X^2}{M_{\rm pl} }
             \Bigg[ \frac{1}{6} \frac{ \langle \sigma v^2 \rangle_{XXX\to XS}}{  (\langle \sigma v\rangle^{(0)}_{XX\to SS})^2 } 
                 \bigg( y_X^3 -  y_X   (y_X^{\text{eq}})^2  \frac{y_S}{ y_S^{\text{eq}}} \bigg)   
   \nonumber\\
 &  +    \frac{1}{2} \frac{ \langle \sigma v^2 \rangle_{XXS\to SS}}{  (\langle \sigma v\rangle^{(0)}_{XX\to SS})^2 } 
                   \bigg( y_X^2 y_S - \frac{y_S^2}{y_S^{\text{eq}}} (y_X^{\text{eq}})^2  \bigg)    
         -  \frac{1}{2} \frac{ \langle \sigma v^2 \rangle_{XSS\to XS}}{  (\langle \sigma v\rangle^{(0)}_{XX\to SS})^2 }  
              \Big( y_X y_S^2 - y_X y_S  y_S^{\text{eq}}  \Big) 
    \nonumber\\
 &  -    \frac{1}{2} \frac{ \langle \sigma v^2 \rangle_{SSS\to XX}}{  (\langle \sigma v\rangle^{(0)}_{XX\to SS})^2 } 
               \bigg( y_S^3 - \frac{y_X^2}{(y_X^{\text{eq}})^2} (y_S^{\text{eq}})^3  \bigg)
          - \frac{1}{6}  \frac{ \langle \sigma v^2 \rangle_{SSS\to SS}}{  (\langle \sigma v\rangle^{(0)}_{XX\to SS})^2 }  
              \Big( y_S^3 - y_S^2  y_S^{\text{eq}}  \Big)  \Bigg] 
    \,,  \label{eq:boltzmann-2}
 \end{align}
where $M_{\rm pl}\equiv (8 \pi G)^{-1/2}=2.44\times 10^{18}$~GeV is the reduced Planck mass, 
\begin{align}
\delta_{\rm XS} \equiv \frac{\langle \sigma v \rangle_{XX\to SS} (T_X)}{\langle \sigma v \rangle^{(0)}_{XX\to SS}} ,   \qquad
\delta_{\rm dof} \equiv \frac{g_*^{1/2}(T) }{ g_*^{1/2} (m_X) } , 
\end{align} and the equilibrium value of $y_i$ is given by 
\begin{equation} \label{eq:norm-yield-eq}
y_i^{\rm eq} (x_i) \equiv y_i^{\rm eq} (x_i; x) =g_i \frac{ \sqrt{90}}{2 \pi^3} M_{\rm pl} \frac{g_*^{1/2}(m_X)}{h_{\rm eff}(T)} m_X \bigg(x \frac{m_i}{m_X} \bigg)^2 
\frac{x}{x_i} \langle \sigma v \rangle^{(0)}_{XX\to SS}  K_2 \bigg(x_i \frac{m_i}{m_X} \bigg) \,.
\end{equation}
Again, it should be noted that in Eq.~(\ref{eq:boltzmann-2}) the terms involving $y_S^{\text{eq}}$'s and relevant to $S\to {\rm SM}_i~ {\rm SM}_i$ and $SS\to    {\rm SM}_i~ {\rm SM}_i$ are functions only of ``$x$", as shown explicitly, while other quantities appearing Eqs.~(\ref{eq:boltzmann-1}) and (\ref{eq:boltzmann-2}) are instead defined as functions of ``$x_S$",  which are not shown explicitly,  before freeze-out. 
In the following section, we will exhibit the evolution of $T_S/T$ as a function of $x$, i.e., a function of the bath temperature $T$. 

The relic abundance is found to be
\begin{equation}
 \Omega_{\rm DM} = \frac{Y_X^\infty s_0 m_X}{\rho_c} 
 \simeq  \frac{1.04 \times 10^9\  {\rm GeV}^{-1}}{ \sqrt{8 \pi g_*(m_X)} M_{\rm pl} h^2} \frac{y_X^\infty}{  \langle \sigma v\rangle^{(0)}_{XX\to SS}} \,,
 \label{eq:abundance}
  \end{equation}
which can be determined by matching the present-day DM relic abundance $\Omega_{\rm{DM}}  =(0.1198 \pm 0.0026)/h^2$  \cite{pdg2018,Ade:2013zuv},
where $Y_X^\infty$ is related to $y_X^\infty = y_X (x\to \infty)$ (see also Eq.~(\ref{eq:yx})), $s_0=2891~\text{cm}^{-3}$ is the visible entropy density today, $\rho_c = 3 H_0^2/ (8\pi G)$ is the critical energy density, and $h\simeq0.678$ is  the Hubble constant $H_0$ of the present day in units of  $100~\text{km}\, s^{-1}  \text{Mpc}^{-1}$.
$y_X^\infty$ is related to  $x=x_f (\equiv m_X/T_f)$ with $T_f$ the freeze-out temperature, and can be understood  as follows.   Well after DM freeze-out, which occurs at $x=x_f$,  the Boltzmann equation in Eq.~(\ref{eq:boltzmann-1}) can be approximated as
\begin{align}
\frac{d y_X}{dx} \approx & - \frac{\delta_{\text{XS}} \delta_{\text{dof}} }{x^2}  y_X^2\,.
\label{eq:approx-xf}
\end{align}
Solving the equation, we get 
\begin{align}
y_X^\infty = \bigg( \int_{x_f}^\infty \frac{\delta_{\text{XS}} \delta_{\text{dof}} }{x^2} dx   \bigg)^{-1}   
=  g_*^{1/2} (m_X) \langle \sigma v\rangle^{(0)}_{XX\to SS} \bigg( \int_{x_f}^\infty \frac{ g_*^{1/2} (T) \  \langle \sigma v\rangle_{XX\to SS}(T_X) }{x^2} dx   \bigg)^{-1} \,,
\label{eq:xf}
\end{align}
where we use the fact that the value of $y_X$ at $x=x_f$ is significantly larger than $y_X^\infty$, and  we can approximate $T_X  \approx T_S$ in the calculation (see Figs.~\ref{fig:relic-not-equal-1}, \ref{fig:relic-not-equal-2}, and \ref{fig:relic-equal} for the temperature dependence in the next section).

\subsection{The Boltzmann moment equation  for   $T_S/T$}\label{sec:boltz-TS}

We consider the case that the DM can be kept in thermal equilibrium with the hidden scalar before DM freeze out, but may be highly decoupled from the SM thermal bath.  
Here we focus on the study about the temperature evolution of the hidden scalar $S$, and then discuss the DM temperature evolution after freeze out. 
The temperature evolution of nonrelativistic $S$ is affected by the following elastic scattering and (species) number changing interactions --- (i) annihilation: $ SS \leftrightarrow \text{SM~SM}$, (ii) elastic scattering: $S +  \text{SM} \leftrightarrow S+\text{SM}$, (iii) cannibalization including $ SSS\leftrightarrow SS$, $ XSS\leftrightarrow XS$, $ XXS\leftrightarrow SS$, $ SSS\leftrightarrow XX$, and $ XXX\leftrightarrow XS$, and (iv) decay: $S \leftrightarrow \text{SM~SM}$.

We adopt the definition of the temperature,
\begin{align}
T_S =\frac{g_S}{n_S(T_S)} \int \frac{d^3 p_S}{(2\pi)^3} \frac{{\bf p}_S^2}{3 E_S}  f_S (T_S) \,,
\end{align} 
which is suitable not only for the nonrelativistic case at low temperatures, $T<m_S$, but also for relativistic case at high temperature, $T \lesssim m_S/0.01$.
The Boltzmann moment equation of the hidden scalar's temperature can be formed by multiplying Eq.~(\ref{eq:boltzmann}) with ${\bf p}_S^2/(3E_S)$ and then 
 integrating over the momentum space. Thus, we arrive at the form of the temperature evolution equation,
\begin{equation}
 \frac{d T_S}{dt} + (2-\delta_H) H T_S 
 = \frac{1}{ n_S(T_S)}  \left[- \left( \frac{d n_S(T_S) }{dt} +3 H n_S (T_S) \right) T_S 
+ g_S \int \frac{d^3 p_S}{(2\pi)^3} \, C \Big[f_S\cdot \frac{{\bf p}_S^2}{3 E_S} \Big] \right] ,  \label{eq:boltz-t}
\end{equation}
where
\begin{align}
\delta_H (T_S) \equiv 1- \frac{1}{n_S(T_S) \, T_S} \int \frac{d^3 p_S}{(2\pi)^3} \frac{{\bf p}_S^2 m_S^2}{3 E_S^3}  f_S (T_S) \,,
\label{eq:deltah}
\end{align} 
which is approximately to be ``0" for nonrelativistic particles or ``1" for ultra-relativistic ones. Here we have denoted the collision term  as $[f_S \cdot {\bf p}_S^2/(3E_S)]$ which is related to $C[f_S]$  
(see Eq. (\ref{eq:generic-collision-1}) or (\ref{eq:generic-collision-2}))  with the replacement
\begin{align}
\Delta - \Delta' \to  \Delta \cdot \frac{{\bf p}_S^2}{3 E_S} - \Delta'  \cdot \frac{{\bf p}_S^{2\prime}}{3 E_S'} \,,
\label{eq:collision-moment}
\end{align}
corresponding to a process $i (\text{initial~state}) \leftrightarrow f (\text{final~state})$  with a prime for the final state.
In the following, the collision term due to various interactions will be discussed term by term in details. 

After the hidden sector is chemically decoupled from the bath, i.e., its number density production rate is less the expanding rate of the Universe, we have $n_h a^3={\rm constant}$ from Eqs.~(\ref{eq:boltz-1}) and (\ref{eq:boltz-2}) if cannibalization can be neglected. Moreover, well after the cannibal epoch, if the nonrelativistic  hidden sector is out of thermal equilibrium with the bath, we have $T_h a^2={\rm constant}$ as read from Eq.~(\ref{eq:boltz-t}).

\vskip1.3cm

\subsubsection{The collision term due to $S S \leftrightarrow \text{SM}_1~\text{SM}_2$}

The collision term resulting from  $S(p_{S,1}) S(p_{S,2}) \leftrightarrow \text{SM}_1 (p_{1'})~\text{SM}_2 (p_{2'})$ is given by
\begin{align}
&  g_S \int \frac{d^3 p_S}{(2\pi)^3} \, C\Big[f_S\cdot \frac{{\bf p}_S^2}{3 E_S}\Big] _{S S \leftrightarrow \text{SM}_1 \text{SM}_2}= 
   \int  \prod_i     d\Pi_{S,i}  \, d\Pi_{i^\prime}    (2\pi)^4 \delta^{(4)} (p_{S,1} +p_{S,2} - p_{1'} - p_{2'}  )
\nonumber \\   
& ~ \times  
 \frac{2}{2! m!}  \frac{{\bf p}_{S,1}^2}{3 E_{S,1}} 
     \Big[ 
          e^{-(E_{1'}+E_{2'})/T} - e^{2\mu_S /T_S}  e^{-(E_{S,1} +E_{S,2})/T_S}
     \Big]  |M|_{S S \to \text{SM}_1 \text{SM}_2}^2 
\nonumber\\
& ~  =
    -      \langle \sigma v \cdot \frac{{\bf p}_S^2}{3 E_S} \rangle_{S S \to \text{SM}_1 \text{SM}_2} (T_S) \,  \big( n_S(T_S) \big)^2
            +   \langle \sigma v \cdot \frac{{\bf p}_S^2}{3 E_S} \rangle_{S S \to \text{SM}_1 \text{SM}_2} (T) \,  \big( n_S^{\text{eq}} (T) \big)^2  \,,
\end{align}
where $m!=2$ for identical final state particles or 1 otherwise, and we have used the energy conservation $E_{1'} +E_{2'} =E_{S,1}+E_{S,2}$, and the relation,
\begin{align}
e^{\mu_S/T_{i}} =\frac{n_S (T_{i})}{n_S^{\text{eq}} (T_{i})}, \quad \text{with $T_i \equiv T$ or $T_S$} \,.
\end{align}
Here the thermal average, for which the detailed description is provided in Appendix~\ref{app:thermal-average}, is given by
\begin{align}
& \langle \sigma v  \cdot \frac{{\bf p}_S^2}{3 E_S} \rangle_{S S \to \text{SM}_1 \text{SM}_2}  (T_i)
  = \frac{1}{48 m_S^4  K_2^2 (m_S/T_i)}  \nonumber\\
     & \times   \int_{4m_S^2}^{\infty} ds
        (\sigma v_{\text{lab}} ) \sqrt{s-4m_S^2} 
    \bigg[ ( s+ 2 m_S^2) K_1 \left(\frac{\sqrt{s}}{T_i} \right)
         +  \left( \frac{s-4m_S^2}{2} \frac{\sqrt{s}}{T_i} + \frac{ 4T_i ( s+ 2 m_S^2)}{\sqrt{s}}  \right) K_2 \left(\frac{\sqrt{s}}{T_i} \right)
    \bigg], 
\label{eq:thermal-average-ann}
\end{align}
where $v_{\rm lab}$ is the relative velocity measured in the laboratory frame where one of the incoming particles is at rest. For the present $S$-wave annihilation, the ratio of $\langle \sigma v \cdot \frac{{\bf p}_S^2}{3 E_S} \rangle /  (\langle \sigma v \rangle T) $  is $0.75 \sim 1.05$
for $x(\equiv m_X/T) > 2$ with $m_X =80$~GeV, and becomes unity  in the limit $x\to \infty$. For a typical case with  $\sigma v$ being constant, the ratio is equal to one. See also discussions in Appendix~\ref{app:thermal-average}.

\vskip1.3cm

\subsubsection{The collision term due to $S \leftrightarrow \text{SM}_1~\text{SM}_2$}

The collision term resulting from  $S(p_S) \leftrightarrow \text{SM}_1 (p_{1'})~\text{SM}_2 (p_{2'})$ is given by
\begin{align}
&  g_S \int \frac{d^3 p_S}{(2\pi)^3} \, C\Big[f_S\cdot \frac{{\bf p}_S^2}{3 E_S}\Big]_{S \leftrightarrow \text{SM}_1 \text{SM}_2}= 
   \int    d\Pi_{S} \prod_i     \, d\Pi_{i^\prime}   (2\pi)^4 \delta^{(4)} (p_{S}  - p_{1'} - p_{2'}  )
             |M|_{S \to \text{SM}_1 \text{SM}_2}^2 
\nonumber \\
& ~~ \times  \frac{1}{m!}  \frac{{\bf p}_{S}^2}{3 E_{S}}  
  \Big[ 
      e^{-(E_1+E_2)/T} - e^{\mu_S /T_S}  e^{- E_S/T_S}
  \Big]  
\nonumber\\
& ~~ = 
-  \Gamma_{S \to \text{SM}_1 \text{SM}_2}
        \left[  T_S  \frac{c(T_S)}{ n_S^{\rm eq} (T_S)} n_S(T_S)      -  T \frac{c(T)}{ n_S^{\rm eq}(T)}  n_S^{\rm eq} (T)  
        \right]
 \,,   \label{eq:collision-decay}
\end{align}
where  $m!=2$ for identical final state particles or 1 otherwise,  $\Gamma_{S \to \text{SM}_1 \text{SM}_2}$ is $S \to \text{SM}_1 \text{SM}_2$ decay width, and
\begin{align}
c(T) = g_S\frac{m_S}{3 T} \int \frac{d^3 p_S}{(2\pi)^3} \frac{{\bf p}_S^2}{E_S^2} e^{-E_S/T} \
= m_S T^2 \frac{g_S}{6\pi^2} \int_{m_S/T}^\infty dy \frac{ (y^2 -m_S^2/T^2)^{3/2}}{y} e^{-y}  \,,
\end{align}
which approaches to $n_S^{\rm eq}(T)$ in a  nonrelativistic limit. The result of Eq.~(\ref{eq:collision-decay}) is also correct if replacing $\Gamma_{S \to \text{SM}_1 \text{SM}_2}$ with the relevant three- (or more-) body decay mode.

\vskip1.3cm

\subsubsection{The collision term due to cannibal annihilations among the hidden sector particles }

The collision term arising from cannibal annihilations among the hidden sector particles contains the following processes: $ SSS\leftrightarrow SS$, $ XSS\leftrightarrow XS$, $ XXS\leftrightarrow SS$, $ SSS\leftrightarrow XX$, and $ XXX\leftrightarrow XS$.   Here, the result for $ XSS\leftrightarrow XS$ will be shown, while for the others can be derived in a similar way.  When  the temperature of the hidden sector drops below $m_{X,S}$, the role of the cannibalization becomes important. If the hidden sector kinetically decouples from the bath at $ T \lesssim m_{X,S}$, its temperature will decrease logarithmically with the cosmic scale factor during cannibalization (see Eq.~(\ref{eq:cannibalization-temp}) for discussion).
The collision term resulting from  $ X(p_X) S(p_{S,1}) S(p_{S,2}) \leftrightarrow  X (p_{X'}) S (p_{S'}) $ is given by
\begin{align}
& g_S  \int \frac{d^3 p_S}{(2\pi)^3}  \,  C\Big[f_S\cdot \frac{{\bf p}_S^2}{3 E_S}\Big]_{X S S \leftrightarrow X S}  
  =  
   \int   d\Pi_{X}  \, d\Pi_{S,1} \,  d\Pi_{S,2} \,  d\Pi_{X'} d\Pi_{S'}
 \nonumber\\
& ~ \times 
   (2\pi)^4 \delta^{(4)} (p_X + p_{S,1} +p_{S,2} -p_{X'} -p_{S'})  |M|_{X S S \to X S}^2 
\nonumber \\
& ~ \times  \frac{1 }{2!}   \Big( \frac{2 {\bf p}_{S,1}^2}{3 E_{S,1}}  -  \frac{ {\bf p}_{S'}^2}{3 E_{S'}}   \Big)
  \Big( 
  e^{(\mu_X + \mu_S) /T_S}  e^{-(E_{X'}+ E_{S'})/T_S} - e^{(\mu_X + 2\mu_S) /T_S}  e^{-(E_X + E_{S,1} +E_{S,2})/T_S}
  \Big)  
\nonumber\\
& ~ \simeq 
 \frac{ m_S (2m_X+m_S) (2m_X+3m_S)}{ 4 (m_X+2m_S)(4m_X+5m_S)}  \langle \sigma v^2 \rangle_{XSS \to XS}  \, 
  \left[ 
    n_X(T_S) \big(  n_S(T_S) \big)^2
-   n_X(T_S)  n_S(T_h)    n_S^{\text{eq}} (T_S) 
     \right]
 \,,
\end{align}
where we have used $p_{S^{(\prime)},i }^\mu= (E_{S^{(\prime)},i}, {\bf p}_{S^{(\prime)},i})$ for the $S$ and $p_{X^{(\prime)}}^\mu = (E_{X^{(\prime)}}, {\bf p}_{X^{(\prime)}})$ for the $X$, and have approximated three initial hidden particles ($XSS$) that annihilate or are produced  in the nonrelativistic limit, i.e., $E_{S,i}\approx m_S, E_X\approx m_X$, such that  
\begin{align} 
 \frac{ {\bf p}_{S,1}^2}{E_{S,1}}  \approx 0, \quad  \frac{ {\bf p}_{S'}^2}{3 E_{S'}}   \approx  \frac{ m_S (2m_X+m_S) (2m_X+3m_S)}{ 2 (m_X+2m_S)(4m_X+5m_S)}.
\end{align}

\vskip1.3cm

\subsubsection{ The collision term due to elastic scattering: $S +  \text{SM} \leftrightarrow S+\text{SM}$}

Here we consider the elastic scattering, $S (p_S) + {\rm SM}(k) \leftrightarrow S (p_{S'}) + {\rm SM}(k')$, 
where $p_{S^{(\prime)}}^\mu=(E_{S^{(\prime)}}, {\bf p}_{S^{(\prime)}})$, $k^{(\prime)\mu}=(\omega^{(\prime)}, {\bf k}^{(\prime)})$ and ``SM" stands for one of the relativistic SM particles that can participate the interaction. The collision term for this elastic scattering takes the following form,
\begin{align}
& g_S  \int \frac{d^3 p_S}{(2\pi)^3}  \,  C\Big[f_S\cdot \frac{{\bf p}_S^2}{3 E_S}\Big]_{S\, \text{SM} \leftrightarrow S\, \text{SM}}=  \sum_{\text{SM}}
\int \frac{d^3p_S}{(2\pi)^3 2E_S}    \frac{d^3 k}{(2\pi)^3 2 \omega}   
       \frac{d^3p_{S'} }{(2\pi)^3 2E_{S'}}  \frac{d^3 k'}{(2\pi)^3 2 \omega'} 
\nonumber\\
& ~ \times 
   (2\pi)^4 \delta^{(4)} (p_S +k -p_{S'} -k')  |M|_{S\, \text{SM} \leftrightarrow S\, \text{SM}}^2 
\nonumber \\
& ~ \times   \frac{1 }{2}   \Big( \frac{{\bf p}_{S}^2}{3 E_{S}}  -  \frac{ {\bf p}_{S'}^2}{3 E_{S'}}   \Big)
  \Big( 
     f_{S'}(T_S) \, f_{\rm SM'}(T) (1 -  f_{\rm SM}(T))
  -  f_{S}(T_S) \, f_{\rm SM}(T) (1 - f_{\rm SM'}(T))
  \Big)  
 \,,
\end{align}
where  the hidden scalar scattering with all relativistic SM fermions is taken into account.  
Under the typical condition $m_S \gg T\sim \omega$,  this term can further reduce to a semi-relativistic Fokker-Planck-type equation \cite{Bringmann:2006mu,Bringmann:2009vf,Gondolo:2012vh,Visinelli:2015eka,Binder:2016pnr,Binder:2017rgn},
\begin{align}
 g_S  \int \frac{d^3 p_S}{(2\pi)^3}  \,  C\Big[f_S\cdot \frac{{\bf p}_S^2}{3 E_S}\Big]_{S\, \text{SM} \leftrightarrow S\, \text{SM}}
 & \simeq 
  \gamma(T) \,   \int   \frac{d^3p_S}{(2\pi)^3 }   \frac{ {\bf p}_S^2}{3 E_S} \frac{\partial}{ \partial {\bf p}_S} \cdot 
        \left( {\bf p}_S f_S(T_S) + E_S T \frac{\partial f_S(T_S)}{\partial{\bf p}_S} \right) \nonumber\\
& \simeq - (2-\delta_H(T_S) ) \gamma  \ n_S(T_S) \,  (T_S -T) ,
\end{align}
where the momentum relaxation rate is given by 
\begin{align}
\gamma (T) =
\sum_f \frac{1}{6 m_S T} \int  \frac{d^3 k}{(2\pi)^3} f_f (T) (1 - f_f (T)) \frac{|{\bf k}|}{\sqrt{{\bf k}^2 +m_f^2}} 
\int_{-4 {\bf k}^2}^0 dt (-t) \frac{d \sigma_{S f \to S f }}{dt} \,, \label{eq:gamma}
\end{align}
for which the sum runs over all relevant relativistic SM species, and the differential elastic scattering cross section is
\begin{align}
\frac{d \sigma_{S f \to S f}}{dt} = \frac{1}{64\pi m_S^2 {\bf k}^2}  |M_{S f \to S f }|^2 \,,
\end{align}
with $|M_{S f\to S f}|^2$ the square of the scattering amplitude summed over initial and final spin states. 
Note that in Eq.~(\ref{eq:gamma}), we have followed the approach given in Ref.~\cite{Gondolo:2012vh} to adopt the $t$-average matrix $(8 {\bf k}^4)^{-1} \int_{-4{\bf k}^2}^0 dt (-t) d\sigma /dt$ due to the fact that the scattering amplitude squared in our case vanishes around $t=0$ in the relativistic limit $m_f\to 0$;  therefore, it is unsuitable to take the result at zero momentum transfer of the $t$-channel as done in Ref.~\cite{Bringmann:2006mu}.

Taking into account the elastic scattering $S f \to S f$ which is dominated by the amplitudes with the SM Higgs or hidden scalar mediated in the $t$-channel, we find the amplitude squared to be  
\begin{align}
  |M_{S f \to S f }|^2 =  2 N_c^f  (4 m_f^2 -t )  \bigg( \frac{g_{SSS}\, g_{S ff}}{t- m_S^2} +  \frac{g_{hSS} \, g_{h ff}}{t- m_h^2} \bigg)^2
 \,,
\end{align}
with $N_c^{f} \equiv 3\, (1)$ for quarks (leptons),  and the couplings shown in Eqs. (\ref{eq:gsss}), (\ref{eq:ghss}),  (\ref{eq:ghff}), and  (\ref{eq:gsff}).
Averaging over $t$ for the scattering amplitude squared, we get the momentum relaxation rate to be 
\begin{align}
\hskip-0.197cm \gamma(T)  \simeq  \sum_{f } \frac{40 N_c^f m_S}{\pi^3} \bigg[ \frac{g_{SSS}\, g_{S ff}}{m_S} + \frac{m_S}{m_h}  \frac{g_{hSS} \, g_{h ff}}{m_h} \bigg]^2
 \bigg[
    \frac{31}{32} \zeta(6)\Big( \frac{T}{m_S}  \Big)^6 + \frac{7}{8}\frac{\zeta(4)}{20} \Big( \frac{m_f}{m_S}\Big)^2 \Big( \frac{T}{m_S} \Big)^4 
  \bigg]  .
\end{align}
Here the transferred momentum $t$ in the denominator of the amplitude squared is neglected  in the calculation  consistent with the requirement  $-t < 4|{\bf k}|^2 \sim T^2 \ll m_S^2 $. Therefore, for the case with the resulting elastic decoupling temperature $\sim m_X$ as shown in the left panel of Figs.~\ref{fig:relic-not-equal-1}, \ref{fig:relic-not-equal-2}, and \ref{fig:relic-equal}, the kinetic transition rate should be overestimated, i.e., the true value of $x_{\rm el} (\equiv m_X/ T_{\rm el}) $ should be less than what is shown (see (\ref{point:xel}) in Sec.~\ref{sec:analysis} for the definition of $x_{\rm el}$). However, such overestimation does not affect our conclusions.

\subsubsection{The temperature evolution equation for the hidden scalar}

After including all interaction terms, we arrive at the Boltzmann moment equation for the temperature of the hidden scalar,
\begin{align}
\frac{d T_S}{dt}  & + (2 - \delta_H) H T_S =  - (2-\delta_H) \gamma (T_S - T)
\nonumber\\
  &  +  \frac{T_S}{n_S (T_S)}
   \bigg( \langle \sigma v \rangle_{SS\to  \sum_{ij} {\rm SM}_i\, {\rm SM}_j} (T_S) n_S^2 (T_S)
     -   \langle \sigma v \rangle_{SS\to  \sum_{ij}{\rm SM}_i\, {\rm SM}_j} (T) (n_S^{\text{eq}} (T) )^2   \bigg)
\nonumber\\
  & -   
   \bigg(   
             \langle \sigma v \cdot \frac{{\bf p}_S^2}{3 E_S} \rangle_{S S \to \sum_{ij} \text{SM}_i \text{SM}_j} (T_S) \,   n_S (T_S)
            -   \langle \sigma v \cdot \frac{{\bf p}_S^2}{3 E_S} \rangle_{S S \to \sum_{ij} \text{SM}_i \text{SM}_j} (T) \,    
                \frac{ \big( n_S^{\text{eq} } (T) \big)^2 }{n_S (T_S)}    \bigg)
 \nonumber\\
 & -  \Gamma_{S}  \bigg(
               \frac{K_1(x \cdot m_S/m_X)}{K_2(x \cdot m_S/m_X)}  (T_S - T\, \delta_\Gamma(x) )  \frac{n_S^{\text{eq}} (T) }{n_S(T_S)}
              -  \frac{K_1(x_S\cdot m_S/m_X)}{K_2(x_S\cdot m_S/m_X)} (1-\delta_\Gamma(x_S) ) T_S
                              \bigg)
\nonumber\\
 &  +
     \frac{(4m_X^2-m_S^2)(16m_X^2 -m_S^2)}{108 m_X(8m_X^2+m_S^2)}
             \langle \sigma v^2 \rangle_{XXX\to XS} 
                    \bigg( \frac{n_X^3 (T_S) }{n_S (T_S) } - \frac{n_X (T_S)  \big(n_X^{\text{eq}} (T_S) \big)^2} {n_S^{\text{eq}} (T_S) }  \bigg)  
\nonumber\\
 & + \frac{(2m_X+3m_S)(2m_X-m_S)}{6 (2m_X+m_S)} 
             \langle \sigma v^2 \rangle_{XXS\to SS} 
                     \bigg( n_X^2 (T_S)  - \frac{ \big(n_X^{\text{eq}} (T_S) \big)^2 n_S (T_S) } {n_S^{\text{eq}} (T_S) }  \bigg)  
\nonumber\\
 &
   + \frac{ m_S(2m_X+m_S)(2m_X+3m_S)}{4(m_X+2m_S)(4m_X+5m_S)}  
   \langle \sigma v^2 \rangle_{XSS\to XS} 
   \bigg( n_X (T_S) n_S (T_S) - n_X (T_S) n_S^{\text{eq}}  (T_S)   \bigg) 
\nonumber\\
 & 
  + \frac{5}{54}  m_S \langle \sigma v^2 \rangle_{SSS\to SS} \bigg( n_S^2 (T_S)  -n_S (T_S) n_S^{\text{eq}}(T_S)   \bigg) 
  \,, \label{eq:ts-t}
\end{align}
where
\begin{align}
\delta_\Gamma (x_i) \equiv \frac{c(T_i)}{ m_S^2 T_i \frac{g_S}{2\pi^2} K_1(x_i \cdot m_S/m_X)}  \,,
\end{align}
which approaches 1 in a nonrelativistic $S$ limit.
If considering the temperature below which the DM and hidden scalar are kinetically decoupled, i.e.,   $T_X \not=T_S$, we need to further include the following two terms to the RHS of Eq.~(\ref{eq:ts-t}),
\begin{align}
- (2-\delta_H) \gamma_S \, (T_S - T_X)
+ \langle \sigma v \rangle_{XX\to SS}(T_X) \frac{ n_X^2 (T_X) }{n_S (T_S)} (T_X - T_S) \,, \label{eq:kd-xs}
\end{align}
where the first and second terms are the kinetic energy-transfer rates by elastic scattering ($XS \leftrightarrow XS$) and by annihilation ($XX \leftrightarrow SS$), respectively. This impact will be discussed in (\ref{point:decay}) of Sec.~\ref{sec:analysis}. Some related results are collected in Appendix~\ref{app:xs-kd}.

By introducing the dimensionless variable, 
\begin{align}
y= \frac{T_S}{T} \,,
\end{align}
the above Boltzmann moment equation for the temperature $T_S$ can be recasted into an alternative form that will used in the analysis:
\begin{align}
\frac{d y}{dx} =
& - \Big( (2 - \delta_H)\frac{\tilde{h}_{\rm eff}(T)}{h_{\rm eff}(T)} -1 \Big) \frac{y}{x}
  -  \frac{a}{x^5} \Big( \frac{\tilde{h}_{\rm eff}(T)}{h_{\rm eff}(T)} \Big) \Big(\frac{m_X}{m_S} \Big)^{4} (y-1) 
  \nonumber\\
&  + \delta_{\rm dof} \frac{y }{ x^2 \langle \sigma v \rangle^{(0)}_{XX\to SS} } 
     \Big( y_S   \langle \sigma v \rangle_{SS\to \sum_{ij} {\rm SM}_i {\rm SM}_j} (x_S)
     -  \frac{( y_S^{\text{eq}} (x) )^2}{y_S}  \langle \sigma v \rangle_{SS\to \sum_{ij} {\rm SM}_i {\rm SM}_j} (x)   \Big)
   \nonumber \\
&  - \delta_{\rm dof} \frac{1}{ x^2 \langle \sigma v \rangle^{(0)}_{XX\to SS} } 
     \Big(y\,  y_S    \widetilde{\langle \sigma v \rangle}_{SS\to \sum_{ij} {\rm SM}_i {\rm SM}_j} (x_S)
     -  \frac{ (y_S^{\text{eq}} (x) )^2}{y_S}  \widetilde{\langle \sigma v \rangle}_{ SS\to \sum_{ij} {\rm SM}_i {\rm SM}_j } (x)   \Big)
   \nonumber \\
&  -x \frac{ \sqrt{90}}{\pi} M_{\rm pl} \frac{g_*^{1/2}(T)}{h_{\text{eff}}(T) } 
    \frac{ \Gamma_{S}}{m_X^2}  
    \bigg(  \frac{K_1(x \cdot \frac{m_S}{m_X})}{K_2(x \cdot \frac{m_S}{m_X})} \frac{ y_S^{\text{eq}} (x) }{ y_S }  (y- \delta_\Gamma (x) ) 
   -         \frac{K_1(x_S\cdot \frac{m_S}{m_X})}{K_2(x_S\cdot \frac{m_S}{m_X})} (1-\delta_\Gamma (x_S) ) y
     \bigg)
\nonumber\\
 &  
  +  \frac{\delta_{\rm dof}}{x^4} \frac{\pi} {\sqrt{90}} \frac{h_{\text{eff}}(T) }{g_*^{1/2}(m_X)} \frac{m_X}{M_{\rm pl} }  \nonumber  \\
&\   \times    \Bigg[
    \frac{(4m_X^2-m_S^2)(16m_X^2 -m_S^2)}{108 m_X(8m_X^2+m_S^2)}
    \frac{ \langle \sigma v^2 \rangle_{XXX\to XS}}{  (\langle \sigma v\rangle^{(0)}_{XX\to SS})^2 } 
    \bigg( \frac{y_X^3 }{y_S} - \frac{  y_X  \big( y_X^{\text{eq}} (x_S) \big)^2}{ y_S^{\text{eq}}(x_S) } \bigg)   
    \nonumber\\
&\   +  \frac{(2m_X+3m_S)(2m_X-m_S)}{6 (2m_X+m_S)} 
    \frac{ \langle \sigma v^2 \rangle_{XXS\to SS}}{  (\langle \sigma v\rangle^{(0)}_{XX\to SS})^2 } 
    \bigg( y_X^2- \frac{y_S}{y_S^{\text{eq}} (x_S) } \big(y_X^{\text{eq}}(x_S) \big)^2  \bigg)   
 \nonumber\\
& \  + \frac{ m_S(2m_X+m_S)(2m_X+3m_S)}{4(m_X+2m_S)(4m_X+5m_S)}  
  \frac{ \langle \sigma v^2 \rangle_{XSS\to XS}}{  (\langle \sigma v\rangle^{(0)}_{XX\to SS})^2 }  
   \Big( y_X y_S - y_X y_S^{\text{eq}}(x_S)  \Big) 
 \nonumber\\
& \   + \frac{5}{54}  m_S 
  \frac{ \langle \sigma v^2 \rangle_{SSS\to SS}}{  (\langle \sigma v\rangle^{(0)}_{XX\to SS})^2 }  
   \Big( y_S^2 - y_S  y_S^{\text{eq}}(x_S)  \Big)
   \Bigg] 
 \,,  \label{eq:y-x}
 \end{align}
where $\widetilde{\langle \sigma v \rangle}_{SS\to \sum_{ij} {\rm SM}_i {\rm SM}_j} (x_k) 
\equiv  \langle \sigma v \cdot \frac{{\bf p}_S^2}{3 E_S} \rangle_{S S \to \sum_{ij} \text{SM}_i \text{SM}_j} (x_k) /T_k 
$ and
 $a (m_X/m_S)^4 /x^4 \equiv (2-\delta_H) \gamma /H(T) $.  In Eq.~(\ref{eq:y-x}),
we do not distinguish $T_X$ from $T_S$.  
An unphysical result may occur when $y_S(x_S) \gg y_S^{\rm eq}(x)$ and other interaction terms become much smaller than that  involving $1-\delta_\Gamma$ which originates from the interaction, $S \leftrightarrow \text{SM~SM}$.
The $(1-\delta_\Gamma)$ term  may result in $y$ to be monotonically increasing with $x$, even though it is highly small and $1-\delta_\Gamma$, which is less than 0.05 for $x>20$, vanishes in nonrelativistic limit. 
Actually, when we consider completely coupled temperature evolutions for $X$ and $S$, such an unphysical result will be suppressed by including the sizable $XS \leftrightarrow XS$ and $X X \leftrightarrow S S$  interactions. For instance,  the terms, which are rewritten from Eq.~(\ref{eq:kd-xs}) and not shown in Eq.~(\ref{eq:y-x}), are given by
 \begin{align}
  - \left[
   (2-\delta_H) \frac{\tilde{h}_{\rm eff}(T)}{h_{\rm eff}(T)}  \frac{\gamma_S}{H(T)} + 
   \delta_{\rm dof} \delta_{\rm XS} \frac{y_X^2(x_X) }{ x^2  y_S(x_S) } 
   \right]
     \left( y - y_X^T \right) \,,
\label{eq:xs-kd-o}
\end{align}
with $y_X^T\equiv T_X/T$.  When the DM freezes out, the effect of the $(1-\delta_\Gamma)$ term is completely washed out due to a much larger DM energy density compared with the hidden scalar one. A more detailed treatment is described as follows.

For most of cases (the exception one will be discussed below) that the hidden sector is decoupled from the bath at temperature $T \approx m_S$ (i.e., $y_S(x_S) > y_S^{\rm eq}(x)$ for $x\gtrsim 1$), the coefficient given in the square bracket of Eq.~(\ref{eq:xs-kd-o}) is always much larger than that of the $S \leftrightarrow \text{SM~SM}$  interaction term involving $1-\delta_\Gamma$.
In other words, the evolution of $T_S$ will closely follow $T_X$ for $x\gtrsim 1$. Moreover, after the time that the hidden sector is decoupled from the SM bath, the DM plays as an effective reservoir with respect to the hidden scalar. Because the dof ratio $g_X/ g_S =3$ and $m_X >m_S$, the $S$ temperature change rate due to the $(1-\delta)$ term can thus be reduced by about $90\%$ and $75\%$ for the cases $m_S=0.8\, m_X$ and $m_S=0.99\, m_X$, respectively.
As such, when the cannibal interaction becomes inactive, both $T_S$ and $T_X$, following $dT_{S,X}/dt + (2-\delta_H) H T_{S,X} \approx 0 $ and  $\propto  a^{-2}$, will approximately evolve with the same temperature. Based on the above reasons, we will neglect $(1-\delta_\Gamma)$ term, i.e., simply take  $\delta_\Gamma =1$, for the case that the hidden sector is decoupled from the bath at $T \approx m_S$. 

Three remarks are in order. First,  we have neglected the reheating of the bath due to the out-of-equilibrium decay of $S$. 
 A thorough treatment of the reheating is beyond the scope of the present work, since the bath temperature is not a suitable variable to take into account the thermal evolution and the total comoving entropy, which is no longer conserved, increases after the out-of-equilibrium $S$ decay occurs.
 Second, the uncertainty due to reheating of the bath can be realized as follows.  After decoupling, compared with the SM radiation energy density $\rho_{\rm SM} \propto a^{-4}$,  the energy density of $S$ evolves as $\rho_S \propto a^{-3}$. Thus, the value of  $\rho_S /\rho_{\rm SM}$ is about  $1/120 $ at  temperature $\sim m_S$, but is increased to be  $\sim(1/120)\times m_S/ T$ at a later time with temperature $T$.  Adopting the sudden-decay approximation, the bath temperature change due to reheating is less than 3\% for the case with $\alpha<5\times 10^{-7}$, and $\sim 10\%$ for $\alpha =1\times 10^{-7}$, where we have used the temperature result ($T_{\rm de}^{\rm out}$) of the out-of-equilibrium $S$ decay from the next section. 
 Third, unlike the other ones,  for the case with $\alpha=1\times 10^{-7}$ and $m_S=0.8m_X$, because the down scattering results in $y_S> y_X$,  the term involving $1-\delta_\Gamma$ can be comparable with  the coefficient given in the square bracket of Eq.~(\ref{eq:xs-kd-o})  in a very short period of time just before DM freeze-out. 
It may result in a small temperature difference between $X$ and $S$ in such a short period of time. However, after that the DM and hidden scalar will still evolve with the almost same temperature, $\propto a^{-2}$ until and after their kinetic decoupling, as the other cases. Here, for simplicity,  we will not consider such effect for this case.

On the other hand,  as for the case that  the $\text{SM SM} \leftrightarrow S$ interaction is strong enough to maintain the hidden scalar in thermal equilibrium with the bath until a temperature, which may be below the DM freeze-out temperature, the cooling rate of the hidden scalar  via $SS \to XX$ will be soon larger than the $\text{SM SM} \to S$  heating rate (described by the $(1-\delta_\Gamma)$ term) in magnitude after the dark matter is kinetically decoupled from the hidden scalar.
To have a more precise estimate for the temperature ($T_{S,{\rm end}}=m_X/x_{S,{\rm end}}$) below that the hidden scalar cannot be in thermal equilibrium with the bath, we will model a term as given in Eq.~(\ref{eq:xs-kd}) to show the possible thermal flow due to the temperature difference $T_S>T_X$, which will occur after $X$ is kinetically decoupled from the hidden scalar.  In the analysis, we improve the numerical result {\it recursively}. The approach is described as follows. We will first use the approximation $\delta_\Gamma=1$ to get the numerical result. From that we then extract the DM kinetic decoupling temperature, $T_X^{\rm kd}$, which will be defined and discussed in the next section. Adopting the obtained $T_X^{\rm kd}$, we use the exact value of $\delta_\Gamma$ and  include  the $X X \leftrightarrow S S$ term (rewritten from the second term of Eq.~(\ref{eq:kd-xs})),
  \begin{align}
  - \delta_{\rm dof} \delta_{\rm XS} \frac{y_X^2(x_X) }{ x^2  y_S(x_S) } 
     \left( y - \frac{x_X^{\rm kd}}{x}   \right)  \theta(x- x_X^{\rm kd}) \,.
\label{eq:xs-kd}
\end{align}
in Eq.~(\ref{eq:y-x}) to have the improved solution for $x_{S,{\rm end}}$, where  $x_X^{\rm kd} \equiv m_X/T_X^{\rm kd}$. Here the $XS \leftrightarrow XS$ term is negligible for this case, and the relation $T_X \simeq T_X^{\rm kd} \cdot  (T/T_X^{\rm kd})^2$ is used when the DM and hidden scalar are kinetically decoupled from each other.

\section{Numerical results for the thermal evolution of the hidden sector}\label{sec:analysis}

We  present the numerical results for  thermodynamic evolutions of the normalized yields (proportional to co-moving number densities) and hidden sector temperatures using two sets of masses for the hidden sector: (i) $m_X=80$~GeV, $m_S=0.8 m_X=64$~GeV, and (ii) $m_X=80$~GeV, $m_S =0.99 m_X=79.2$~GeV, where the latter one is the nearly degenerate case.
These two sets of the hidden masses are capable of generating one-step cascade DM annihilation spectra that provide a good fit to the observed GC gamma-ray excess which will be further discussed in the next section.  

\begin{figure}[h!]
\begin{center}
\vskip-0.35cm
\includegraphics[width=0.372\textwidth]{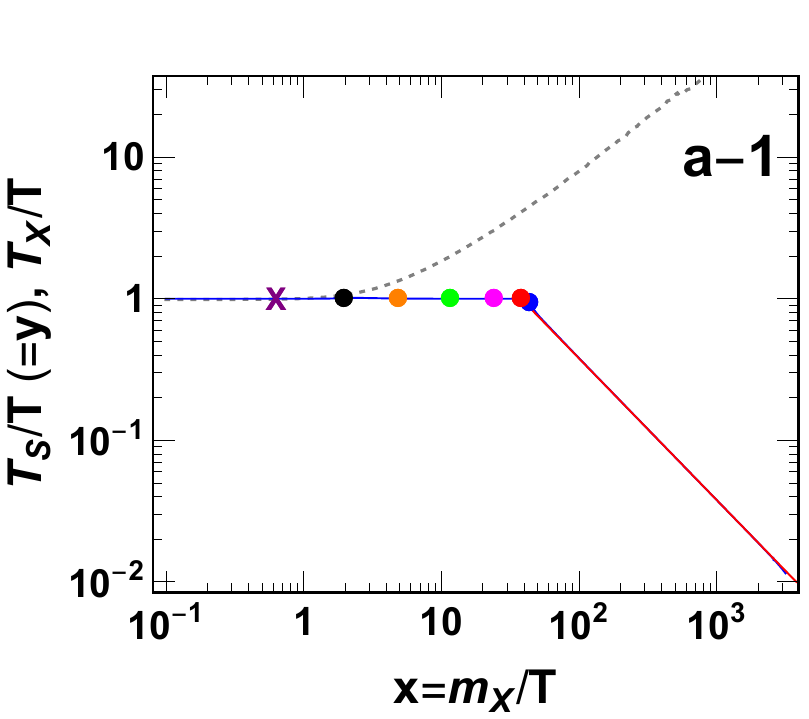}\hskip0.5cm
\includegraphics[width=0.372\textwidth]{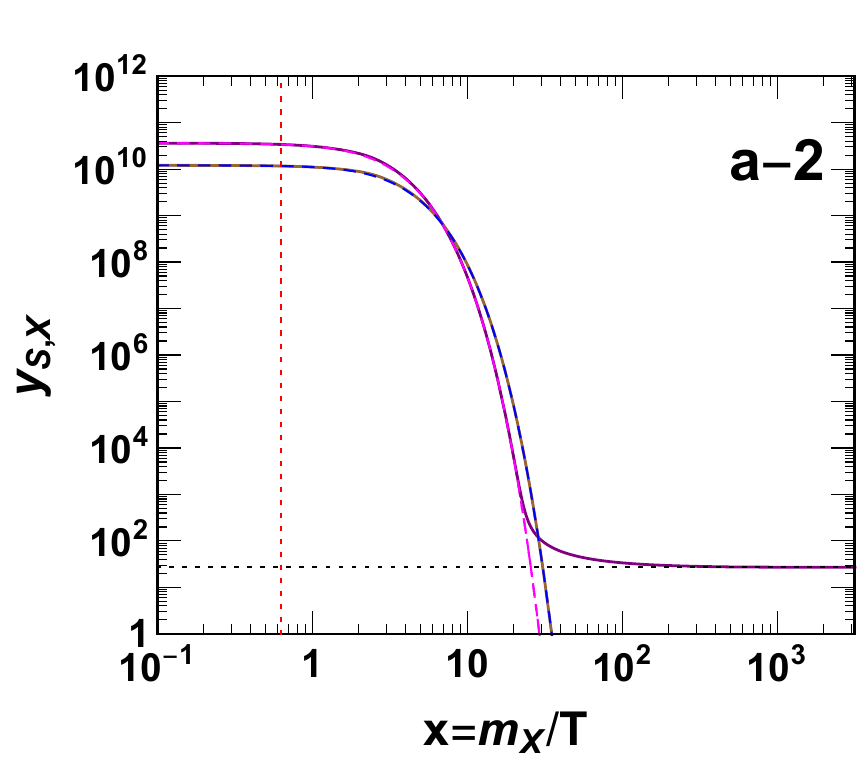}\\
\vskip-0.35cm
\includegraphics[width=0.372\textwidth]{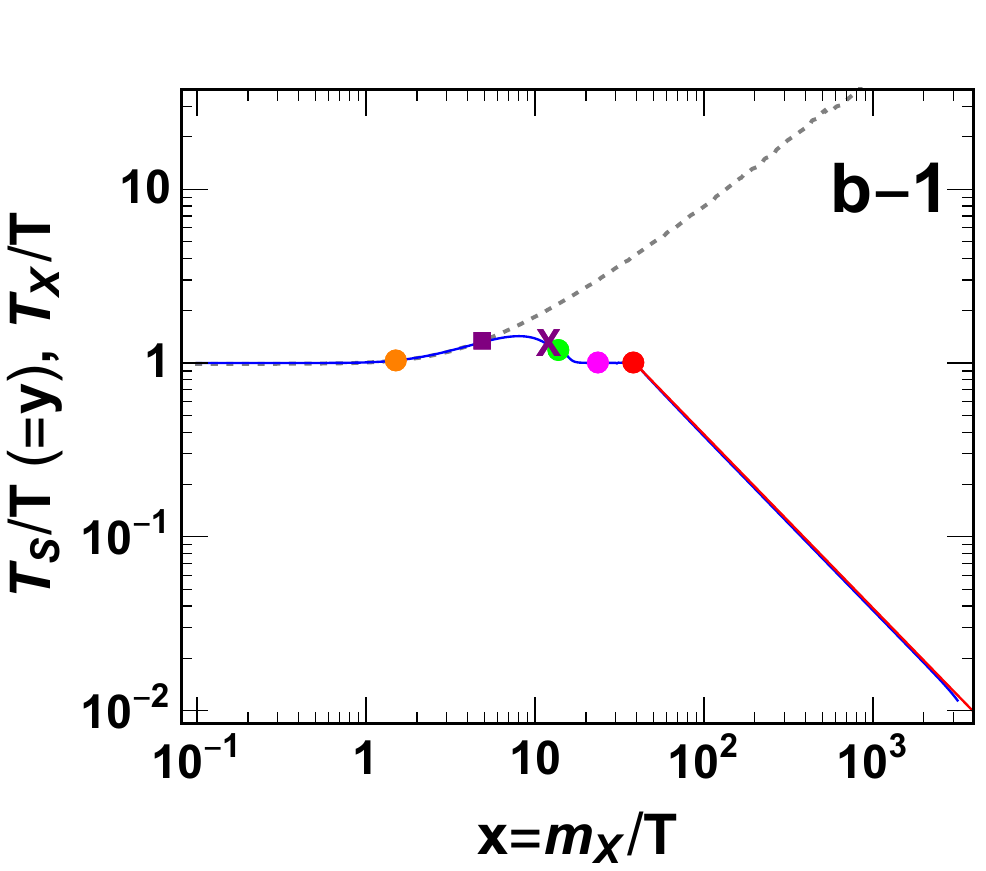}\hskip0.5cm
\includegraphics[width=0.372\textwidth]{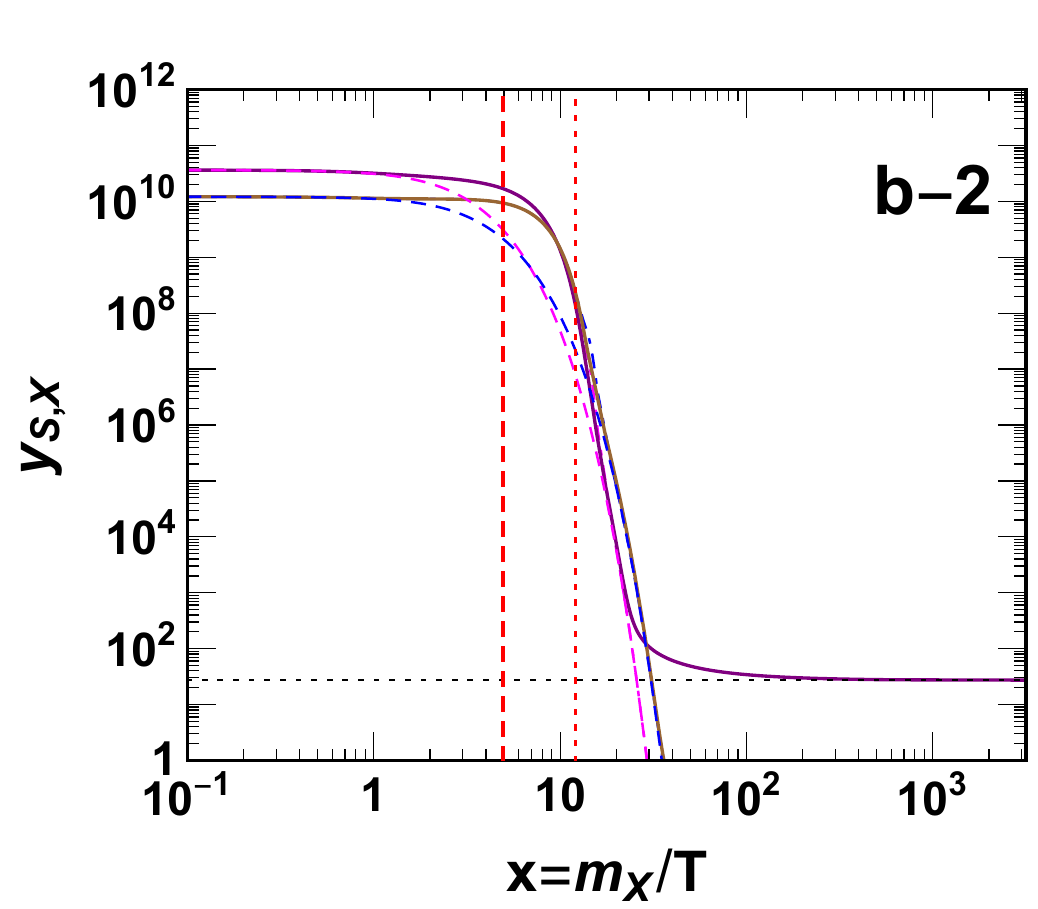}\\
\vskip-0.35cm
\caption{Left panel:  $T_S/T$ and $T_X/T$ versus $m_X/T$, where $T$ is the bath temperature.  As $T$ is below that denoted by the red dot, 
$T_X/T$ follows the red line. 
The DM freeze-out temperature is denoted by the magenta dot.  
For $T$ below that denoted by black and orange dots, the hidden sector cannot keep the thermal equilibrium with the bath via the inverse $S$ annihilation and $S$-SM elastic scattering, respectively, while for $T$ below that denoted by the blue dot, 
$S$ is completely thermally decoupled from the bath. 
 When $T$  is below that denoted by the purple ``X" (corresponding to the vertical dotted (red) line in the right panel),  the heating rate due to the inverse decay of $S$ can be larger than that needed to keep the hidden sector in thermal equilibrium with the bath.  The purple square (corresponding to the vertical dashed (red) line in the right panel) stands for the temperature below which the $S$ undergoes an out-of-equilibrium decay.
When $T$ falls below that denoted by the green dot, the heating rate of the cannibal process in the hidden sector is less than $H$.
The conservation of the comoving hidden sector entropy is described by the dotted curve (see Eq.~(\ref{eq:hidden-entropy})). Right panel:
$y_X$ (purple solid line) and $y_S$ (brown solid line) as functions of $x$, where the magenta and  blue dashed (or dashed-dotted) lines show the corresponding yields if following Boltzmann suppression with $T_{S,X}=T$ (or with their true temperatures). The horizontal line denotes the asymptotic DM yield, $y_X^\infty $.  Here we use $m_X=80$~GeV, $m_S=64$~GeV, and,  in (a) and (b), separately adopt $\alpha= 1\times 10^{-5}$ and $1\times 10^{-6}$. The value $g_{\rm dm}$ is determined to have the correct relic abundance.
}
\label{fig:relic-not-equal-1}
\end{center}
\end{figure}

\begin{figure}[h!]
\begin{center}
\vskip-0.35cm
\includegraphics[width=0.371\textwidth]{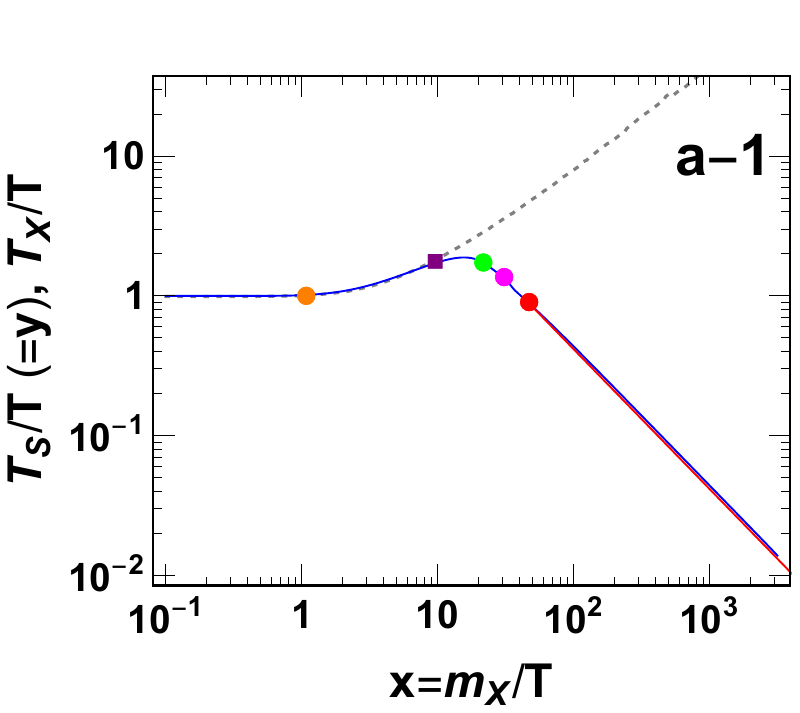}\hskip0.5cm
\includegraphics[width=0.371\textwidth]{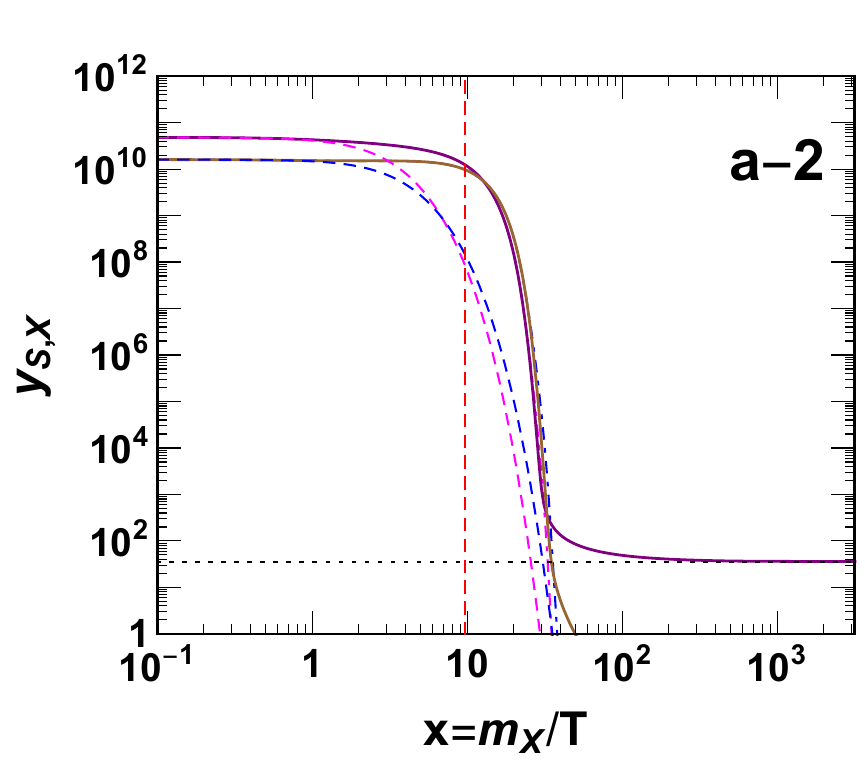}\\
\vskip-0.33cm
\includegraphics[width=0.371\textwidth]{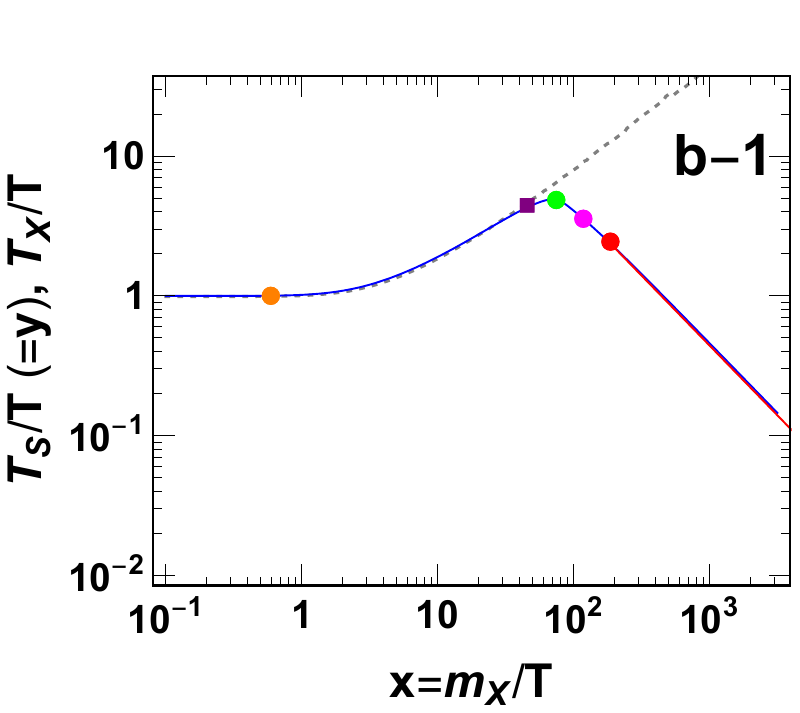}\hskip0.5cm
\includegraphics[width=0.371\textwidth]{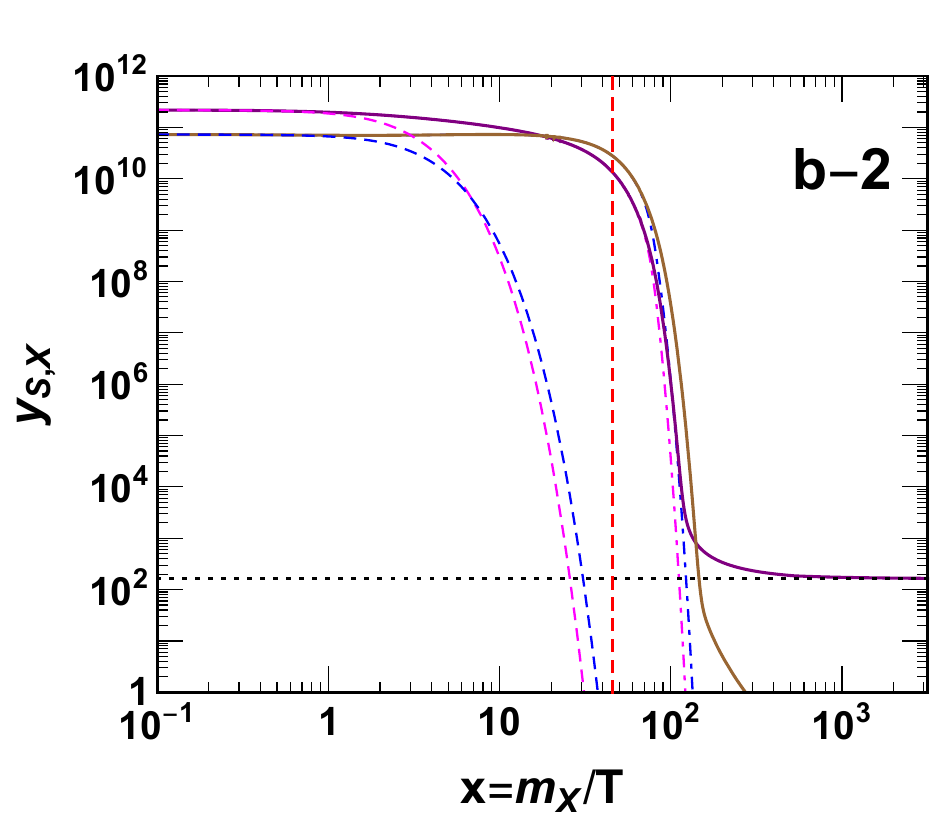}\\
\vskip-0.33cm
\caption{ Same as Fig.~\ref{fig:relic-not-equal-1}, but using $\alpha= 5\times 10^{-7}$, and $1\times 10^{-7}$ for (a), and (b), respectively.
}
\label{fig:relic-not-equal-2}
\end{center}
\end{figure}

\begin{figure}[t!]
\begin{center}
\vskip-0.35cm
\includegraphics[width=0.371\textwidth]{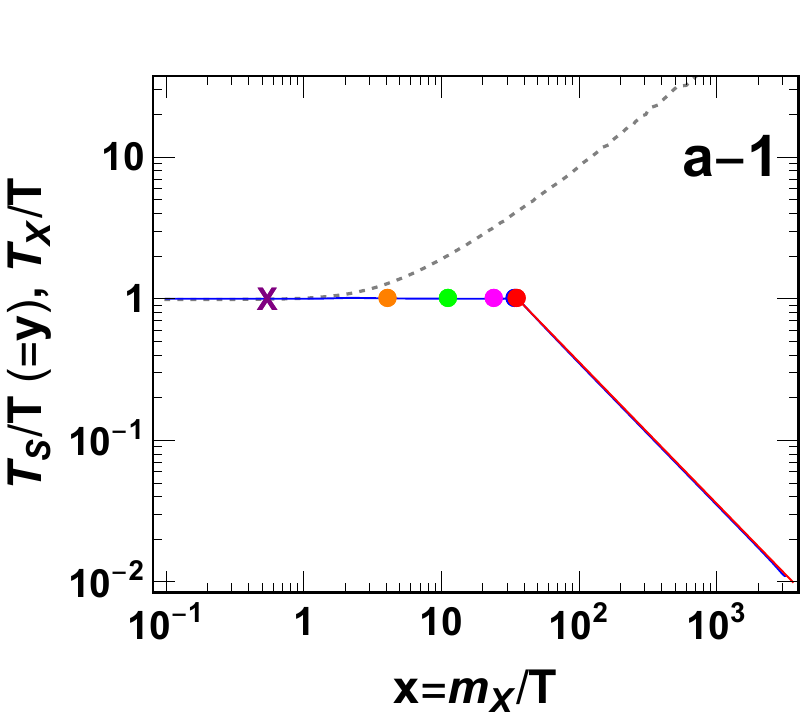}\hskip0.5cm
\includegraphics[width=0.371\textwidth]{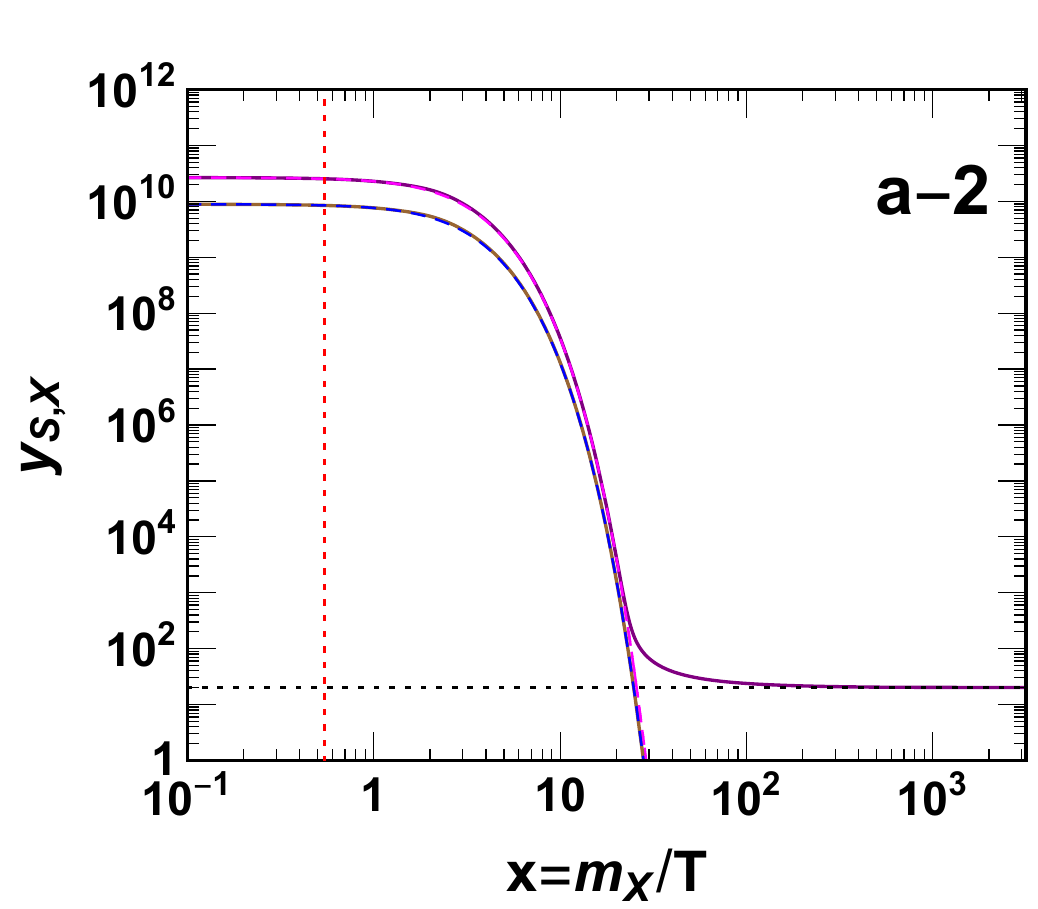}\\
\vskip-0.35cm
\includegraphics[width=0.371\textwidth]{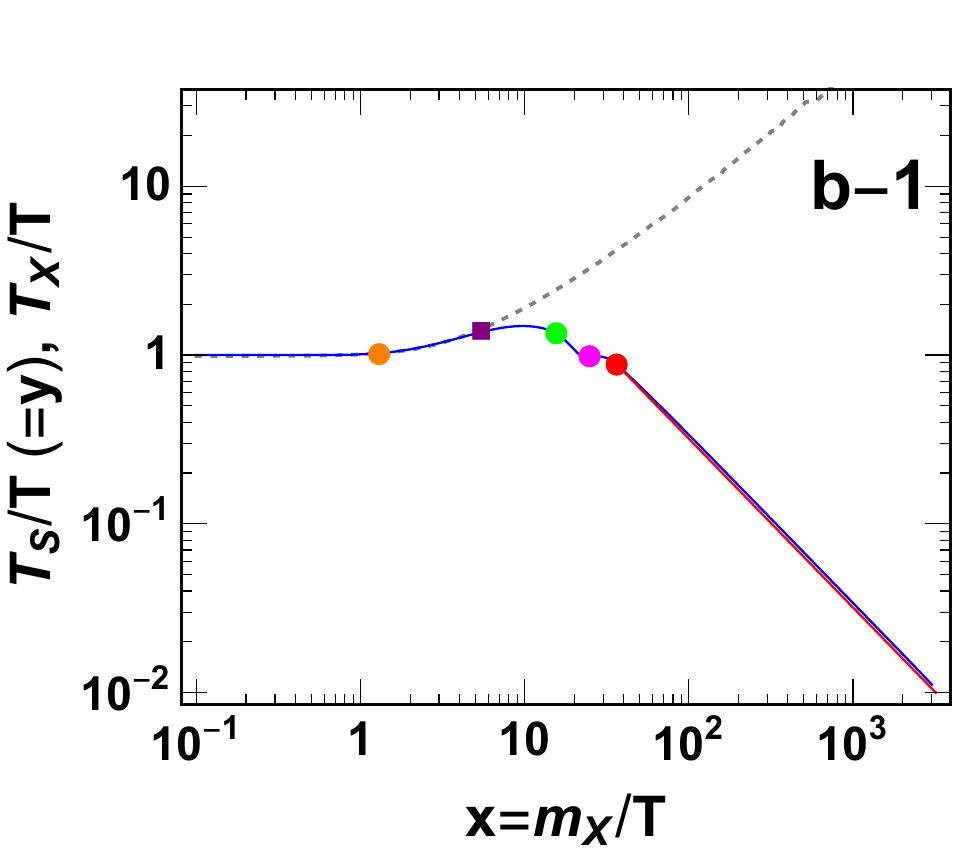}\hskip0.5cm
\includegraphics[width=0.371\textwidth]{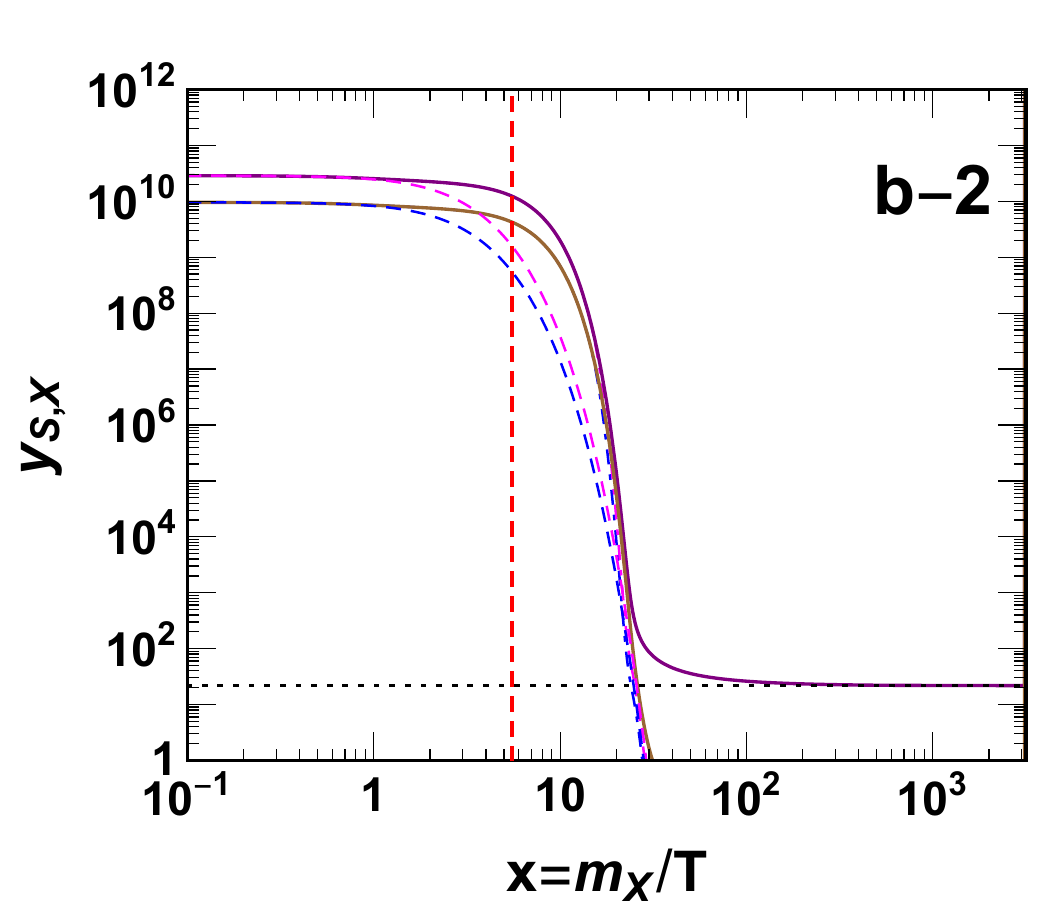}\\
\vskip-0.35cm
\includegraphics[width=0.371\textwidth]{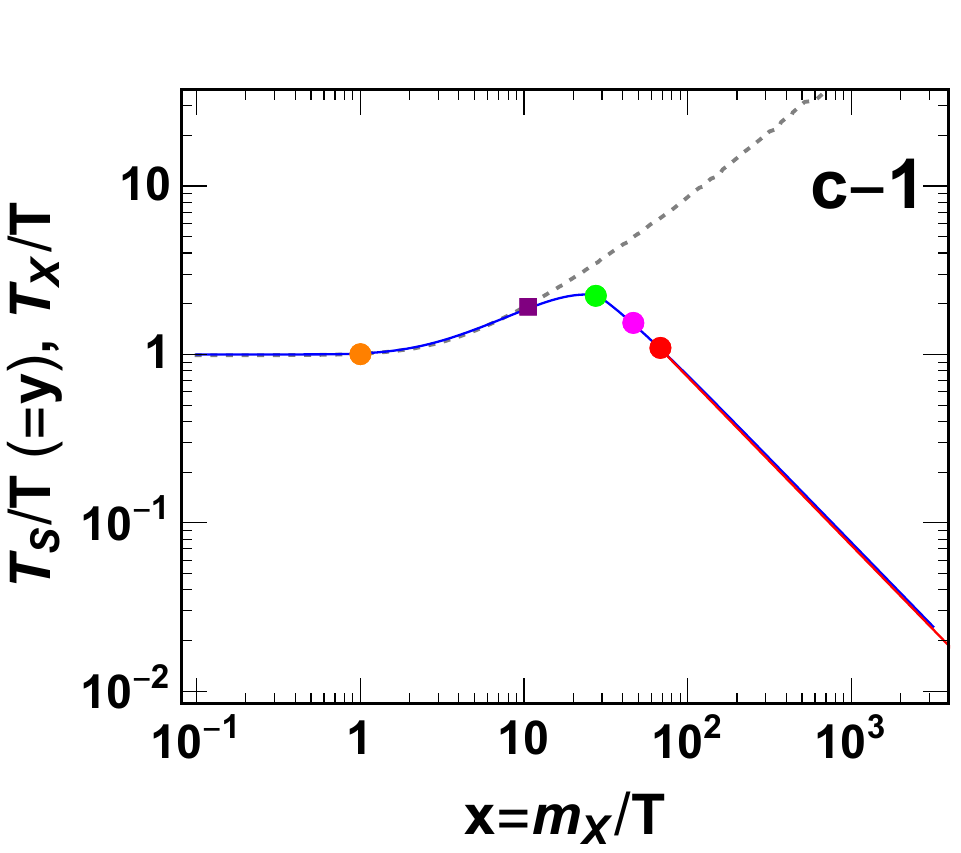}\hskip0.5cm
\includegraphics[width=0.371\textwidth]{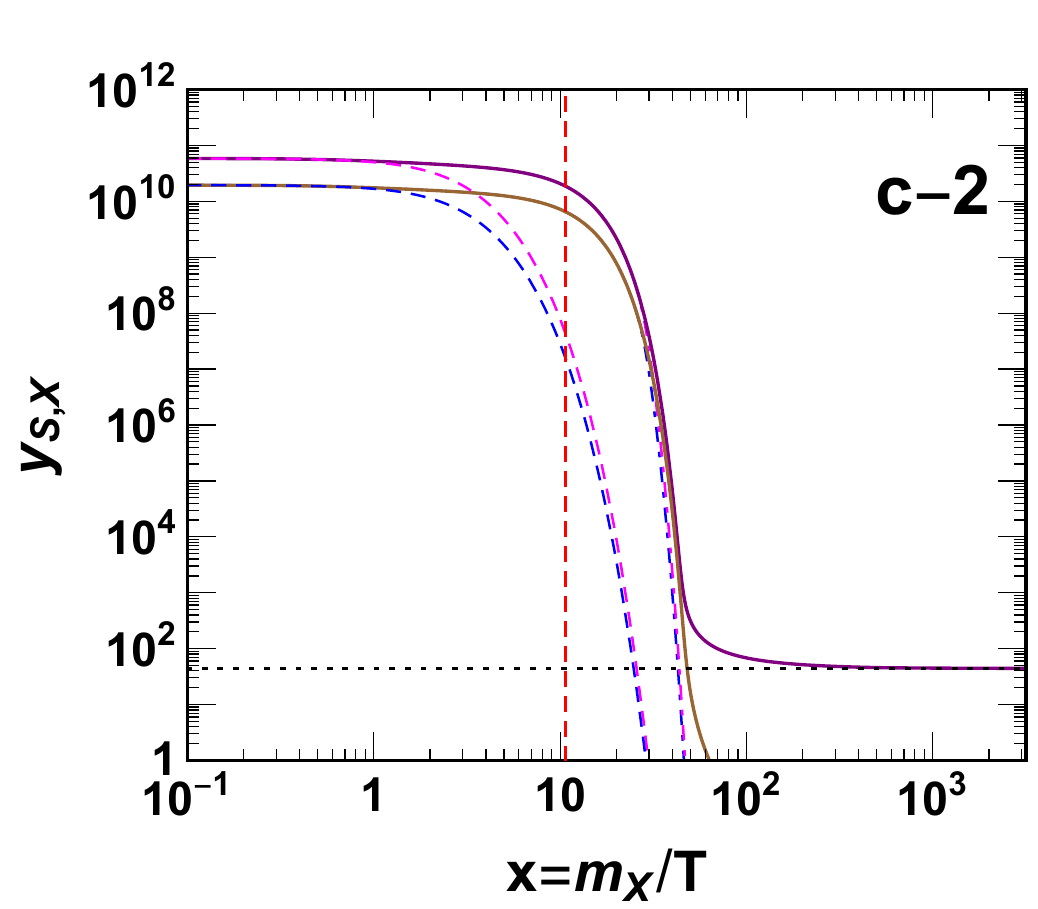}\\
\vskip-0.35cm
\includegraphics[width=0.371\textwidth]{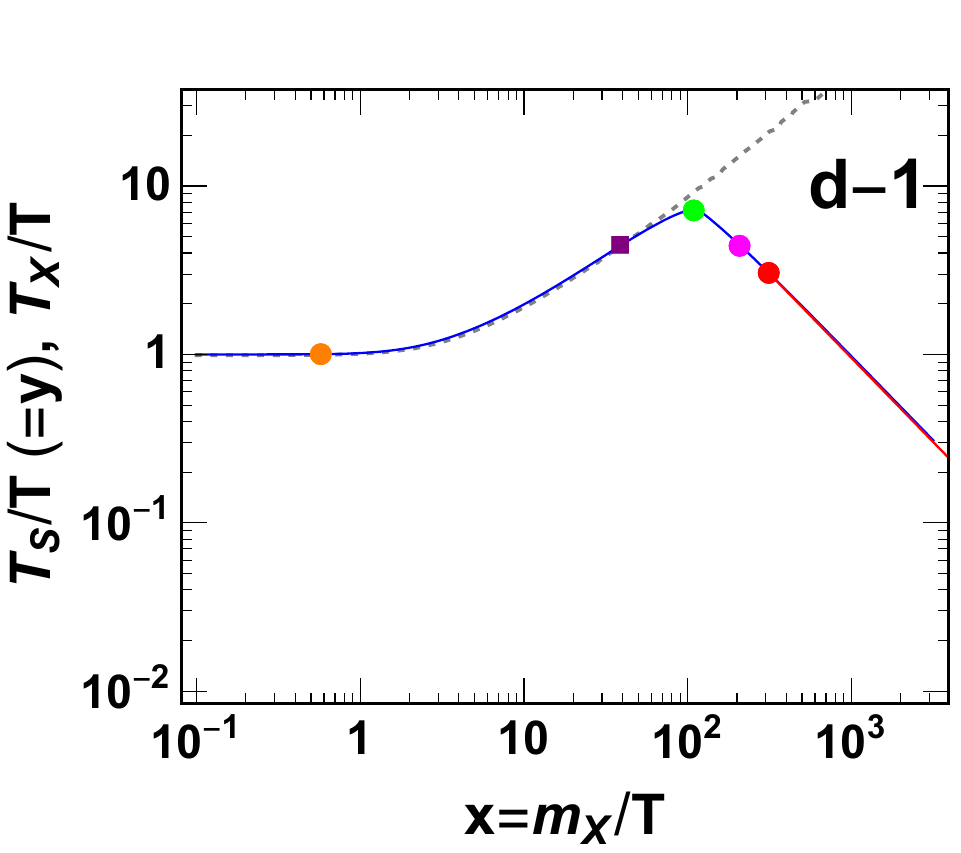}\hskip0.5cm
\includegraphics[width=0.371\textwidth]{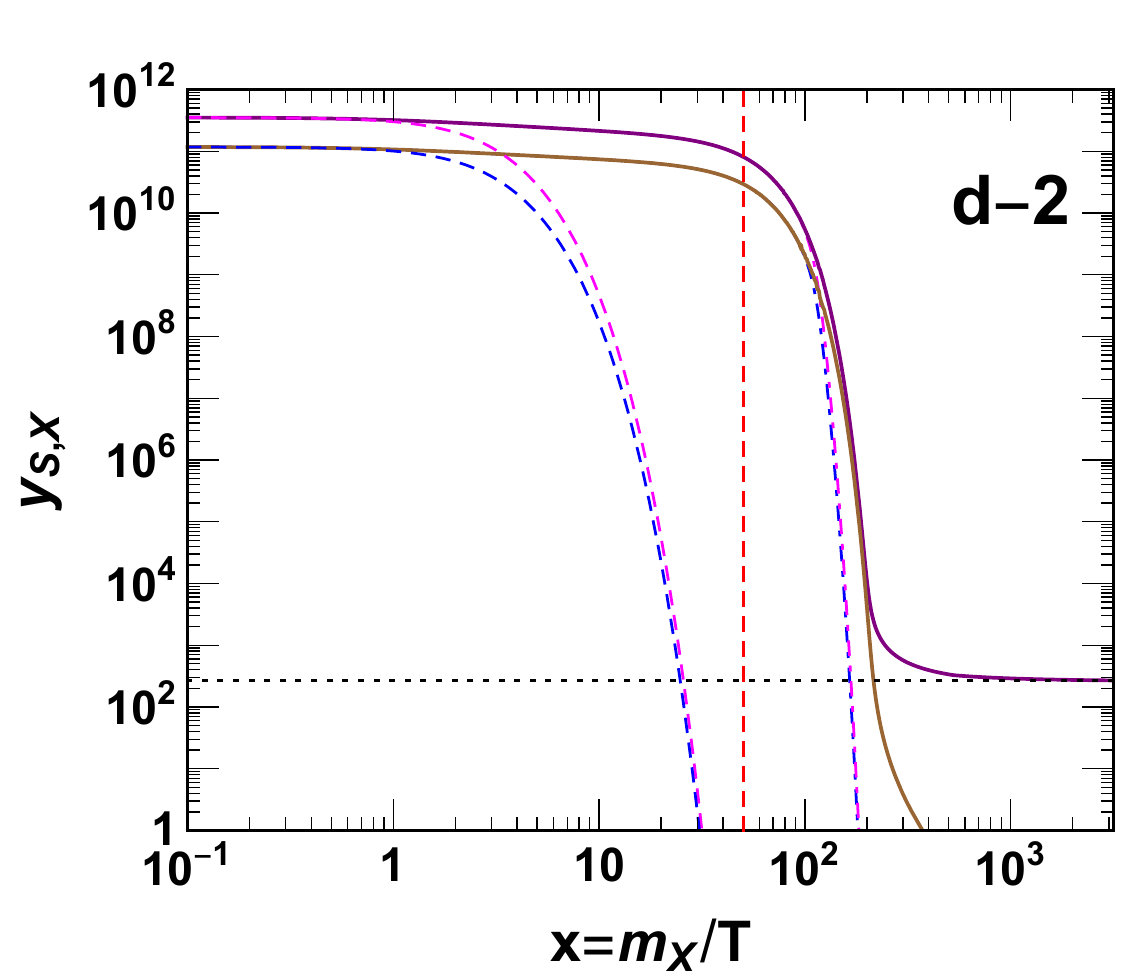}\\
\caption{ Same as Fig.~\ref{fig:relic-not-equal-1}, but using $m_X=80$~GeV and $m_S=0.99\, m_X$ and,  in (a), (b), (c), and (d), separately adopting $\alpha= 1\times 10^{-5},  1\times 10^{-6},  5\times 10^{-7}$, and $1\times 10^{-7}$. 
}
\label{fig:relic-equal}
\end{center}
\end{figure}

To illustrate how this secluded DM model could be highly decoupled from the SM bath, for these two mass sets, we take small mixing angle $\alpha$, i.e. to be (1) $1\times 10^{-5}$, (2) $1\times 10^{-6}$, (3) $5\times 10^{-7}$, and (4) $1\times 10^{-7}$, respectively. Meanwhile, in the analysis, we set the parameter $g_{\rm dm}$, of which the value is relevant to  $XX \to SS$ annihilation cross section, to account for  the observed DM relic abundance  (see Eq.~(\ref{eq:abundance})). Our results are shown in Figs.~\ref{fig:relic-not-equal-1}, \ref{fig:relic-not-equal-2}, and \ref{fig:relic-equal}.
  
 The  equilibrium number densities, described by the Boltzmann equations given in Eqs. (\ref{eq:boltzmann-1}) and (\ref{eq:boltzmann-2}),  can be maintained by  interactions, including $XX \leftrightarrow SS$, $SS \leftrightarrow \text{ SM SM}$,  $S \leftrightarrow \text{ SM SM}$, and $3 \leftrightarrow 2$ hidden sector cannibalization. 
 The temperature evolution of the hidden scalar is affected by the kinetic energy transfer by interacting with the bath and with the DM. Such a kinetic energy transfer can be generated from the hidden scalar number changing interactions and  from the elastic scattering.
 
The main results, categorized in terms of  temperature scales  relevant to the transition phases during the evolution of the hidden sector,  are summarized as follows.

\begin{enumerate} [(i)]

\item $x_f \equiv m_X /T_{\rm f}$ is the usual freeze-out temperature variable. For $T<T_{\rm f}$, the comoving DM number density tend to be conserved.  From Eq.~(\ref{eq:boltz-1}), we can estimate  the freeze-out temperature, below which 
 the DM production rate from $SS \to XX$  is overtaken by the dilution rate, giving the relation
\begin{align}
\langle \sigma v \rangle_{XX\to SS} \frac{(n_X^{\rm eq} (T_S))^2}{(n_S^{\rm eq} (T_S))^2}  \, (n_S (T_S))^2  \lesssim 3 H(T)  n_X (T_X) \,, \label{eq:freeze-out}
\end{align}
where $T_X=T_S$  before freeze out are functions of $T$, and $T_S$ can be determined by solving numerically the Boltzmann equations, Eqs.~(\ref{eq:boltzmann-1}), (\ref{eq:boltzmann-2}) and (\ref{eq:y-x}).
 In Figs.~\ref{fig:relic-not-equal-1}, \ref{fig:relic-not-equal-2}, and \ref{fig:relic-equal}, $x_f$ is denoted by the magenta dot in each plot of the left panel,  while the corresponding asymptotic yield $y_X^\infty$, described by Eq.~(\ref{eq:approx-xf}),  is depicted by the horizontal dotted line on the right panel. 

For $T>T_{\rm f}$, the rate on the left hand side (LHS) of Eq.~(\ref{eq:freeze-out}) is larger than the expansion rate, resulting in the detailed balance
$\langle \sigma v \rangle_{XX\to SS}(T_X)\,  (n_X (T_X))^2 = \langle \sigma v \rangle_{SS\to XX}(T_S) \,  (n_S (T_S))^2$.
 This implies that
\begin{align}
\frac{n_X^2}{(n_X^{\text{eq}})^2}  = \frac{n_S^2}{(n_S^{\text{eq}})^2} \,,
\end{align}
so that $X$ and $S$ (with $T_X=T_S$) have the same chemical potential, $\mu_X=\mu_S$,  i.e., the hidden sector is in chemical equilibrium, for which the chemical potential can be non-zero if the hidden scalar undergoes an out-of-equilibrium decay before the DM freezes out (see (\ref{item:decay}) for related discussions).

 In the right panel of Figs.~\ref{fig:relic-not-equal-1}, \ref{fig:relic-not-equal-2}, and \ref{fig:relic-equal}, the dashed curves exhibit the results with the Boltzmann suppression  (i.e., with $\mu_X=\mu_S=0$) for the hidden sector if $T_{X,S}=T$, while dot-dashed curves follow the Boltzmann suppression at their true temperatures, where curves in magenta and blue colors are for the DM and hidden scalar, respectively.

\item $x_{\rm ann,2}\equiv m_X/T_{\rm ann,2}$, plotted as the black dot  in the left panel of Fig.~\ref{fig:relic-not-equal-1},  corresponds to the bath temperature $T=T_{\rm ann,2}$ set by 
\begin{align}
& (n_S^{\rm eq} (T))^2 \langle \sigma v \cdot \frac{{\bf p}_S^2}{3 E_S} \rangle_{\small S S \to \sum_i \text{SM}_i \text{SM}_i} (T)  \nonumber\\
& \qquad\qquad  \simeq (2 - \delta_H) H(T)  n_S (T_S) T_S
+ \frac{T_S}{a^3}  \left[ \frac{d \big( n_S(T_S) a^3 \big)}{dt} \right]_{SS\leftrightarrow \sum_{ij} {\rm SM}_i {\rm SM}_j} ,
\label{eq:temp-ann}
\end{align}
 where $T_S$, the temperature of $S$, is a function of $T$ determined by Eq.~(\ref{eq:ts-t}), and the second term of the RHS is described by  Eq.~(\ref{eq:boltz-t}).  Here, considering the relevant channels $S S \leftrightarrow  \sum_{ij} {\rm SM}_i~{\rm SM}_j $,  the value of  $(T_S/a^3) \, d(n_S a^3)/dt =T_S (d n_S /dt +3 H n_S)  \propto dy_S/dt$ is about zero for $x\lesssim x_{{\rm ann},2}$, if the thermal equilibrium between the hidden scalar and bath can be maintained by this annihilation reaction.
 When the bath temperature falls below $T_{\rm ann,2}$, 
 the kinetic energy changing rate of the hidden scalar due to variations of its temperature (the first term of RHS) and density  (the second term of RHS) during the cosmic expansion becomes larger in magnitude than the heating rate transported from the bath via annihilations $\sum_{ij}{\rm SM}_i~ {\rm SM}_j \to S S$, such that this kind of interactions cannot play the role to keep thermal equilibrium with the SM bath for $T < T_{\rm ann,2}$. 
 In Figs.~\ref{fig:relic-not-equal-1}(b-1), \ref{fig:relic-not-equal-2}, and \ref{fig:relic-equal}, $x_{\rm ann,2}$ is too small to be visible in plots.

\item \label{point:xel}
$x_{\rm el} \equiv m_X/ T_{\rm el}$, sketched as the orange dot in the left panel of Figs.~\ref{fig:relic-not-equal-1},  \ref{fig:relic-not-equal-2}, and \ref{fig:relic-equal}, denotes the temperature below which the hidden sector is elastically decoupled from the bath. For $T > T_{\rm el}$,  the heating rate of the hidden sector, which gains  energy by the elastic scattering $S~{\rm SM} \to S~{\rm SM}$,  is larger in magnitude than the cooling rate due to the Hubble expansion,
\begin{align}
(2-\delta_H) \gamma\,  n_S \,  (T_S)\,  T \gtrsim (2 - \delta_H)\,  H(T) \, n_S (T_S) \,  T_S \,,
\end{align}
where $(2-\delta_H) \gamma T$ is the kinetic energy-transfer rate from the relativistic SM particles to a hidden scalar particle via elastic scattering. Unlike the inverse hidden scalar annihilation into relativistic SM particles, of which the rate is reduced by the Boltzmann suppression of the $S$ number density (with $\mu_S=0$) for $T \lesssim m_S$, the kinetic energy-transfer rate of $S~{\rm SM} \to S~{\rm SM}$ is proportional to the relativistic SM density which is not suppressed. Therefore, in most cases, we have $x_{\rm el}> x_{\rm ann, 2}$. Two remarks are in order. First, $x_{\rm el} < x_{\rm ann, 2}$ may occur, if the annihilation cross section is largely enhanced by the resonant $s$-channel SM-Higgs exchange with $s\sim (2 m_S)^2 \sim m_h^2$.
Second,  compared with the elastic scattering, because the $2\to 2$ annihilation is much more sensitive to $S$-$f$-$\bar{f}$ coupling, which is proportional to  $s_\alpha$, a small enough mixing angle $\alpha$ may result in the $2\to2$ annihilation decoupling to occur significantly before elastic decoupling.

\item \label{point:decay}
$x_{\rm de}\equiv  m_X/T_{\rm de}$ is respectively denoted by a purple ``X"  and by the vertical dotted (red) line in the left and right panels of Figs.~\ref{fig:relic-not-equal-1}(a), \ref{fig:relic-not-equal-1}(b) and \ref{fig:relic-equal}(a). $T_{\rm de}$ denotes the temperature below which not only the requirements, $n_S(T_S) = n_S^{\rm eq}(T)$ and $T_S=T$, need to be satisfied, but also the heating rate, generated from the inverse decay: $\text{SM~SM} \to S$, is larger in magnitude than the rate needed to keep the $S$ particles in kinetic equilibrium with the bath during the cosmic expansion,
 \begin{align}  
  \Gamma_{S}  
               \frac{K_1(x \cdot m_S/m_X)}{K_2(x \cdot m_S/m_X)}  n_S^{\text{eq}} (T)  \, \delta_\Gamma \, T
       \gtrsim
 (2 - \delta_H) H n_S(T_S)  T_S 
+ \frac{T_S}{a^3} \left[ \frac{d \big( n_S(T_S) a^3 \big)}{dt} \right]_{S \leftrightarrow \sum_{ij} {\rm SM}_i {\rm SM}_j} 
 \,,
 \label{eq:tde}
\end{align}  
where, on the RHS, the first term is the rate for temperature variation with an unchanged comoving number density of $S$,
while the second term, describing the thermal energy rate due to a change of the $S$ comoving number density,  results from $S \leftrightarrow \text{SM~SM}$ interactions, for which its contribution to  $(T_S/a^3) \, d(n_S a^3)/dt = T_S \big(d n_S(T_S) /dt +3 H n_S (T_S) \big) $ is relatively negligible when $T\approx T_{\rm de}$ for the case shown in Figs.~\ref{fig:relic-not-equal-1}(a) and \ref{fig:relic-equal}(a).
As such, if $T_{\rm de} > \text{min} (T_{\rm el}, T_{\rm ann,2})$ is satisfied, the hidden scalar (as well as the dark matter) can be still maintained in thermal equilibrium with the bath (i.e., $n_S(T_S)=n_S^{\rm eq}(T)$ and $T_S=T$) until a later time, $x_{\rm S, end} (\equiv m_X/ T_{\rm S,end})$, which could be larger than $x_f$. For this case, to have a more precise estimate, we have included a term given in Eq.~(\ref{eq:xs-kd}) to show the possible thermal flow due to the temperature difference $T_S>T_X$, which occurs after $X$ is kinetically decoupled from the hidden scalar. 
$x_{\rm S, end}$, denoted as the blue dot in Figs.~\ref{fig:relic-not-equal-1}(a-1) and \ref{fig:relic-equal}(a-1), is determined by
\begin{align}
& \Gamma_{S}  
               \frac{K_1(x \cdot m_S/m_X)}{K_2(x \cdot m_S/m_X)}  n_S^{\text{eq}} (T)  \, \delta_\Gamma \, T
   -   \langle \sigma v \rangle_{XX\to SS}(T_S) \frac{ n_X^{\text{eq}} (T_S) }{ n_S^{\text{eq}} (T_S) } \, n_S^2 (T_S) \, T_S
    - (2 -\delta_H) \gamma_S  n_S(T_S) T_S 
 \nonumber\\ 
 &\quad \gtrsim
  (2 - \delta_H) H n_S(T_S)  T_S 
  +    \langle \sigma v \rangle_{XX\to SS}(T_X) \, n_X^2 (T_X) \, T_S ,
\end{align}
where the second and third terms of the left hand side are respectively the rates originating from $SS \to XX$ and $X S\to X S$  collision terms of the Boltzmann moment equation, while the second term of the RHS is the contribution from $\big(d n_S(T_S) /dt +3 H n_S (T_S) \big) T_S$, resulting mainly from $XX\to SS$ due to the fact that $n_X(T_X) >n_X^{\rm eq}(T_X)$ after the DM freezes out.
The DM temperature after kinetic decoupling follows $T_X(a)\simeq T_X^{\rm kd} \cdot (a_X^{\rm kd}/a)^2$, with $a$ being the cosmic scale factor and $a_X^{\rm kd}$ being the corresponding value at $T_X=T_X^{\rm kd}$. For the cases shown in  Figs.~\ref{fig:relic-not-equal-1}(a) and \ref{fig:relic-equal}(a), this relation can be rewritten as  $T_X =T_X^{\rm kd} (T/T_X^{\rm kd})^2 $.
After DM kinetically decouples from the hidden scalar, the evolution of $T_X/T$ is sketched as the red line in the left panel of Figs.~\ref{fig:relic-not-equal-1}, \ref{fig:relic-not-equal-2},  and \ref{fig:relic-equal}, where  $ T_X^{\rm kd} \equiv m_X/x_X^{\rm kd}$, depicted as the red dot, is the DM kinetic decoupling temperature, featuring $ T_X^{\rm kd} \leq T_{\rm f}$. 
A detailed discussion for $ T_X^{\rm kd}$ will be given in Appendix \ref{app:xs-kd}.
 Numerically, we obtain $x_{\rm S, end} \approx  x_X^{\rm kd}$, as seen from Figs.~\ref{fig:relic-not-equal-1}(a-1)  and \ref{fig:relic-equal}(a-1).

On the other hand, provided that  the relation given in Eq.~(\ref{eq:tde}) is satisfied but with $T_{\rm f} \ll T_{\rm de} < \text{min}(T_{\rm el}, T_{\rm ann,2})$, i.e.,  $x_f \gg x_{\rm de} > \text{max}(x_{\rm el}, x_{\rm ann,2})$, the hidden sector may be kinetically decoupled from the bath at $T \lesssim \text{min}(T_{\rm el}, T_{\rm ann,2})$, such that at  $x \gtrsim x_{\rm de}^{\rm out}$ (see (\ref{item:decay}) for the definition), the $S$ first undergoes an out-of-equilibrium decay with a rate much larger than its inverse production rate  $\text{SM~SM} \to S$ due to the fact that $n_S \gg n_S^{\rm eq}$, resulting in the RHS of  Eq.~(\ref{eq:tde}) to be less than zero,
 \begin{align}  
 (2 - \delta_H) H n_S(T_S)  T_S 
& + \frac{T_S}{a^3}  \left[ \frac{d \big( n_S(T_S) a^3 \big)}{dt} \right]_{S \leftrightarrow \sum_{ij} {\rm SM}_i {\rm SM}_j}  \nonumber\\ 
& \simeq  (2 - \delta_H) H(T) n_S(T_S) T_S
 -    \Gamma_{S}  
               \frac{K_1(x_S \cdot m_S/m_X)}{K_2(x_S \cdot m_S/m_X)} n_S(T_S)  T_S< 0
 \,,
 \label{eq:tde-2}
\end{align}  
It is interesting to note that, as the time evolves, the $n_S$ Boltzmann equation gives $n_S(T_S) \to n_S^{\rm eq}(T)$, which is displaced in Figs.~\ref{fig:relic-not-equal-1}(b-2) and \ref{fig:relic-equal}(b-2), so that as shown in Fig.~\ref{fig:relic-not-equal-1}(b-1) the requirement of Eq.~(\ref{eq:tde}) is possible to be met and $x_{\rm end}$ can thus exist.

\item $x_{\rm c} \equiv m_X/ T_c$ corresponds to the bath temperature $T=T_c$, illustrated by the green dot in the left panel of  Figs.~\ref{fig:relic-not-equal-1}, \ref{fig:relic-not-equal-2},  and \ref{fig:relic-equal}, and described by 
\begin{align}
\sum_{i,j,k\equiv S,X}  K_{3\to2} \langle \sigma v^2 \rangle_{3\to 2}\,   n_i^{\rm eq}(T_S) n_j^{\rm eq} (T_{S})  n_k^{\rm eq} (T_{S})
\simeq 
  2 H(T) n_S^{\rm eq}(T_S) \, T_S \,,  \label{eq:Tc}
\end{align}
where $K_{3\to 2}$ shown in Eq.~(\ref{eq:ts-t}) is the kinetic energy released in a relevant $3\to 2$ process involving nonrelativistic $S$ and $X$. 
For $T > T_c$, this number changing interaction maintains the hidden sector, which is undergoing cannibalism, in kinetic equilibrium and in chemical equilibrium with $\mu_{X,S}=0$:   $n_X (T_S) \to n_X^{\rm eq}(T_S)$, $n_S (T_S) \to n_S^{\rm eq}(T_S)$. 
We are  interested in the cases, as given in Figs.~\ref{fig:relic-not-equal-1}(b), \ref{fig:relic-not-equal-2}(a), \ref{fig:relic-not-equal-2}(b) and \ref{fig:relic-equal}(b), \ref{fig:relic-equal}(c), \ref{fig:relic-equal}(d), that the hidden sector is decoupled from the thermal bath and evolves with different temperature independently, before it becomes nonrelativistic.  For these cases with $T \lesssim m_{S,X}$, the total comoving entropy density of the hidden sector tends to be conserved before the $S$ decay occurs. Moreover, during the cannibal process, the entropy density ratio for the SM, $s_{\rm SM} = (2\pi^2 /45) h_{\rm SM}^{\rm eff}(T) T^3$, to the hidden sector, $s_{h} = (2\pi^2 /45) h_{\rm h}^{\rm eff}(T_S) T_S^3$, is constant, where $h_{\rm SM}^{\rm eff}$ and $h_{\rm h}^{\rm eff}$ are the effectively  relativistic degrees of freedom of the SM and hidden sector, respectively. Thus, we find 
\begin{align}
\frac{T_S}{T} =\Big(\frac{s_h}{s_{\rm SM}} \Big)^{1/3} 
   \Bigg(    \frac{h_{\rm SM}^{\rm eff}(T) }{ h_{\rm h}^{\rm eff}(T_S)}  
   \Bigg)^{1/3} \,, \label{eq:hidden-entropy}
\end{align}
where
\begin{align}
h_{\rm h}^{\rm eff}(T_S) 
 & \simeq  \frac{45}{ (2\pi^2)^2} \sum_{h\equiv S,X}  g_h
        \Big( \frac{m_h}{T_S} \Big)^3 \Big[ K_1\Big(\frac{m_h}{T_S} \Big) + 4 \frac{T_S}{m_h} K_2\Big(\frac{m_h}{T_S} \Big) \Big]    \hskip0.5cm \text{(for $T_h \lesssim 100\, m_X$)} \\
  &\simeq  \frac{45}{ 2\pi^2} \frac{1}{(2\pi)^{3/2}} \sum_{h\equiv S,X}  g_h \Big( \frac{m_h}{T_S} \Big)^{5/2} e^{- m_h /T_S}  
       \hskip2.8cm \text{(for $T_h \lesssim 0.05\, m_X$)} \,,
\end{align}
and $s_{\rm SM}/ s_h \simeq 30$ for decoupling at $T\sim m_X$. From this scenario of entropy conservation, the temperature ratio $T_h/T$ increases due to cannibalization and follows the dotted gray curve, illustrated on the left panel of Figs.~\ref{fig:relic-not-equal-1}, \ref{fig:relic-not-equal-2},  and \ref{fig:relic-equal}. The hidden sector temperature will deviate from the dotted curve earlier if the out-of-equilibrium decay of $S$ takes place before the end of cannibalization.
As time evolves such that $T <T_c$, the cannibal process is inactive, and the out-of-equilibrium number densities of the hidden sector starts to be exponentially depleted (see Figs.~\ref{fig:relic-not-equal-2}(a-2), \ref{fig:relic-not-equal-2}(b-2) and \ref{fig:relic-equal}(c-2),  \ref{fig:relic-equal}(d-2)).

For $S$ with a lifetime longer than the inverse Hubble rate and during its epoch of cannibalization, the conservation of the total comoving entropy for the hidden sector gives $s_h a^3  \simeq (\rho_h /T_S) a^3 \simeq$ constant. Therefore, the comoving number density of hidden sector as well as its temperature decreases logarithmically with the scale factor, i.e. logarithmically with the bath temperature parameter $x$,
\begin{align}
\frac{m_X n_X + m_S n_S}{s} \sim T_S \sim \frac{m_S}{\log a^3/ a_{\rm out,h}^3} \sim \frac{m_S}{\log x^3/ x_{\rm out,h}^3} \,,
\label{eq:cannibalization-temp}
\end{align} 
where $s=s_{\rm SM}+s_h$, and $a_{\rm out,h}$ (the cosmic scale factor) and $x_{\rm out,h}$ (the bath temperature parameter) correspond to the values  at which the hidden sector starts to be out of equilibrium with the bath. The logarithmic dependence of the comoving number densities for $X$ and $S$ can be seen from Figs.~\ref{fig:relic-not-equal-1}(b-2), \ref{fig:relic-not-equal-2}(a-2), \ref{fig:relic-not-equal-2}(b-2) and \ref{fig:relic-equal}(b-2), \ref{fig:relic-equal}(c-2), \ref{fig:relic-equal}(d-2).

\item \label{item:decay}
$x_{\rm de}^{\rm out} \equiv  m_X/T_{\rm de}^{\rm out}$,  denoted by the purple square in the left panel and by the vertical dashed (red) line in the right panel of Figs.~\ref{fig:relic-not-equal-1}, \ref{fig:relic-not-equal-2}, and \ref{fig:relic-equal},  is the temperature below which the $S$ undergoes an out-of-equilibrium decay, i.e.,  the second term in the RHS of Eq.~(\ref{eq:tde}) (see also Eq.~(\ref{eq:tde-2})) is much larger than the term in the LHS of Eq.~(\ref{eq:tde}) in magnitude due to the fact that $n_S(T_S)  \gg n_S^{\rm eq}(T) $ at $T=T_{\rm de}^{\rm out}$.
For this case, corresponding to a much smaller mixing angle as that given with $\alpha =5\times 10^{-7}$ or $1 \times 10^{-7}$ in this paper, because $n_S(T_S)\, T_S  \gg n_S^{\rm eq}(T)\, T $  at a later time with  $x_f \gg x > \text{max}(x_{\rm el}, x_{\rm ann,2})$, we thus approximately define $x_{\rm de}^{\rm out}$ from Eq.~(\ref{eq:tde}) to satisfy
 \begin{align}  
 (2 - \delta_H) H(x_{\rm de}^{\rm out})
 \approx   \Gamma_{S}  
               \frac{K_1(x_{S,{\rm de}}^{\rm out} \cdot m_S/m_X)}{K_2(x_{S,{\rm de}}^{\rm out} \cdot m_S/m_X)}  
 \,,
 \label{eq:tde-3}
\end{align}
where $x_{S,{\rm de}}^{\rm out}$ is the value of $x_S$, corresponding to $x=x_{\rm de}^{\rm out}$.
When $x= x_{\text{de}}^{\rm out}$, the hidden scalar thus starts to undergo out-of-equilibrium decay at the cosmological time, $(2H)^{-1} \approx \Gamma_S^{-1}   \big[ \frac{K_1(x_{S,{\rm de}}^{\rm out} \cdot m_S/m_X)}{K_2(x_{S,{\rm de}}^{\rm out} \cdot m_S/m_X)}  \cdot  2/(2 - \delta_H)\big]^{-1} \approx \Gamma_S^{-1}$. See also the related discussion in Sec~\ref{sec:model}. 

Because the number changing interactions between $S$ and $X$ affect the $S$ number density for a longer time interval,  thus the estimation of the value of $x_{\rm de}^{\rm out}$ needs to be further improved.
For simplicity, here we neglect the cannibal interaction between $S$ and $X$. Such an interaction results in a logarithmic dependence of the hidden sector comoving number densities on the temperature variable  $x$. 
In the plots, we will use the definition for  $x= x_{\rm de}^{\rm out}$ which satisfies $y_S (x_{\rm de}^{\rm out}) / y_S (1) \simeq e^{-1}$  \cite{Yang:2018fje} with the initial value $y_S (1) =y_S^{\rm eq} (1)$. The value of $x= x_{\rm de}^{\rm out}$ is estimated as follows.
For the case with $m_X-m_S$ sizable enough ({\it e.g.} $m_S=0.8m_X$),  the down-scattering rate, $XX\to SS$, can be significantly  larger than the up-scattering rate, $SS\to XX$,
such that after a sufficient time at $ x= x_{\text{de}}^{\rm out} \gg 1$, we have $y_X^{\rm eq} /y_S^{\rm eq} \ll 1$ and $y_X\ll y_S$. Therefore we set the {\it effective} initial $S$ yield  to be $y_S^{\rm in}(1) \approx y_X(1) + y_S(1) \approx  y_X^{\rm eq}(1) + y_S^{\rm eq} (1) = \kappa y_S^{\rm eq} (1)$ with $ \kappa \equiv  [3(m_X/m_S)^2 K_2(1)/K_2(m_S/m_X)+1] $, and approximate Eq.~(\ref{eq:boltzmann-2}) as  
\begin{equation}
\frac{d y_S}{dx} \approx - C x  y_S \,,
 \label{eq:decaywidth-2}
\end{equation}
with
\begin{equation}
C \equiv \frac{ \sqrt{90}}{\pi} M_{\rm pl} \frac{g_*^{1/2}(T)}{h_{\text{eff}}(T) } \frac{\Gamma_{S}}{m_X^2} 
   \approx \frac{ \sqrt{90}}{\pi} M_{\rm pl} \frac{g_*^{1/2}(T_{\rm f})}{h_{\text{eff}}(T_{\rm f}) } \frac{\Gamma_{S}}{m_X^2}  \,.
   \label{eq:c}
\end{equation}
Here the approximation in the last step of Eq.~(\ref{eq:c}) is reasonable because $g_*^{1/2}/h_{\text{eff}}$, which is 0.109 for $T=m_X/20$ and 0.125 for $T=m_X/150$, weakly depends on $T_{\rm f}$  in the present study. 
Solving this equation, we obtain
\begin{equation}
\frac{y_S(x_{\rm de}^{\rm out})}{y_S^{\rm in}(1)} \simeq e^{-\frac{C}{2} [ (x_{\rm de}^{\rm out})^2 -1]} 
\defeq \frac{e^{-1}}{\kappa} \,,\label{eq:width-sol2}
\end{equation}
and  $x_{\rm de}^{\rm out} \simeq \sqrt{1+ 2(1+ \ln\kappa)/C}$. In terms of the cosmic time variable of the radiation dominated epoch,
\begin{equation}
t  \simeq  x^2  \sqrt{\frac{ 45}{2\pi^2}} \frac{M_{\rm pl} }{m_X^2}\frac{g_*^{1/2}(T_{\rm f})}{h_{\text{eff}}(T_{\rm f}) } \simeq \big( 2H(T) \big)^{-1} \label{eq:s-lifetime} \,,
\end{equation}
 the solution of the normalized yield can be rewritten as 
$y_S \simeq \kappa y_{S}^{\rm eq}(1)   e^{-\Gamma_{S} t}$. If $\kappa=1$, we have $t (\approx (2H)^{-1} ) = \Gamma_S^{-1}$ at $x=x_{\text{de}}^{\rm out}$, consistent with the that given in Eq.~(\ref{eq:tde-3}).
For the case with $m_X \approx m_S$  ({\it e.g.} $m_S=0.99m_X$), summing Eqs.~(\ref{eq:boltzmann-1}) and (\ref{eq:boltzmann-2}), we have
\begin{equation}
\frac{d (y_X+y_S)}{dx} \approx - C x  y_S \,. \label{eq:decaywidth-1}
\end{equation}
Using the initial conditions: $y_X(1) = y_X^{\rm eq}(1)$, $y_S(1) = y_S^{\rm eq}(1)$, and $y_X^{\rm eq}(1) / y_S^{\rm eq}(1) = g_X/g_S =3$, and approximating $y_X+y_S\approx 4 y_S$,
we have
\begin{equation}
\frac{y_S(x_{\rm de}^{\rm out})}{y_S(1)} \simeq e^{-\frac{C}{8} [ (x_{\rm de}^{\rm out})^2 -1] } 
\defeq e^{-1} \,, \label{eq:width-sol1-1}
\end{equation}
with $x_{\rm de}^{\rm out} \simeq \sqrt{1+ 8/C}$. The solution can be given by
$ y_S \simeq y_{S}^{\rm eq}(1) e^{-\Gamma_{S} t/4 }$.
Note that a longer-lived $S$ will result in a larger $x_f$, i.e., a larger $y_X^\infty$, so that, to have a correct relic density, the dark matter annihilation cross section is generally boosted above the conventional WIMP value. This point will be further discussed in the next section.
\end{enumerate}

For the case that the hidden sector is kinetically decoupled from the bath at $T \sim m_{X,S}$, the ending value ($x_c$) of cannibalization depends on the magnitude of $x_{\rm de}^{\rm out}$ since the number density of $S$ is exponentially depleted during decay. In Figs.~\ref{fig:relic-not-equal-2}(b) and \ref{fig:relic-equal}(d), we show the cases with $x_{\rm de}^{\rm out} \sim x_c$, for which,  when $S$ decays out of equilibrium, the $X$ and $S$ densities are exponentially depleted, instead of following the Boltzmann suppression with a zero chemical potential (see the dot-dashed curves in Figs.~\ref{fig:relic-not-equal-2}(b-2) and \ref{fig:relic-equal}(d-2)). Note that, for this case, $X$ and $S$ are still in chemical equilibrium but with non-zero chemical potential before freeze-out.
   Moreover, it is also interesting to note that for $T<T_c$, we have $T_X=T_S$ and $T_S = T_S^{\rm c} \cdot (a^{\rm c}/a)^2$ even after thermal decoupling, where $a$ is the cosmic scale factor and $a^{\rm c}$ is its corresponding value at $T_S^{\rm c} =T_S(T=T_c)$.

\section{Discussions}\label{sec:discussions}

 Since we have considered the secluded vector dark matter model with DM mass $\sim {\cal O}\text{(80 GeV)}$ as an example to exhibit the thermodynamic evolution of the hidden sector, the related parameters in this model should be very likely constrained by the astrophysical and cosmological measurements.
Therefore, before making conclusion, let us discuss the parameter space that can fit to the excess of GeV-scale gamma-rays emitted from the GC region and evade constraints from dwarf spheroidal observation, cosmic microwave background, direct detection, and big bang nucleosynthesis.

The {\em differential gamma-ray flux} from the one-step cascade DM annihilations is described by
\begin{eqnarray} 
\label{eq:gammaflux}
\frac{d \Phi_\gamma}{dE} = \frac{\langle \sigma v\rangle_{\rm LV}}{8\pi m_X^2} 
\sum_f   {\rm Br}(S\to f)
\Bigg(\frac{dN_\gamma^f}{dE}\Bigg)_X \,  \frac{1}{\Delta\Omega}
 \underbrace{ \int_{\Delta\Omega}  \int_{\rm l.o.s.} ds \rho^2(r(s,\psi))d\Omega }_{\text{J-factor}}  ,
\end{eqnarray}
where $\langle \sigma v\rangle_{\rm LV}$ is the DM annihilation cross section in the low-velocity limit (consistent with $T\to 0$), $(dN_\gamma^f / dE)_X$ is the prompt gamma-ray spectrum produced per annihilation with final state $f$ in the DM rest frame, and
the J-factor is the integral along the line of sight (l.o.s.) and over the region of interest (ROI) denoted by the solid angle $\Delta\Omega$. 
We use a Galactic DM density distribution which is a function of $r$, the distance to the GC, and parametrized by a generalized Navarro-Frenk-White (gNFW) profile  \cite{Navarro:1995iw,Navarro:1996gj},
 \begin{equation}
 \label{eq:gNFW}
 \rho(r)=\displaystyle \rho_{\odot} \left(\frac{r}{r_\odot}\right)^{-\gamma} \left(\frac{1+r/r_s}{1+r_\odot /r_s}\right)^{\gamma-3} \,,
 \end{equation}
 where we adopt  $r_s=20$~kpc, $r_\odot=8.5$~kpc, $\gamma=1.2$ and $\rho_\odot=0.4$~GeV/cm$^3$ as the canonical inputs. Here ``$-\gamma$" is the inner log slope of the halo density near the GC, and $\rho_\odot $ is the local DM density at a distance of $r_\odot$ from the GC.  
 The gamma-ray spectrum in the DM rest frame can be expressed in terms of that given in the rest frame of the metastable mediator ($S$) by means of one-step Lorentz boost \cite{Elor:2015tva} (see also Eq.~(10) in Ref.~\cite{Yang:2017zor}), where we use PPPC4DMID result \cite{Cirelli:2010xx,Ciafaloni:2010ti} to described the gamma spectra that are generated from the final state SM particle pair in the $S$ decay at rest.  As for the parameter region with $m_V < m_S <2m_V$ (with $V\equiv W$ or $Z$), the three-body decay channel $S\to V V^* \to V f_1 \bar{f}_2 $ is kinematically open and becomes much more important when $m_S$ is close to $2 m_V$ (see also Fig.~\ref{fig:BrS}). In the $S$ rest frame, the gamma-ray spectrum generated from three-body decay channels can be obtained by boosting the gamma-ray spectra produced from $V$ at rest and from $V^*$ at rest, respectively. Because the description for this part, relevant to the parameter region of the gamma-ray line emission, is sophisticated and does not affect the conclusion of this paper, we will thus defer the details in a future study. 
 \begin{figure}[t!]
  \begin{center}
   \includegraphics[width=0.4\textwidth]{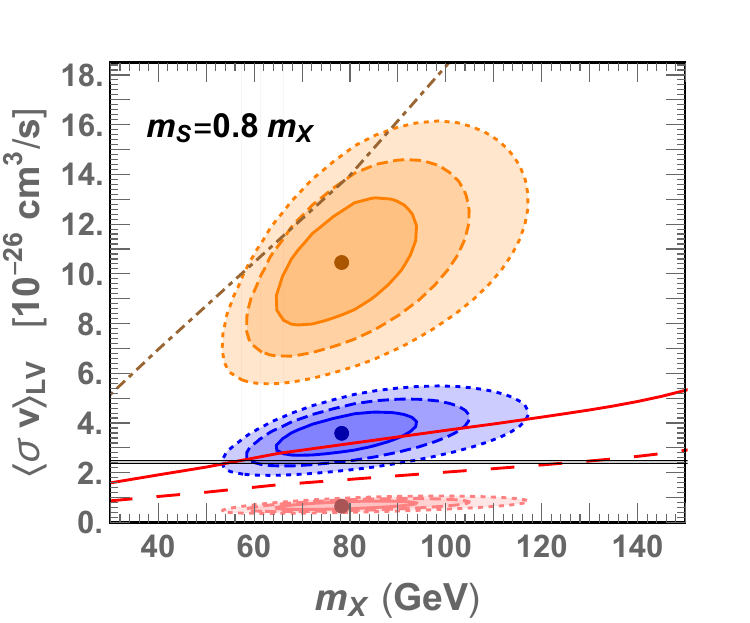}\hskip0.9cm
   \includegraphics[width=0.4\textwidth]{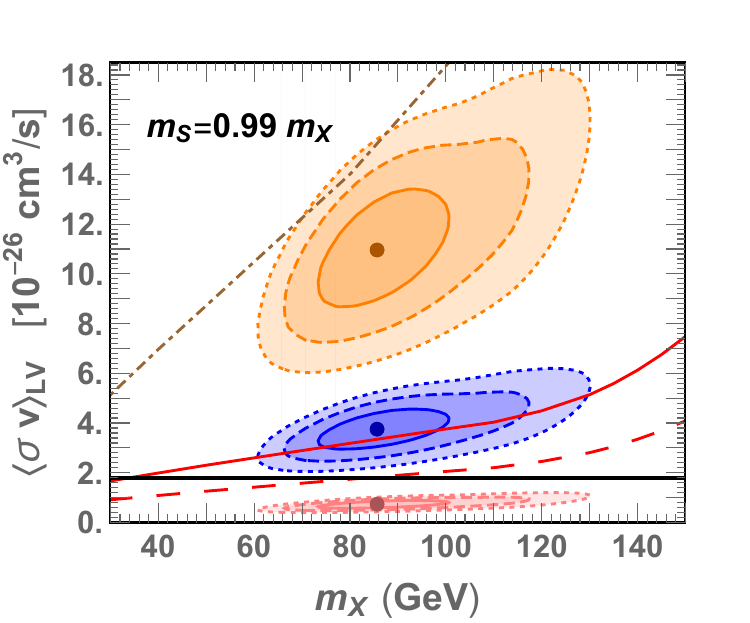} 
\caption{The GC allowed regions in the dark matter mass $m_X$ and low-velocity annihilation cross section $\langle \sigma v\rangle_{\rm LV}$ plane, where the best-fit point is denoted as the dot, and the regions with solid, dashed and dotted boundaries, satisfy $p$-value $\geq$ 0.3, 0.15, and 0.05, respectively.  The regions with color to be blue, orange, and pink refer to $(\rho_\odot, \gamma) = (\text{0.4 GeV/cm}^3, 1.2),  (\text{0.25 GeV/cm}^3, 1.15)$, and $ (\text{0.85 GeV/cm}^3, 1.25)$.  In conventional WIMP DM, the relic density is accounted for by the narrow gray range, while for the nonconventional WIMP, the correct relic density could be extended to the upper region of the gray range. The 95\% C.L. upper bound and projected limit from  Fermi-LAT observations of dSphs  are denoted as the solid  and long-dashed red lines, respectively, while the Planck  CMB 95\% C.L. upper limit is depicted as dot-dashed brown line. }
\label{fig:gcmxsv}
\end{center}
\end{figure}

\begin{figure}[t!]
\begin{center}
\includegraphics[width=0.39\textwidth]{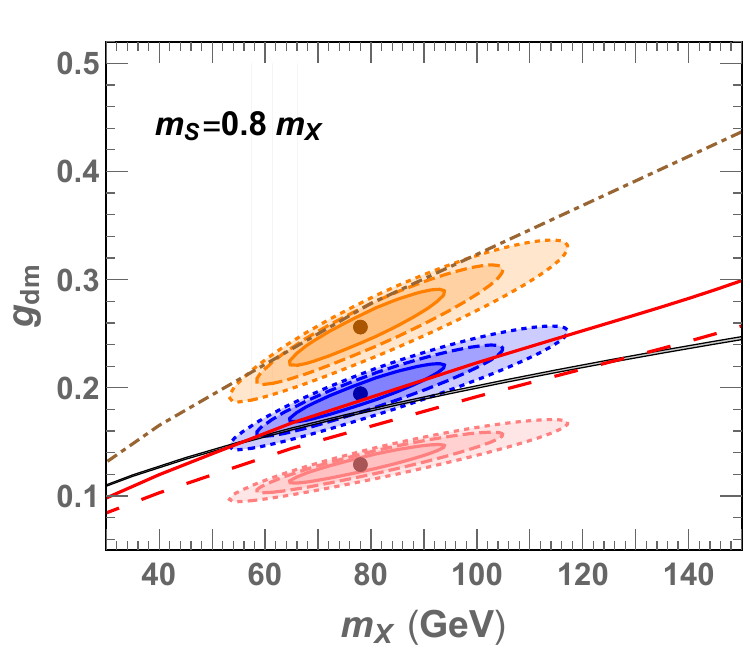}\hskip0.8cm
\includegraphics[width=0.39\textwidth]{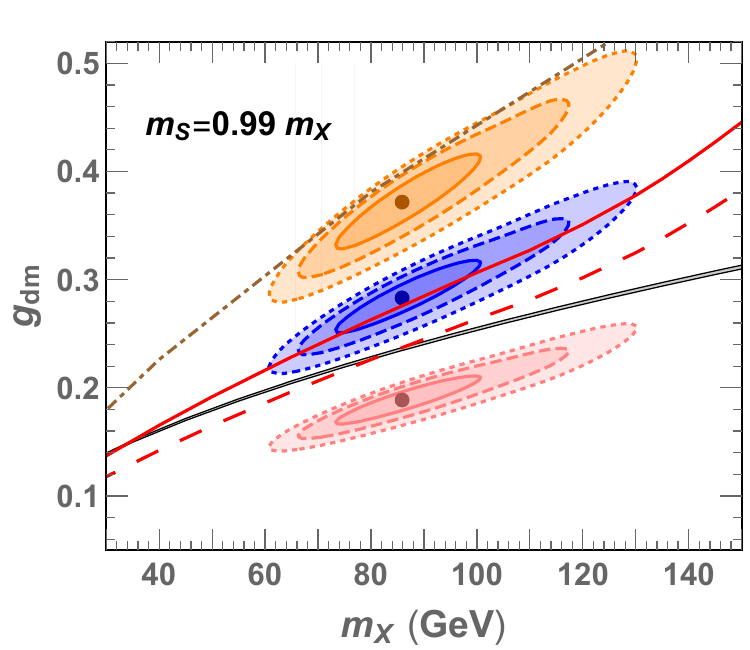} \\
\caption{Same as Fig.~\ref{fig:gcmxsv}, but in the $(m_X, g_{\rm dm})$ plane.}
\label{fig:lxmx}
\end{center}
\end{figure}

Using the GC excess result  extracted by Calore, Cholis, and Weniger (CCW) \cite{Calore:2014xka} from the study of  Fermi-LAT Pass 7 data, 
where  the gamma-ray spectrum covers the energy range between 300 MeV to 500 GeV in the $40^\circ \times 40^\circ$ square ROI around the Galactic center with latitude $|b|\leq 2^\circ$ masked out,
we do the goodness-of-fit with a $\chi^2$ test statistic for the DM mass and annihilation cross section $\langle\sigma v\rangle_{\rm LV}$. 
 In Fig.~\ref{fig:gcmxsv},  two ratio values of $m_S=0.8 m_X$ and $m_S=0.99 m_X$ are used to show the GC excess result, where, taking $\rho_\odot = \text{0.4 GeV/cm}^3$ and  $\gamma=1.2 $, the blue regions  with solid, dashed and dotted boundaries respectively satisfy p-value $\geq$ 0.3, 0.15, and 0.05, corresponding to $\chi^2/{\rm dof}=$ 24.9/22, 28.8/22, and 33.9/22. The best fit is denoted as the blue dot with p-value = 0.46 or 0.42, corresponding to $\chi^2_{\rm min}/{\rm dof}=$ 22.0/22 or 22.7/22, for the case of $m_S=0.8 m_X$ or $m_S=0.99 m_X$. 

Further allowing variation of  $\rho_\odot \in$ [0.25, 0.85] GeV/cm$^3$ and $\gamma \in [1.15, 1.25]$, the value of $\langle \sigma v\rangle_{\rm LV}$ can be raised (or lowered) extremely by a factor of 2.94 (or 0.194). For illustration, in Fig.~\ref{fig:gcmxsv}, we also show the GC allowed region in orange color corresponding to $\rho_\odot = \text{0.25 GeV/cm}^3$ and  $\gamma=1.15$, while that in pink color corresponding to $\rho_\odot = \text{0.85 GeV/cm}^3$ and  $\gamma=1.25$.  In Fig.~\ref{fig:lxmx}, the GC fit together with other constraints is redrawn on the ($m_X, g_{\rm dm}$) plane, where  a larger $g_{\rm dm}$ is needed to account for the data for the nearly degenerate case because $\langle\sigma v\rangle_{\rm LV}$ vanishes in the limit $m_S \to m_X$.

We remark that a newer Pass 8 Fermi data set was analyzed in Ref.~\cite{Linden:2016rcf}, in which the authors showed that the considerable difference between Fermi Pass 7 and Pass 8 data appears only at low energies which might be due to the modeling for the point sources in various datasets \cite{TheFermi-LAT:2017vmf,Linden:2016rcf}. 

\begin{figure}[t!]
\begin{center}
\includegraphics[width=0.67\textwidth]{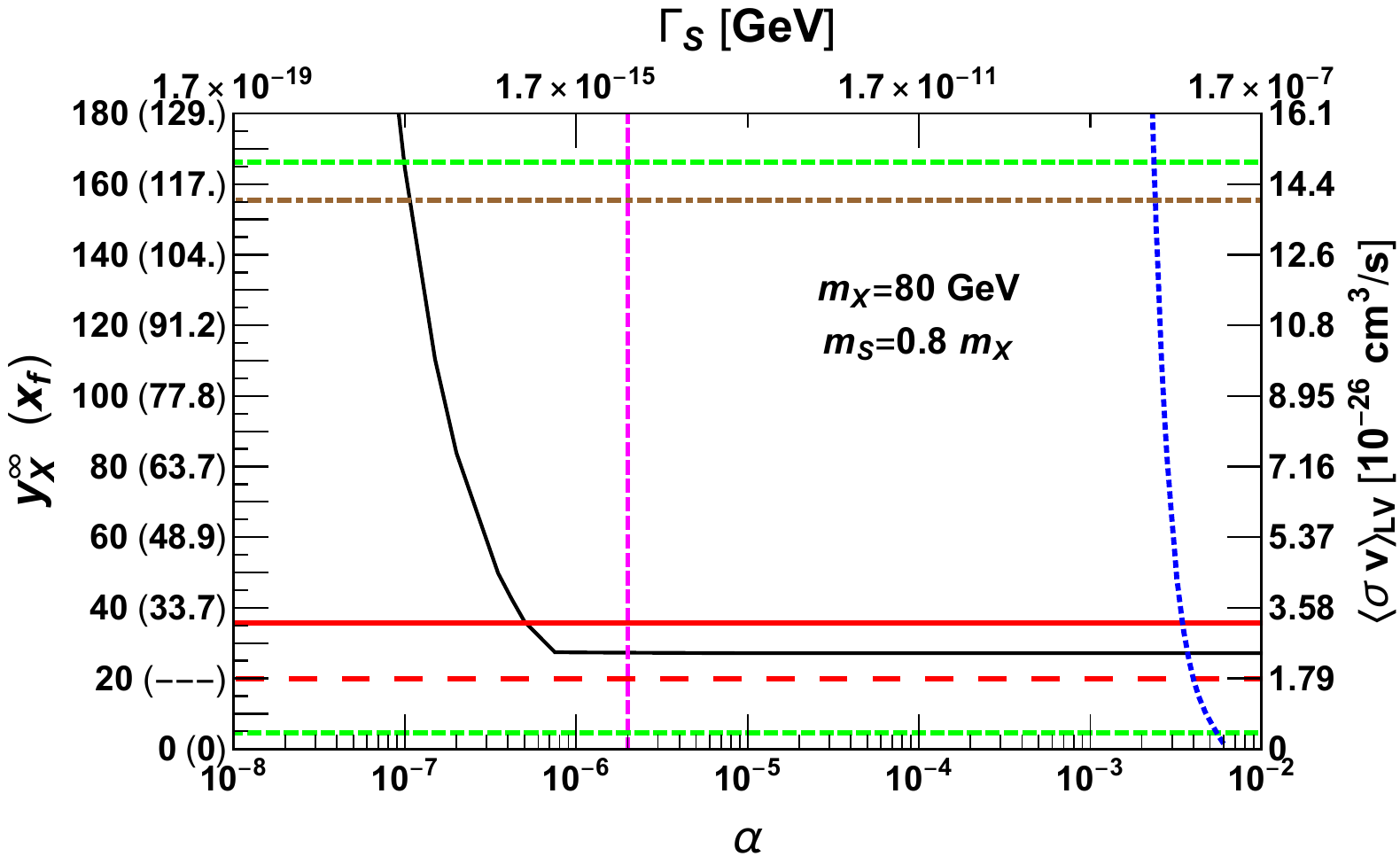}\\
\vskip0.32cm
\includegraphics[width=0.67\textwidth]{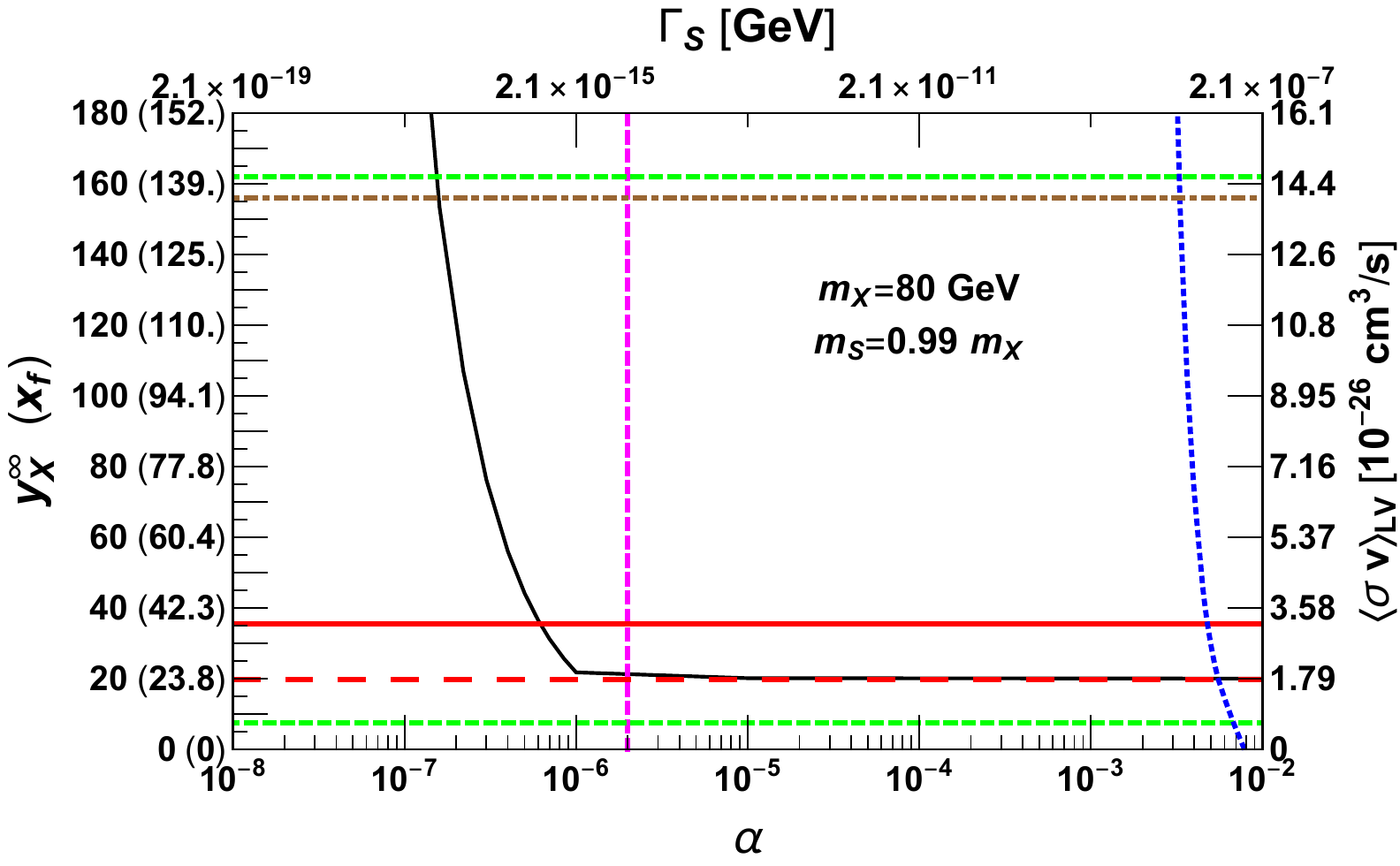} 
\caption{  
Contour of having the correct relic abundance (black curve) on the ($\alpha-y_X^\infty$) plane.
 On the top,  the value of the $S$ decay width  $\Gamma_S$, which is a function of $\alpha$, is labeled.
Here for a given $\alpha$, we fine-tune the value of  $g_{\rm dm}$ from the numerical analysis of Boltzmann equations to obtain $y_X^\infty$, which matches the observed relic abundance determined by Eq.~(\ref{eq:abundance}). 
From each set of allowed parameters, we get $x_f$ from the relation given by Eq.~$(\ref{eq:freeze-out})$, and further have the corresponding value $\langle \sigma v\rangle_{\rm LV} (\equiv  \langle \sigma v\rangle^{(0)}_{XX\to SS})$ from  Eq.~(\ref{eq:abundance}) or from the value of  $g_{\rm dm}$. The dependence of $\langle \sigma v\rangle_{\rm LV}$ on $\alpha$ is negligible.  
The 95\% C.L. upper limit from CMB is denoted as the dot-dashed (brown) line.
The 95\% C.L. upper limit and project sensitivity for dSphs observations are shown as horizontal solid (red) and long-dashed (red) lines, respectively.
The RHS of the dotted (blue) curve corresponds to projected reach by LZ. When $\alpha$ is larger than that denoted by the vertical dashed (magenta) line, the dark sector is well in the chemical and thermal equilibrium with the bath before freeze out.   The region between the two horizontal short-dashed (green) lines provides a good fit to the GC gamma-ray excess. 
}
\label{fig:alpha-xf}
\end{center}
\end{figure}

In Fig.~\ref{fig:gcmxsv},  the relic density of the conventional WIMP dark matter is accounted for by the narrow gray range, while above the gray range the nonconventional WIMP scenario, showing a boosted annihilation cross section, can be satisfied.  The result can be also easily read from Fig.~\ref{fig:alpha-xf}, where the nonconventional WIMP scenario corresponds to  a small mixing angle $\alpha \lesssim 2 \times 10^{-6}$, for which the hidden sector has kinetically decoupled from the thermal bath before it becomes nonrelativistic.   The resulting  nonconventional WIMP DM annihilation cross section that can account for the correct relic density is significantly boosted above the conventionally thermal WIMP value for $\alpha \lesssim 6 \times 10^{-7}$ (or $\alpha \lesssim 1 \times 10^{-6}$) if $m_X= 80$~GeV, $m_S=64$~GeV  (or $m_X=80$~GeV, $m_S=79.2$~GeV).

Fig.~\ref{fig:gcmxsv}  shows constraints from the Fermi gamma-ray observations of dwarf spheroidal galaxies (dSphs) and the measurement of the cosmic microwave background (CMB).  For the dSphs constraint, we have performed a combined likelihood analysis using the 6-year Fermi-LAT data of 28 confirmed and 17 candidate dSphs for gamma-ray energies within 500~MeV to 500~GeV  \cite{Fermi-LAT:2016uux,FermiLatDesData}. 
In the likelihood analysis, we adopt the spectroscopically determined nominal J-factor for the individual target along with its error when possible, or use a predicted value from the distance
scaling relationship with an uncertainty of 0.6 dex, otherwise  \cite{Fermi-LAT:2016uux}.  
See the detailed description in Ref.~\cite{Yang:2018fje} for the likelihood analysis. We also show the dSphs projection sensitivity denoted by the dashed red line by assuming that the 15-year data can be collected from 60 dSphs. For the CMB constraint which is complementary to that determined from dSphs observations, Planck sets a bound from temperature and polarization data (TT, TE, EE+lowP) at recombination to be  \cite{Ade:2015xua}
\begin{align}
f_{\rm eff} (m_X) \frac{\langle \sigma v \rangle_{\rm CMB}}{m_X} < 4.1 \times 10^{-28} \  \text{cm}^3 \text{s}^{-1}  \text{GeV}^{-1} \,, 
\end{align}
where $\langle \sigma v \rangle_{\rm CMB} \simeq \langle \sigma v\rangle_{\rm LV}$ for s-wave DM annihilation, and the efficiency factor is 
\begin{align}
f_{\rm eff} (m_X) =\frac{1}{2 m_X} \int_0^{m_X}  E dE \bigg[ 2 f_{\rm eff}^{e^-}(E)  \bigg( \frac{d N_{e^-}}{dE}\bigg)_X + f_{\rm eff}^\gamma (E) \bigg( \frac{dN_{\gamma}} {dE}\bigg)_X \bigg] \,.
\end{align}
Here, we use $f_{\rm eff}^{\gamma, e^-}(E)$ curve results suited for the ``3 keV" baseline prescription shown in Ref.~\cite{Slatyer:2015jla}. Moreover, as the previous study for the GC excess, $(d N_{\gamma, e^-}/ dE)_X$  generated from the one-step cascade DM annihilation is the photon/electron energy spectrum that can be obtained by boosting the spectra provided in PPPC4DMID.  The current bound obtained from the CMB analysis seems to be much weaker than that from the Fermi-LAT dSphs data (see also Fig.~\ref{fig:alpha-xf}).

In the present model, compared with the SM, we have 4 additional parameters, $m_X, m_S, \alpha$ and $g_{\rm dm}$.  As shown in Figs.~\ref{fig:relic-not-equal-1}, \ref{fig:relic-not-equal-2}, and \ref{fig:relic-equal},
having the chosen masses for the DM and hidden scalar, and giving the magnitude of $\alpha$,  we can fine-tune the value of  $g_{\rm dm}$  in the numerical analysis of Boltzmann equations to obtain $y_X^\infty$, which matches the observed relic abundance determined by Eq.~(27).  Note that, in Eq.~(27), $\langle \sigma v\rangle^{(0)}_{XX\to SS}$ is a function of $g_{\rm dm}$, and its $\alpha$-dependence is negligible in our study.

In order to have a more comprehensive understanding of the phenomenological constraints on the secluded DM that could exhibit a boosted DM annihilation cross section, as discussed in the previous section, using the two ratio values of $m_S=0.8 m_X$ and $m_S=0.99 m_X$ with  $m_X=80$~GeV, we display the correct relic abundance as the black curve on the  ($\alpha, y_X^\infty$) plane in Fig.~\ref{fig:alpha-xf}.  Since the $S$ decay width  $\Gamma_S$ is a function of $\alpha$, we label its corresponding values on the top of the plots.
On the other hand, for a obtained $y_X^\infty$, we can get $x_f$ from the relation given by Eq.~$(\ref{eq:freeze-out})$, and further have the corresponding value $\langle \sigma v\rangle_{\rm LV} (\equiv  \langle \sigma v\rangle^{(0)}_{XX\to SS})$ from  Eq.~(\ref{eq:abundance}) or from the value of  $g_{\rm dm}$. The dependence of $\langle \sigma v\rangle_{\rm LV}$ on $\alpha$ is weak and thus neglected in the plots.  All of the corresponding quantities are labeled in the plots.
The range favored by observed features of the GC excess with variation of  $\rho_\odot \in$ [0.25, 0.85] GeV/cm$^3$ and $\gamma \in [1.15, 1.25]$ is given in between the two horizontal dot-dashed (purple) lines.

As shown in Fig.~\ref{fig:alpha-xf}, the LZ projected sensitivity can only reach the RHS of the dotted (blue) curve corresponding to the thermal WIMP region with $\alpha \gtrsim 3.7 \times10^{-3}$ for $(m_X, m_S)=$ (80~GeV, 64~GeV), or for  $\alpha \gtrsim 5.5 \times10^{-3}$ for $(m_X, m_S)=$ (80~GeV, 79.2~GeV).  If  $\alpha$ is larger than the value denoted by the vertical dashed (magenta) line which corresponds to $T_{\rm de} = \text{min}(T_{\rm el}, T_{\rm ann,2})$, the hidden sector particles can be well in the chemical and thermal equilibrium with the bath before freeze out. Nevertheless, the case of $(m_X, m_S)=$ (80~GeV, 64~GeV) with $\alpha \lesssim 6 \times10^{-7}$, or $(m_X, m_S)=$ (80~GeV, 79.2~GeV) with $\alpha \lesssim 1 \times10^{-6}$, clearly exhibits the boosted annihilation cross section capable of accounting for the correct relic density. The 95\% C.L. limit from CMB is denoted as the dot-dashed (brown) line, while the 95\% C.L. limit and project sensitivity for dSphs observations are shown as horizontal solid (red) and long-dashed (red) lines, respectively.
The boosted annihilation cross section is thus stringently constrained by the current dSphs observations. The secluded DM scenario can be further tested by the dSphs projection.

Finally, we discuss the bound on the lifetime of the hidden scalar from the big bang nucleosynthesis constraint, i.e., a lower bound on the mixing coupling $\alpha$.
It has been shown and discussed in Refs.~\cite{Kawasaki:1999na,Kawasaki:2000en,deSalas:2015glj,Hasegawa:2019jsa} that the late-time entropy production by the massive particle decay could induce cosmological effects at the time $\sim \Gamma_S^{-1}  \sim 1$ sec. We consider the case with a long-lived massive $S$ which starts to decouple from the SM bath at the temperature below $m_S$. After decoupling, the energy density of the nonrelativistic hidden scalar then decreases as $\rho_S \propto a^{-3}$, while the SM radiation energy density $\rho_{\rm SM}$ scales as $a^{-4}$. As a result, $\rho_S /\rho_{\rm SM}$, scaling linearly with $a$,  is about $1/120 $ at $T \sim m_S$, but becomes $\sim(1/120)\times m_S/ T$ at a later time with temperature $T$. In other words, the universe can be rapidly dominated by the nonrelativistic hidden sector particles if $S$ is long-lived. When the hidden scalar decays out-of-equilibrium into SM particles, the universe becomes radiation-dominated again and the SM bath experiences the reheating due to the large entropy injection. 
While photons and charged leptons are quickly thermalized during reheating, the weakly interacting neutrinos are slowly produced in the bath. 
 Since neutrinos decouple from the thermal bath after $T \lesssim1.5$~MeV, they would not be well thermalized if  the reheating temperature $T_{\rm RH} \sim {\cal O}(1)$~MeV. See the example shown in Fig.~3 of Ref.~\cite{Kawasaki:2000en}, where each neutrino follows the perfect Fermi-Dirac distribution very well for $T_{\rm RH}=10$~MeV,  while the distributions are not in thermal equilibrium form for $T_{\rm RH}=2$~MeV (See also Fig.~4 in Ref.~\cite{Hasegawa:2019jsa}, where neutrino self-interaction and oscillation are included). 
If $T_{\rm RH}\lesssim 7$~MeV, the effective number of neutrino species $N_{\rm eff}$ becomes smaller than three (see Fig.~4 in Ref.~\cite{Kawasaki:2000en} or Fig.~1 in Ref.~\cite{Hasegawa:2019jsa} for reference).
Note that in our model there are no additional relativistic particles present before or after BBN, although such particles can large the value of $N_{\rm eff}$.
 The deficit of the neutrino distribution functions due to the insufficient thermalization will decrease the interaction rate between proton and neutron, so that the helium nucleon fraction $Y_p \equiv 4n_{\rm He}/ n_b$, and the deuterium ratio $D/H$ are thus enhanced.
 
Following Refs.~\cite{Kawasaki:2000en,Hasegawa:2019jsa}, we define the reheating temperature of the SM bath to be $\Gamma_S = 3 H(T_{\rm RH})$. Using the approximation,
\begin{align}
H(T_{\rm RH}) =  \left(   \frac{g_{\rm eff} \pi^2}{90} \right)^{1/2} \frac{T_{\rm RH}^2}{M_{\rm pl}} \,,
\end{align}
$T_{\rm RH}$ can be related to the decay width of $S$ as
\begin{align}
T_{\rm RH} \simeq  0.7 \left( \frac{\Gamma_S}{{\rm sec}^{-1}}\right)^{1/2} {\rm MeV},
\label{bbn-time}
\end{align}
where we have used $g_{\rm eff} =43/4$.  Here, we quote  a lower bound $T_{\rm RH} \gtrsim 4.1$~MeV at 95\% C.L., corresponding to $m_S=$10~GeV$-$100~TeV, from the $Y_p +D/H$ analysis in the case of 100\% hadronic decay (see Figs.~12 and 13 in  Ref.~\cite{Hasegawa:2019jsa}),  which is suitable for our model. Further considering the neutrino self-interaction and oscillation, the same reheating bound is also required by $N_{\rm eff} = 3.15\pm 0.23$ from Planck report  \cite{Ade:2015xua} (see Fig.~1 in Ref.~\cite{Hasegawa:2019jsa}). Therefore, from Eq.~(\ref{bbn-time}), we can obtain the BBN constraint on the $S$ width to be $\Gamma^{-1} \lesssim 0.03$~sec.
 As such, we have $\alpha \gtrsim 1.13 \times10^{-10}$ for $(m_X, m_S)=$ (80~GeV, 64~GeV), or $\alpha \gtrsim 1.03 \times10^{-10}$ for $(m_X, m_S)=$ (80~GeV, 79.2~GeV).

\section{Conclusions}\label{sec:conclusions}

Using the secluded vector dark matter model, we have presented a comprehensive study on thermodynamic evolutions of the hidden sector particles from the first principle.
We have solved numerically the coupled Boltzmann moment equations for number densities and temperature evolutions of the hidden sector particles. Our formalism can be easily extended to a general secluded dark matter model.

Taking two mass sets: (i) $m_X=80$~GeV, $m_S=0.8 m_X=64$~GeV, and (ii) $m_X=80$~GeV, $m_S =0.99 m_X=79.2$~GeV, we have shown the detailed thermodynamics for which, while the dark matter in thermal equilibrium with the hidden scalar is secluded from the visible sector with small interaction rates in agreement with the limit from the direct detection and collider experiments, the hidden sector can be either in thermal equilibrium or out of equilibrium  with the bath before the DM freezes out.  The results are briefly summarized as below. More details about the thermodynamics of the hidden sector have been given in Sec.~\ref{sec:analysis}.

For the case satisfying $T_{\rm de} \geq \text{min}(T_{\rm el}, T_{\rm ann,2})$, the kinetic decoupling of elastic scattering $S\, \text{SM} \leftrightarrow S\, \text{SM}$ and/or annihilation $SS \leftrightarrow \text{SM SM}$ occurs only when the bath temperature is below $T_{\rm de}$ at which the heating rate of the hidden sector generated from the inverse decay $\text{SM~SM} \to S$ starts to overcome the dilution rate due to the cosmic expansion. As such, the nonrelativistic hidden sector can keep thermal equilibrium with the bath until freeze-out. Therefore, the DM is consistent with the conventional WIMP scenario, but can easily evade the searches from the colliders and direct detections ({\it e.g.} projected LZ measurement) for a small mixing angle $ 2\times 10^{-6} \lesssim \alpha \lesssim  4  \times 10^{-3} $ as in the present model.

On the other hand, for the case that 
 the hidden sector starts to be kinetically decoupled from the thermal bath at $T\sim m_{X,S}$ due to its weak couplings to the SM particles,
the nonrelativistic  hidden sector will first undergo a cannibal epoch, during which the total comoving entropy density of the hidden sector is approximately conserved before $S$ decays out of equilibrium. When out-of-equilibrium $S$ decay occurs, the hidden sector particles $X$ and $S$ are still in chemical equilibrium, but their densities, instead of following Boltzmann suppression with zero chemical potential, are exponentially depleted with non-zero chemical potential until freeze-out.   
We have shown that having a small mixing angle $\alpha \lesssim 6 \times 10^{-7}$ (or $\alpha \lesssim 1 \times 10^{-6}$) which corresponds to $m_X= 80$~GeV, $m_S=64$~GeV  (or $m_X=80$~GeV, $m_S=79.2$~GeV), the secluded DM annihilates into ``long-lived" hidden mediators which later decay out of equilibrium with the bath, such that the resulting nonconventional WIMP-like DM annihilation cross section accounting for the observed relic density is  boosted above the conventionally thermal WIMP value.

For the experimental constraints, we have shown the parameter space which yields a good fit to the GC excess data and is compatible with the LZ projected sensitivity, BBN bound, Planck CMB measurement and Fermi dSphs observation.   Moreover, we expect that Fermi-LAT 15-yr dSph observations can explore the parameter region of the correct relic density described not only by the nonconventional WIMP scenario but also,  if the DM and hidden scalar are not well degenerate, by the conventional WIMP one.

\acknowledgments \vspace*{-1ex}
 This work was supported in part by the Ministry of Science and Technology, Taiwan, under Grant Nos. 105-2112-M-033-005 and 108-2112-M-033-002.

\appendix

\section{The partial decay widths of the hidden mediator $S$}\label{app:s-width}

The main partial decay widths of the hidden scalar $S$ with mass $\lesssim$ 130~GeV are given by
\begin{align}
 \Gamma(S \rightarrow \bar{f} f) 
 & =K_{f} N_c^f  \frac{m_S}{8\pi} g_{Sff}^2  \left( 1- \frac{4 m_f^2}{m_S^2} \right)^{3/2} \theta(m_S-2m_f) \; ,  \label{eq:S-width-fermion} \\
\Gamma(S \rightarrow g g) & =K_{g}  \frac{\alpha_s^2}{2  \pi^3 m_S}\left|\sum_{q\equiv {\rm quark}} m_q  g_{Sqq} f_S \left( \frac{4 m_q^2}{m_S^2}\right)   \right|^2 \; , \label{eq:S-width-gg}  \\
\Gamma (S \rightarrow VV^*) & = \frac{3 G_F^2 m_V^4}{16 \pi^3} m_S s_\alpha^2 \delta_V' R_T(x) \; \theta(m_S-m_V) \; ,
\label{HVV-3body}
\end{align}
where $N_c^{q(\ell)} \equiv 3\, (1)$ for the quark (lepton), $G_F =(\sqrt{2} v_H^2)^{-1}$,  the coupling $g_{Sff}=-s_\alpha m_f/v_{H} $,  $K_{q}=1+5.67\alpha_s(\mu)/\pi$ and $K_{g}=1+(215/12)\alpha_s(\mu)/\pi$ are the NLO QCD corrections \cite{Djouadi:2005gj}, $\delta'_W=1$, $\delta_Z' = \frac{7}{12} - \frac{10}{9}
\sin^2\theta_W+ \frac{40}{27}\sin^4\theta_W$,
\begin{eqnarray}
R_T(x) & = & \frac{3(1-8x+20x^2)}{(4x-1)^{1/2}} \arccos \left( \frac{3x-1}
{2x^{3/2}} \right) -\frac{1-x}{2x} (2-13x+47x^2) \non - \frac{3}{2}(1-6x+4x^2) \log x  \,,
\label{HVVrt}
\end{eqnarray}
with $x\equiv m_V^2/m_S^2$  \cite{Keung:1984hn,Djouadi:2005gi}, and $f_S (\tau) =[1+ (1 -\tau) f( \tau)]$  with
\begin{equation}
f (\tau)  =  \left\{ \begin{array}{lr} \text{arcsin}^2 \sqrt{\tau^{-1}}
        \, , \ \ & \tau \geq 1 \\ -\frac{1}{4} \left[ \log \frac{1+\sqrt{1-\tau}}{1-\sqrt{1-\tau}} - i \pi\right]^2 \, , & \tau < 1 \end{array} \right. \; .
\label{eq:S-partial-width-f} 
\end{equation}
Here, we will take the scale $\mu=m_S/2$.

\section{$2\to 2$ annihilation cross sections}\label{app:annXS}

\subsection{The annihilation process for $XX \to SS$} \label{app:XX2SS}

In this secluded DM case, the relic density is determined by the thermally averaged annihilation cross section $\langle \sigma v_{\text{M\o l}} \rangle_{XX\to SS}$ 
which is also relevant to the indirect detection searches, where $v_{\text{M\o l}}$ is the M{\o}ller velocity.  In the text, we have used $\sigma v \equiv \sigma v_{\text{M\o l}}$ for simplicity. The value of $\langle \sigma v_{\text{M\o l}} \rangle_{XX\to SS}$ equals to $\langle \sigma v_{\rm lab}\rangle_{XX\to SS}$ which is the result calculated in the rest frame of one of the incoming particles. 

\begin{figure}[t!]
\begin{center}
\includegraphics[width=0.76\textwidth]{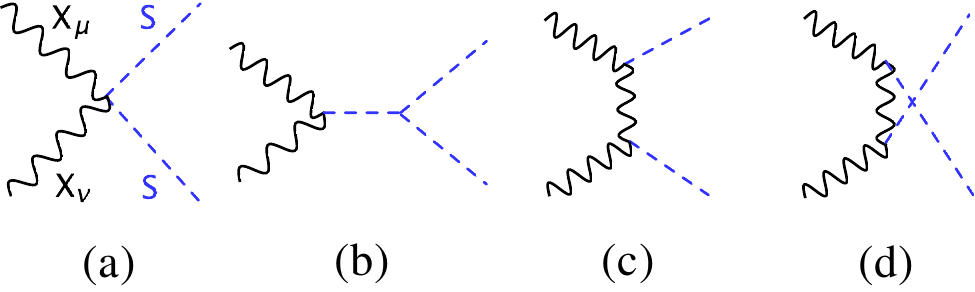}
\caption{Feynman diagrams that dominantly contribute to the DM annihilation cross section.}
\label{fig:vdm_ann}
\end{center}
\end{figure}

The diagrams for the $XX\to SS$ process are depicted in Fig.~\ref{fig:vdm_ann}, where the $s$-channel annihilation via $h$ is negligible and does not shown. The resulting cross section is given by
\begin{align}
& \!\!\!  (\sigma v_\text{lab})_{XX\to SS} \nonumber\\
&  = 
\frac{c_\alpha^2 g_{\rm dm}^2 \sqrt{s-4 m_S^2}}
 {288 \pi  m_X^4 \sqrt{s}  (s-2 m_X^2)  \big( (s-m_S^2)^2 + \Gamma_S^2 m_S^2 \big)}
\times \Bigg[   
\frac{ 2c_\alpha^2 g_{\rm dm}^2  \big( (s- m_S^2)^2 + \Gamma_S^2 m_S^2 \big) }{m_S^4-4 m_S^2 m_X^2+m_X^2 s} \nonumber\\
 &
 \times \Big(3 m_S^8-20 m_S^6 m_X^2+m_S^4  (46 m_X^4+6 m_X^2 s)
     -4 m_S^2 \left(14 m_X^6+5 m_X^4 s\right)+48 m_X^8+6 m_X^6 s+4
   m_X^4 s^2 \Big)
   \nonumber\\
   & 
   +  g_{SSS}^2 m_X^2  (12 m_X^4-4 m_X^2  s+s^2 ) 
   -4   g_{SSS} c_\alpha g_{\rm dm}  m_X (m_S^2-s)  \big(m_S^2 (2  m_X^2+s )-m_X^2 (6 m_X^2+s ) \big)
   \nonumber\\
     & -
     \frac{8 c_\alpha g_{\rm dm} {\rm Arcoth} \left(\frac{s-2 m_S^2 }{ \sqrt{s-4 m_S^2} \sqrt{s-4 m_X^2}} \right) }{(s - 2 m_S^2)  \sqrt{s-4 m_S^2} \sqrt{s-4 m_X^2}}
     \bigg( c_\alpha g_{\rm dm} \Big( (s-m_S^2)^2 + \Gamma_S^2 m_S^2 \Big)   
   \nonumber\\
   & \ \times
   \Big(3 m_S^8-2 m_S^6 (6 m_X^2+s )+2 m_S^4 (4 m_X^4+5 m_X^2 s)
   -4 m_S^2 m_X^2  (4  m_X^4+s^2 ) +24 m_X^6 (s-2  m_X^2 )
   \Big)
   \nonumber\\
   &  
    - g_{SSS} m_X (s-2 m_S^2) (s-m_S^2)
   \Big(m_S^2 (2 m_X^2+s) (m_S^2-4 m_X^2)+2 m_X^2 (12 m_X^4-2 m_X^2  s+s^2) \Big) 
   \bigg)
     \Bigg] ,
   \end{align}
 where $s$ is the center-of-mass energy squared, and
\begin{align}
g_{SSS}=  -\frac{ 3 c_\alpha^3  m_S^2}{v_S} + \frac{3 s_\alpha^3 m_S^2}{v_{H}} \,. \label{eq:gsss}
\end{align}
Using the above result, the thermally averaged annihilation cross section for $T\lesssim  3 m_X$ can be obtained by calculating \cite{Gondolo:1990dk},
\begin{equation}
\langle \sigma v_{\text{lab}} \rangle_{XX\to SS} = \frac{1}{8m_X^4 T K_2^2 (m_X/T)} \int_{4m_X^2}^{\infty}
(\sigma v_{\text{lab}} )_{XX\to SS}  ( s-2 m_X^2)  (s-4m_X^2)^{1/2} K_1 (\sqrt{s}/T) ds, \label{eq:thermal-average}
\end{equation}
with $K_{1,2}$ being the modified Bessel functions. At the indirect detection, we can take the approximation in the low-velocity limit, i.e, 
$\langle \sigma v_{\text{lab}} \rangle_{XX\to SS} =(\sigma v_{\text{lab}} )_{XX\to SS} $ with the replacement $s=4m_X^2$.

\subsection{The annihilation process for $SS \to \text{SM SM}$} \label{app:XX2SMSM}

\begin{figure}[t!]
\begin{center}
\includegraphics[width=0.8\textwidth]{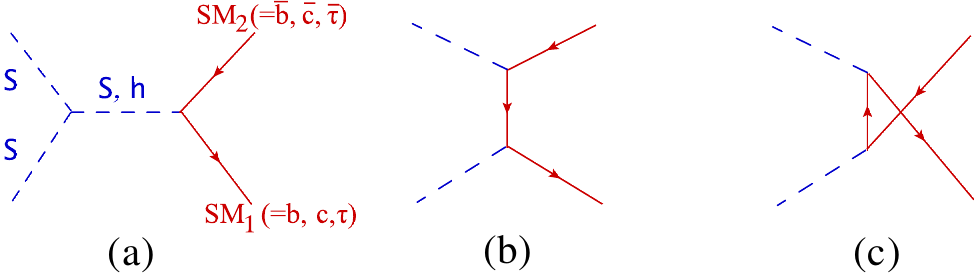}
\caption{Feynman diagrams for the annihilation of the hidden scalar pair into SM particles. The results are relevant to the chemical equilibrium between the hidden sector and thermal bath in the early Universe. }
\label{fig:ss2smsm}
\end{center}
\end{figure}

The diagrams for the hidden scalar $S$ annihilation into the SM particles are depicted in Fig.~\ref{fig:ss2smsm}. The resulting annihilation cross section is given by
\begin{align}
 ( \sigma v_\text{lab} )_{SS\to \bar{f} f}
&= N_c^f
\frac{\sqrt{s-4 m_f^2} }{8 \pi  \sqrt{s} \left(s-2 m_S^2\right)}
\Bigg( \frac{g_{Sff}^2 \, g_{SSS}^2 \left(s-4 m_f^2\right)}{(s-m_S^2)^2 + \Gamma_S^2 m_S^2 }
 +\frac{g_{hSS}^2 g_{hff}^2 \left(s-4 m_f^2\right)}{(s-m_h^2)^2+ \Gamma_h^2 m_h^2 }
\nonumber\\
& +\frac{2 g_{Sff}^4 \left(m_f^2 \left(4 s-8 m_S^2\right)+2 m_S^4-4 m_S^2 s+s^2\right) 
\ln \left(\frac{s-2 m_S^2+\sqrt{s-4 m_f^2} \sqrt{s-4 m_S^2}}{s-2 m_S^2 - \sqrt{s-4 m_f^2} \sqrt{s-4 m_S^2}}\right)}{\sqrt{s-4 m_f^2} \sqrt{s-4 m_S^2} \left(s-2 m_S^2\right)}
   \nonumber\\
&+\frac{2  g_{SSS} \, g_{hSS} \, g_{Sff} \,   g_{hff} (s-4 m_f^2) \left(\Gamma_h \Gamma_S m_h m_S+\left(m_h^2-s\right) \left(m_S^2-s\right)\right)}
{\left( (s-m_h^2)^2+ \Gamma_h^2 m_h^2 \right) \left( (s-m_S^2)^2 + \Gamma_S^2 m_S^2  \right) }
\nonumber\\
&-\frac{4 g_{Sff}^3 g_{SSS} m_f \sqrt{s-4 m_f^2} \left(s-m_S^2\right) 
 \ln \left(\frac{s-2 m_S^2+\sqrt{s-4 m_f^2} \sqrt{s-4 m_S^2}}{s-2 m_S^2-\sqrt{s-4 m_f^2} \sqrt{s-4 m_S^2}}\right)}
 {\sqrt{s-4 m_S^2} \left( (s-m_S^2)^2 + \Gamma_S^2 m_S^2  \right)}
\nonumber\\
&+\frac{4 g_{Sff}^2 \, g_{hSS} \, g_{hff}\,  m_f \sqrt{s-4 m_f^2} \left(m_h^2-s\right) 
\ln \left(\frac{s-2 m_S^2 + \sqrt{s-4 m_f^2} \sqrt{s-4 m_S^2}}{s-2 m_S^2-\sqrt{s-4 m_f^2} \sqrt{s-4 m_S^2}}\right)}
  {\sqrt{s-4 m_S^2} \left( (s-m_h^2)^2+ \Gamma_h^2 m_h^2 \right)}
\nonumber\\
&-\frac{2 g_{Sff}^4 \left(8 m_f^4-4 m_f^2 m_S^2+m_S^4\right)}{m_f^2 \left(s-4 m_S^2\right)+m_S^4}
\Bigg) \,,
\end{align}
where $N_c^f\equiv 3\, (1)$ for $f\equiv$ quarks (leptons), $g_{SSS}$ is shown in Eq.~(\ref{eq:gsss}), and the remaining couplings are
\begin{align}
g_{hSS}=
&
-  \frac{c_\alpha^2 s_\alpha (2 m_S^2 + m_h^2)}{v_S} - \frac{c_\alpha s_\alpha^2 (2 m_S^2 + m_h^2)}{v_{H}} \,,  \label{eq:ghss}\\
g_{hff} = 
& \frac{m_f}{v_H} c_\alpha \,, \label{eq:ghff}\\
g_{Sff} =
&-\frac{m_f}{v_H} s_\alpha  \label{eq:gsff}\,.
\end{align}
On can further apply Eq.~(\ref{eq:thermal-average}) to obtain thermally averaged value of the annihilation cross section. This result is relevant to the chemical equilibrium between the hidden sector and thermal bath in the early Universe.

\section{  The thermal average  $\langle \sigma v \cdot \frac{{\bf p}_S^2}{3 E_S} \rangle_{S S \to \text{SM}_1 \text{SM}_2}$}\label{app:thermal-average}

In this paper,  for the process  $S(p_S) \, S(p_{S'}) \to \text{SM}_1 \, \text{SM}_2$ that follows the Maxwell-Boltzmann distribution,  its thermal average, $\langle \sigma v_{\text{M\o l}}  \cdot \frac{{\bf p}_S^2}{3 E_S} \rangle$, at temperature $T_i$ in the cosmic comoving frame is defined by
\begin{align}
\langle \sigma v_{\text{M\o l}}  \cdot \frac{{\bf p}_S^2}{3 E_S} \rangle (T_i)
=
\frac{\int \sigma v_{\text{M\o l}}  \cdot \frac{{\bf p}_S^2}{3 E_S} e^{-E_S/T_i} e^{-E_{S'}/T_i} d^3p_S d^3p_{S'}}{\int e^{-E_S/T_i} e^{-E_{S'}/T_i} d^3p_S d^3p_{S'}} \,,
\end{align}
where the M{\o}ller velocity is given by 
\begin{align}
 v_{\text{M\o l}} =\frac{\sqrt{ (p_S\cdot p_{S'})^2 - m_S^2}}{E_S E_{S'}} \,.
 \label{app:moller}
\end{align}
For a typical case that $\sigma v_{\text{M\o l}} =(\sigma v)_0$ is constant, because
\begin{align}
 \int \frac{d^3 p_S}{(2\pi)^3} \frac{{\bf p}_S^2}{3 E_S}   e^{-E_S/T_i}  = T_i  \int \frac{d^3 p_S}{(2\pi)^3}   e^{-E_S/T_i}    \,,
\end{align} 
we thus have $\langle \sigma v_{\text{M\o l}}  \cdot \frac{{\bf p}_S^2}{3 E_S} \rangle =T_i \langle \sigma v_{\text{M\o l}}  \rangle = T_i (\sigma v)_0$. For a general case, we can first rewrite the momentum-space volume element to be 
\begin{align}
d^3p_S d^3p_{S'} = 4 \pi | {\bf p}_S| E_S dE_S \,  4 \pi | {\bf p}_{S'}| E_{S'} dE_{S'} \frac{1}{2} d\cos\theta \,,
\end{align}
with $\theta$ being the angle between ${\bf p}_S$ and ${\bf p}_{S'}$. As seen from Eq.~(\ref{app:moller}) that $ v_{\text{M\o l}} E_S E_{S'}$ is Lorentz invariant, we can relate the M{\o}ller velocities in two different frames with and without a prime to be
\begin{align}
\sigma v_{\text{M\o l}} = \sigma v_{\text{M\o l}}^\prime \frac{ E_S^\prime E_{S'}^\prime}{ E_S E_{S'}}
   =  \sigma v_{\text{M\o l}}^\prime \frac{ 1- {\bf v}_S \cdot {\bf v}_{S'} }{1- {\bf v}_S^\prime \cdot {\bf v}_{S'}^\prime } \,,
 \label{app:moller-2}
\end{align}
where the last step uses the fact that $p_S \cdot p_{S'} = E_S E_{S'} (1- {\bf v}_S \cdot {\bf v}_{S'} )$ which is Lorentz invariant.
Thus, we can use the M{\o}ller velocity given in the laboratory frame, which is equivalent to the rest frame of one of the incoming particles, to obtain the M{\o}ller velocity defined in the cosmic comoving frame; for simplicity,  in the following we will use $v_\text{lab}$ for the former and  $v_{\text{M\o l}}$ for the latter. The resulting relation is 
\begin{align}
\sigma v_{\text{M\o l}} 
   =  \sigma v_\text{lab} (1- {\bf v}_S \cdot {\bf v}_{S'} )  
   =  \sigma v_\text{lab} \left(1-  \frac{|{\bf p}_S| \, | {\bf p}_{S'}|}{E_S E_{S'}} \cos\theta \right)  \,.
 \label{app:moller-3}
\end{align}
As such, we get 
\begin{align}
\langle \sigma v_{\text{M\o l}}  \cdot \frac{{\bf p}_S^2}{3 E_S} \rangle (T_i)
\approx
\langle \sigma v_{\text{lab}}  \cdot \frac{{\bf p}_S^2}{3 E_S} \rangle (T_i)
 \,,
\end{align}
where we have used the fact that $\int_{-1}^{+1} \cos\theta \, d\cos\theta =0$. In order to calculate $\langle \sigma v_{\text{lab}}  \cdot \frac{{\bf p}_S^2}{3 E_S} \rangle (T_i)$, we further change integration  variables from $E_S, E_{S'}, \theta$ to $E_-, E_ +, s$, given by 
\begin{align}
E_- =E_S - E_{S'}, \quad E_+ = E_S + E_{S'}, \quad s=2m_S^2 + 2E_S E_{S'} -2 |{\bf p}_S| \,  |{\bf p}_{S'}| \cos\theta \,,
\end{align}
with the integration region
\begin{align}
           |E_-| \leq \left( 1- \frac{4m_S^2}{s} \right)^{1/2} (E_+^2 -s)^{1/2}, 
\quad  \sqrt{s} \leq E_+ \leq \infty ,
\quad 4m_S^2 \leq s \leq \infty \,.
\end{align}
Using and new variables and calculating the thermal average in terms of modified Bessel functions of the second kind, we obtain 
\begin{align}
& \langle \sigma v_{\text{lab}}  \cdot \frac{{\bf p}_S^2}{3 E_S} \rangle (T_i)
  = \frac{1}{288 m_S^4  K_2^2 (m_S/T_i)}   \int_{4m_S^2}^{\infty} ds
        (\sigma v_{\text{lab}} ) \frac{\sqrt{s}}{T_i}  \sqrt{s-4m_S^2}  \nonumber\\
   &\qquad  \qquad \qquad \qquad \times \left\{ ( s - 16 m_S^2) K_2 \left(\frac{\sqrt{s}}{T_i} \right)
         +  2( s+ 2 m_S^2)  \left[ K_4 \left(\frac{\sqrt{s}}{T_i}  \right)  -  3\frac{T_i}{\sqrt{s}} K_3 \left(\frac{\sqrt{s}}{T_i} \right) \right]
   \right\}
    \nonumber\\
  &= \frac{1}{48 m_S^4  K_2^2 (m_S/T_i)}  \nonumber\\
     & \times   \int_{4m_S^2}^{\infty} ds
        (\sigma v_{\text{lab}} )  \sqrt{s-4m_S^2} 
    \bigg[ ( s+ 2 m_S^2) K_1 \left(\frac{\sqrt{s}}{T_i} \right)
         +  \left( \frac{s-4m_S^2}{2} \frac{\sqrt{s}}{T_i} + \frac{ 4T_i ( s+ 2 m_S^2)}{\sqrt{s}}  \right) K_2 \left(\frac{\sqrt{s}}{T_i} \right)
    \bigg], 
\label{eq:thermal-average-t}
\end{align}
where we have used the recursive relation in the last step,
\begin{align}
K_{n+1} (x) = K_{n-1} (x) + \frac{2 n}{x} K_n (x) \,.
\end{align}

\section{$3 \to 2$ annihilations}\label{app:3-2}

\begin{figure}[t!]
\begin{center}
\includegraphics[width=0.65\textwidth]{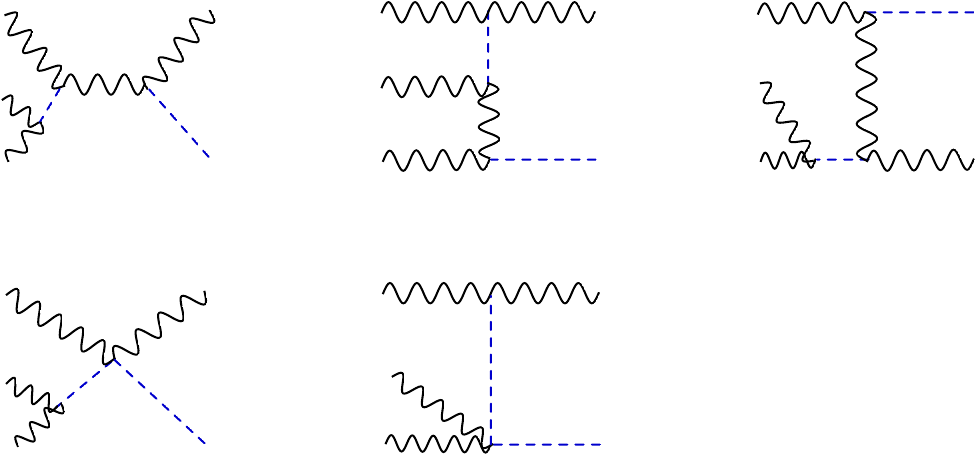}
\caption{Feynman diagrams for the annihilation $XXX\to XS$.  }
\label{fig:xxx2xs}
\end{center}
\end{figure}

\begin{figure}[t!]
\begin{center}
\includegraphics[width=0.65\textwidth]{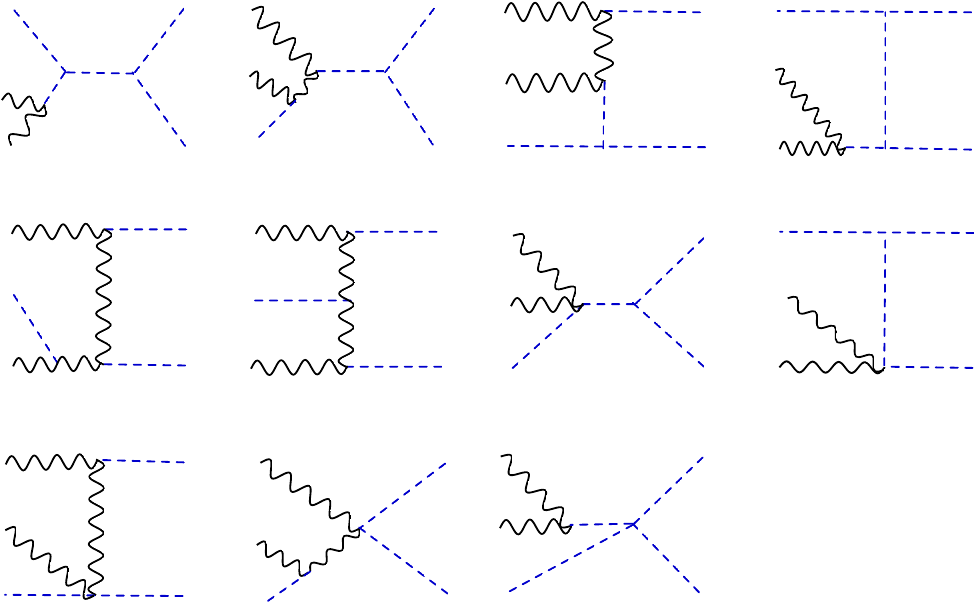}
\caption{Feynman diagrams for the annihilation $XXS\to SS$. }
\label{fig:xxs2ss}
\end{center}
\end{figure}

\begin{figure}[t!]
\begin{center}
\includegraphics[width=0.65\textwidth]{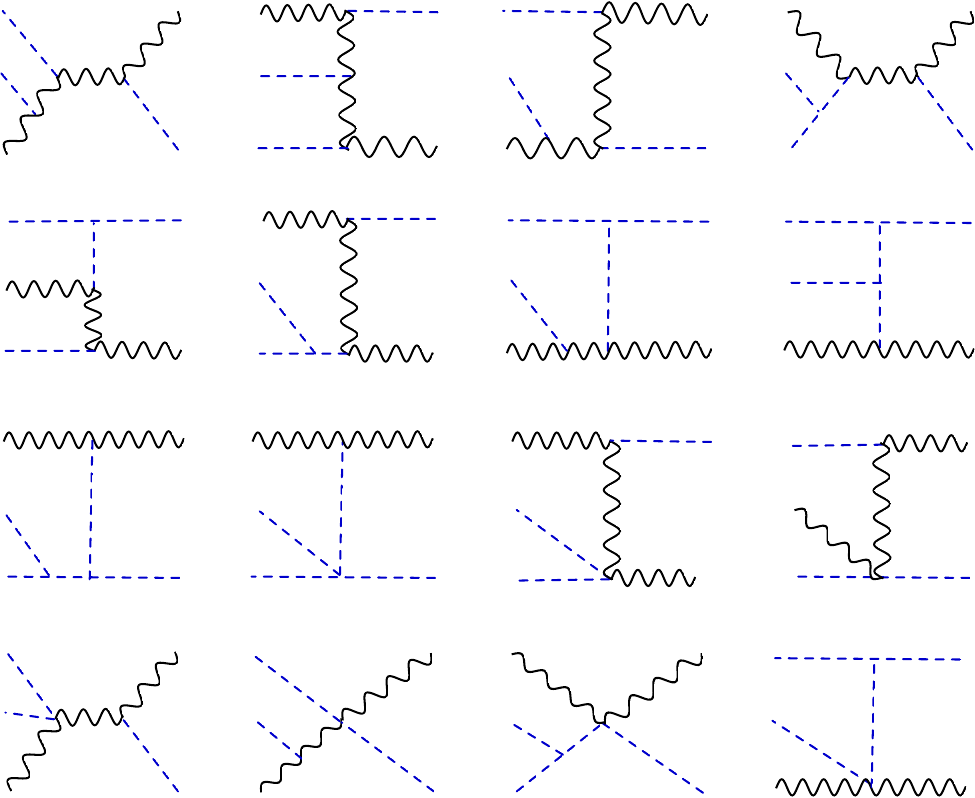}
\caption{Feynman diagrams for the annihilation $XSS\to XS$.  }
\label{fig:xss2xs}
\end{center}
\end{figure}

\begin{figure}[t!]
\begin{center}
\includegraphics[width=0.55\textwidth]{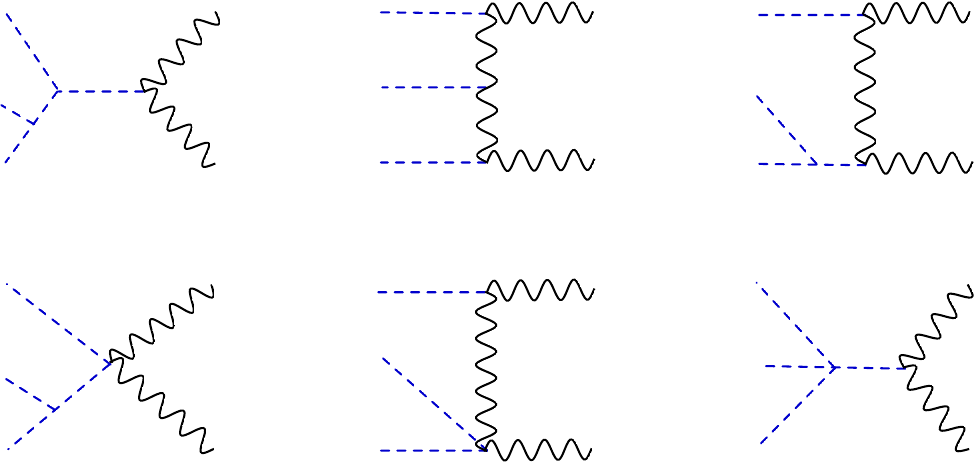}
\caption{Feynman diagrams for the annihilation $SSS\to XX$.  }
\label{fig:sss2xx}
\end{center}
\end{figure}

\begin{figure}[t!]
\begin{center}
\includegraphics[width=0.55\textwidth]{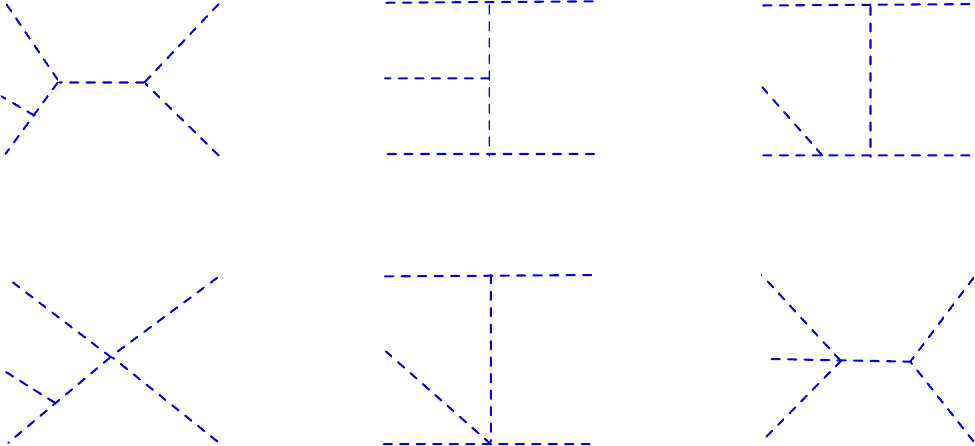}
\caption{Feynman diagrams for the annihilation $SSS\to SS$. }
\label{fig:sss2ss}
\end{center}
\end{figure}

We consider a thermally averaged cannibal annihilation cross section for $a (p_a)\, b (p_b)\,  c (p_c) \to d (p_d)\,  e(p_e)$, where all particles resides in a hidden sector and keep the same temperature, $T_X=T_S=T_h$, during the interaction. The generic form defined through this  paper is given by
\begin{align}
 \langle \sigma v^2 \rangle (T_h) & = 
  \frac{1}{m!}  \frac{1}{n_a^{\text{eq}}(T_h)  n_b^{\text{eq}}(T_h)  n_c^{\text{eq}}(T_h)  }
  \int   \frac{d^3 p_a}{(2\pi)^3 2 E_a}    \frac{d^3p_b}{(2\pi)^3 2E_b}   \frac{d^3p_c}{(2\pi)^3 2E_c}  
           \frac{d^3 p_d}{(2\pi)^3 2 E_d}    \frac{d^3p_e}{(2\pi)^3 2E_e}  
\nonumber\\
& ~ \times 
   (2\pi)^4 \delta^{(4)} (p_a +p_b +p_c -p_d -p_e)  |M|_{abc \to de}^2 
     e^{-(E_a + E_b +E_c)/T_h} 
\nonumber\\ 
&  =   \frac{1}{n_a^{\text{eq}}(T_h)  n_b^{\text{eq}}(T_h)  n_c^{\text{eq}}(T_h)  }
  \int   \frac{d^3 p_a}{(2\pi)^3 2 E_a}    \frac{d^3p_b}{(2\pi)^3 2E_b}   \frac{d^3p_c}{(2\pi)^3 2E_c}  
    \sigma v^2 e^{-(E_a + E_b +E_c)/T_h} 
\,,
\end{align}
where $m\equiv 2$ if the final state particles, $d$ and $e$, are the same particle species, otherwise $m\equiv 1$. Here, the sum for the amplitude squared, $|M|_{abc \to de}^2$,  has 
been taken over all internal degrees of freedom of the initial and final states.
In the nonrelativistic limit, $E_a\approx m_a, E_b\approx m_b, E_c\approx m_c$, the cross section is approximately given by \cite{Berlin:2016gtr}
\begin{align}
\sigma v^2 =\frac{ \left[ (m_a +m_b +m_c)^4 -2 (m_a +m_b +m_c)^2 (m_d^2 + m_e^2) + (m_d^2 -m_e^2)^2 \right]^{1/2}}
{m! \, 64\,  \pi m_a m_b m_c (m_a +m_b +m_c)^2}  \overline{|M|^2}_{abc \to de}   \,,
\end{align}
where $\overline{|M|^2}_{abc \to de} $ is the amplitude squared but with the initial state spin-averaged.
To calculate $3\to 2$ thermally averaged annihilation cross sections for the nonrelativistic hidden sector particles with a temperature below their masses, i.e.,   $T_h <m_{S,X}  $, we take the low-velocity approximation, $\langle \sigma v^2\rangle \simeq \sigma v^2$ and neglect its subleading corrections of order $T_h/m_{X,S}$. We show the diagrams in Figs.~\ref{fig:xxx2xs}, \ref{fig:xxs2ss}, \ref{fig:xss2xs}, \ref{fig:sss2xx}, and \ref{fig:sss2ss}, and summarize all the relevant results as below,
\begin{align}
\langle \sigma v^2 \rangle_{XXX\to XS}  & \simeq \frac{[(16m_X^2-m_S^2) (4m_X^2 -m_S^2)]^{1/2}}{1152 \pi m_X^7 } 
 2 g_{\rm dm}^6 m_X^2 \nonumber  \\
  &  \times \Bigg(
  \frac { m_S^4} {108 m_X^6} - \frac {7  m_S^2} {18  m_X^4} + \frac {1007 } {108 m_X^2}  + \frac {228 } {m_S^2 + 2 m_X^2} 
     - \frac {324   m_S^2} {\left (m_S^2 + 2 m_X^2 \right)^2} + \frac {27
        m_S^2} {64 (2 m_X - m_S)^4} \nonumber \\
    &+ \frac {27   m_S^2} {64 (m_S + 2 m_X)^4} + \frac {6937 } {384  m_S (2 m_X - m_S)} 
      - \frac {6937 } {384 m_S (m_S + 2 m_X)}  \nonumber\\
      & + \frac {1037} {128 (2  m_X - m_S)^2} + \frac {1037 } {128 (m_S + 2  m_X)^2} 
      - \frac {3  m_S} {8 (2  m_X - m_S)^3} + \frac { 3   m_S} {8 (m_S + 2 m_X)^3} \Bigg)
      \,, \nonumber \\
  \end{align}
\begin{align}
 \langle \sigma v^2 \rangle_{XXS\to SS}  & \simeq \frac{[(2m_X-m_S) (2m_X +3 m_S)]^{1/2}}{128 \pi m_X^4 m_S (2m_X +m_S)} 
  g_{\rm dm}^6  m_X^2 \nonumber \\
& \times\Bigg( \frac{1971   m_S^2}{16 m_X^4}+\frac{172 m_S^2}{243 (m_S+m_X)^4}
  +\frac{64   m_S^2}{243 (2 m_X-m_S)^4}+\frac{176  }{9 m_S^2}-\frac{371 m_S}{8 m_X^3} 
   \nonumber\\
&   -\frac{15619  }{36 m_S m_X}+\frac{1387627  }{26244 m_S  (m_S+m_X)}
     -\frac{6586685  }{6561 m_S (2 m_X-m_S)} +\frac{15541  }{9 m_S (m_S+2 m_X)}
   \nonumber\\
  & +\frac{528025  }{34992 (m_S+m_X)^2}+\frac{408049}{2187 (2 m_X-m_S)^2}+\frac{3707  }{9 (m_S+2
   m_X)^2}+\frac{1961   m_S}{729(m_S+m_X)^3}
    \nonumber\\
  & -\frac{8656   m_S}{729 (2 m_X-m_S)^3} +\frac{8529  }{16 m_X^2}  
   \Bigg) \,, 
  \end{align}
\begin{align}
  \langle \sigma v^2 \rangle_{XSS\to XS}  & \simeq \frac{ [3(2m_X+m_S) (2m_X +3 m_S)]^{1/2}}{128 \pi m_X^3 m_S (m_X +2m_S)^2} 
2 g_{\rm dm}^6  m_X^2
    \nonumber\\
&  \Bigg(  \frac {2187  m_S^2} {16 m_X^4} + \frac {16  (205 m_S + 191 m_X)} { 3 m_S\left (- m_S^2 + 2 m_S m_X +  2 m_X^2 \right)} 
    + \frac {96 \left (2  m_S^2 +   m_S m_X \right)} {\left (- m_S^2 + 2 m_S m_X +   2 m_X^2 \right)^2}  \nonumber\\
 &   + \frac { 44  m_S^2} {(  m_S + m_X)^4} + \frac {36  m_S^2} {(   m_S + 2 m_X)^4} 
    - \frac {891  m_S} {2 m_X^3} - \frac { 15785 } {8 m_S m_X} + \frac {15585 } {4 m_S (  m_S + m_X)}  \nonumber \\
 &   + \frac { 295 } {72 m_S (2 m_S +   m_X)} - \frac {43858 } {9 m_S (   m_S + 2 m_X)}  
    + \frac {4967 } {4 (  m_S + m_X)^2} + \frac {25 } {3 (2 m_S +    m_X)^2} 
    \nonumber\\
   & + \frac {12657 } {4 (   m_S + 2 m_X)^2} 
    + \frac { 278  m_S} {( m_S + m_X)^3} - \frac {650  m_S} {(  m_S + 2 m_X)^3} + \frac {2451 } {4 m_X^2}\Bigg) \,,
 \end{align}
\begin{align}
\langle \sigma v^2 \rangle_{SSS\to XX} & \simeq \frac{\sqrt{9-4m_X^2/m_S^2}}{384 \pi m_S^3 m_X^2}  g_{\rm dm}^6 m_X^2
 \nonumber\\
  & \times \Bigg(  \frac {432  m_X^6} {m_S^8} - \frac {2160 m_X^4} {m_S^6} + \frac {729  m_S^4} {16 m_X^6} 
  + \frac {3672  m_X^2} {m_S^4} + \frac {891   m_S^2} {4 m_X^4} - \frac {1944 } {m_S^2} - \frac {729  } {4 m_X^2}
  \Bigg) \,, \\
 \langle \sigma v^2 \rangle_{SSS\to SS}  & \simeq \frac{\sqrt{5}}{384 \pi} \frac{18225}{16 m_S^5} \Bigg(\frac{g_{\rm dm} m_S}{m_X}\Bigg)^6  
    \,. 
\end{align}


\section{The kinetic decoupling of the DM from the hidden scalar}\label{app:xs-kd}

In this Appendix, we discuss the epoch that the DM and hidden scalar starts to be kinetically decoupled from each other.
For this period of time, which is well after $3\to 2$ annihilation decoupling,  the DM temperature follows the Boltzmann equation,
\begin{align}
&  \frac{d T_X}{dt}  + (2 - \delta_H^X) H T_X 
= \frac{1}{n_X(T_X)} \bigg[ - \left( \frac{d n_X(T_X) }{dt} +3 H n_X (T_X) \right) T_X
+ g_X \int d\Pi_X \, C \Big[f_X \cdot \frac{{\bf p}_X^2}{3 E_X} \Big]   \bigg]
\nonumber\\
&
\simeq    (2 - \delta_H^X) \gamma_X (T_S - T_X) 
 + T_X \bigg[ 
  \langle \sigma v \rangle_{XX\to SS}(T_X) \,  n_X (T_X) 
 -    \langle \sigma v \rangle_{XX\to SS}(T_S)  \frac{(n_X^{\rm eq} (T_S) )^2}{n_X (T_X)}  \frac{ n_S ^2(T_S)}{ (n_S^{\rm eq} (T_S) )^2 }  
   \bigg]
 \nonumber\\
&\quad\,  -
  \langle \sigma v \cdot  \frac{{\bf p}_X^2}{3E_X} \rangle_{XX\to SS}(T_X) \,  n_X (T_X) 
 +   \langle \sigma v \cdot  \frac{{\bf p}_X^2}{3E_X} \rangle_{XX\to SS}(T_S)  \frac{(n_X^{\rm eq} (T_S) )^2}{ n_X(T_X)} \frac{ n_S^2 (T_S) }{ (n_S^{\rm eq} (T_S) )^2 }  
  \,,
 \label{app:boltz-t-X}
\end{align}
where $\delta_H^X \approx 1$ is the same form as $\delta_H$ (defined in Eq.~(\ref{eq:deltah})) but with $T_S$ replaced by $T_X$, and  the $XS \to XS$ momentum relaxation rate is given by 
\begin{align}
\gamma_X (T_S) 
& =
 \frac{1}{768 g_X \pi^3 m_X^3 T_S} \int _{m_S}^\infty d E_S \,   f_S (T_S)  \, (1 + f_S (T_S))
\int_{-4 {\bf p}_S^2}^0 dt (-t)  |M_{X S \to X S }|^2  \nonumber\\
& \simeq 
   \mathop{\hspace{-10ex}  |M_{X S \to X S }|^2_{t=0} }_{\hspace{11.5ex}s=m_X^2+2m_X m_S+m_S^2}
     \frac{1}{12 \pi^3 g_X}  \frac{m_S^2}{m_X} \frac{T_S^2}{m_X^2} 
  \bigg( 1 + 3 \frac{T_S}{m_S} + 3 \frac{T_S^2}{m_S^2} \bigg)  e^{-m_S/T_S}
\,.  \label{app:gammaX}
\end{align}
Based on the fact that $-t \ll 4 |{\bf p}|^2 \sim T \ll m_X^2, m_S^2$, we neglect $t$ in the calculation of the amplitude squared to obtain the approximate form of $\gamma_X$.  Thus, the elastic scattering amplitude squared and summed over all internal degrees of freedom of initial and final spin states is given by
\begin{align}
 |M_{X S\to X S}|^2 
& \simeq   
   \mathop{\hspace{-10ex}  |M_{X S \to X S }|^2_{t=0} }_{\hspace{11.5ex}s=m_X^2+2m_X m_S +m_S^2 } \nonumber\\
& \simeq
  4 g_X g_{\rm dm}^2 \cos^2\alpha 
   \frac{ \big( g_{\rm dm}  \cos\alpha  \, m_S^4 - g_{SSS} m_X (4m_X^2 -m_S^2) \big)^2}{ m_S^4 (4m_X^2 -m_S^2)^2} \,.
\end{align}
Here the elastic scattering process $X S\to X S$ contains the amplitudes with a hidden scalar $S$ mediated in the $t$-channel and with $X$ in the $s/u$-channel. The contribution is dominated by the former one. We define the DM kinetic decoupling temperature $T_X^{\rm kd}$ below which the kinetic energy injection rate transferred by the elastic scattering $ X S\to X S$ and/or by the annihilation $SS\to XX$  to the DM falls below the diluting rate arising from the Hubble expansion. $T_X^{\rm kd}$ approximately satisfies
\begin{align}
(2-\delta_H^X) \gamma_X (T_S) \,  n_X(T_X)\, T_S 
& +  \langle \sigma v \rangle_{XX\to SS}(T_S)   \frac{ (n_X^{\rm eq} (T_S) )^2}{ (n_S^{\rm eq} (T_S) )^2}  \, (n_S (T_S) )^2 \, T_S 
\nonumber\\
& \simeq  (2-\delta_H^X)  H(T) \,  n_X(T_X)\,  T_X
 -  \langle \sigma v \rangle_{XX\to SS}(T_X)  \,  n_X^2 (T_X) \, T_X \,,
\end{align}
where  the approximation $ \langle \sigma v \cdot  \frac{{\bf p}_X^2}{3E_X} \rangle_{XX\to SS}(T_S) \simeq  T_S\, \langle \sigma v \rangle_{XX\to SS} (T_S)$ with an error less than 10\% is used for the present nonrelativistic s-wave annihilation, and the second term of the RHS arises from the contribution of $XX\to SS$ to $\big(d n_X(T_X) /dt +3 H n_X (T_X) \big) T_X$ for $T<T_f$.  
Here, before kinetic decoupling, $T_{X}=T_S$ are functions of $T$ and can be determined from solving the Boltzmann moment equations. The DM temperature after kinetic decoupling satisfies $T_X(a)\simeq T_X^{\rm kd} \cdot (a_X^{\rm kd}/a)^2$, where $a$ is the cosmic scale factor and $a_X^{\rm kd}$ is its corresponding value at $T_X=T_X^{\rm kd}$. The $T_X/T$ evolution after kinetic decoupling is sketched as the red line in the left panel of Figs.~\ref{fig:relic-not-equal-1}, \ref{fig:relic-not-equal-2},  and \ref{fig:relic-equal},
where  $ T_X^{\rm kd} \equiv m_X/x_X^{\rm kd}$ is depicted as the red dot.

Similarly, if we consider the temperature below that the DM and hidden scalar are decoupled from each other, the following terms need to be included in the RHS of Eq.~(\ref{eq:ts-t}), which is the temperature evolution equation of $S$,
\begin{align}
& \frac{1}{ n_S(T_S)}  \left[- T_S \left( \frac{d n_S(T_S) }{dt} +3 H n_S (T_S) \right)_{SS\leftrightarrow XX} 
+ g_S \int d\Pi_S \, C \Big[f_S\cdot \frac{{\bf p}_S^2}{3 E_S} \Big]_{XS\leftrightarrow XS, SS\leftrightarrow XX} \right]
\nonumber\\
&=  
  T_S \bigg[ 
    \langle \sigma v \rangle_{XX\to SS}(T_S)  (n_X^{\rm eq} (T_S) )^2  \frac{ n_S (T_S)}{ (n_S^{\rm eq} (T_S) )^2 }  
 -  \langle \sigma v \rangle_{XX\to SS}(T_X) \,  \frac{n_X^2 (T_X) }{ n_S(T_S)} 
   \bigg]
  + (2 -\delta_H) \gamma_S (T_X - T_S)
 \nonumber\\
&\quad\,  -
  \langle \sigma v \cdot  \frac{{\bf p}_S^2}{3E_S} \rangle_{SS\to XX}(T_S) \,  n_S (T_S) 
 +   \langle \sigma v \cdot  \frac{{\bf p}_S^2}{3E_S} \rangle_{SS\to XX}(T_X)  \frac{ (n_S^{\rm eq} (T_X) )^2 }{n_S(T_S)}  \frac{ n_X^2 (T_X) }{ (n_X^{\rm eq} (T_X) )^2 } 
 \nonumber\\
& \overset{\text{\tiny  $n_X(T_X) \gg n_S(T_S)$}}{\simeq} 
     - (T_S-T_X)   \langle \sigma v \rangle_{XX\to SS}(T_X) \,  \frac{n_X^2 (T_X) }{ n_S(T_S) }
     - (2-\delta_H)  \gamma_S (T_S -T_X)
 \,, \label{app:kd-xs}
\end{align}
where  the $XS \to XS$ momentum relaxation rate is given by 
\begin{align}
\gamma_S (T_X) 
 \simeq &
  \frac{m_X^4}{m_S^3} g_{\rm dm}^2 \cos^2\alpha     
       \frac{T_X^2}{m_X^2} 
   \frac{ \big( g_{\rm dm}  \cos\alpha  \, m_S^4 - g_{SSS} m_X (4m_X^2 -m_S^2) \big)^2}{ m_S^4 (4m_X^2 -m_S^2)^2}
  \nonumber \\
 &  \times   \bigg( 1 + 3 \frac{T_X}{m_X} + 3 \frac{T_X^2}{m_X^2} \bigg)  e^{-m_X/T_X}
     \,,
\end{align}
and the following approximations have been used in the last step: 
\begin{align}
\delta_H \simeq 1,  \qquad
 \langle \sigma v \cdot  \frac{{\bf p}_S^2}{3E_S} \rangle_{SS\to XX}(T_S)  (n_S^{\rm eq} (T_S) )^2
 \simeq 
 T_S \langle \sigma v  \rangle_{XX \to SS}(T_S)  (n_X^{\rm eq} (T_S) )^2 \,.
\end{align}

\end{document}